\documentclass[review]{elsarticle}
\usepackage{lineno,hyperref}
\usepackage{color}
\usepackage{amsmath}
\usepackage{diagbox}

\usepackage[font=footnotesize,labelfont=bf,labelsep=endash]{caption}

\newcounter{chem}
\newcounter{temp}

\usepackage{mhchem}

\newenvironment{chequation}{%
  \setcounter{temp}{\value{equation}}%
  \setcounter{equation}{\value{chem}}%
}{%
  \setcounter{chem}{\value{equation}}%
  \setcounter{equation}{\value{temp}}%
}

\usepackage{longtable}
\usepackage{subcaption}
\usepackage{graphicx}

\usepackage[english]{babel}

\modulolinenumbers[5]

\journal{Combustion and Flame}

\begin{document}

\begin{frontmatter}

\title{Strain Rate and Pressure 
Effects on Multi-branched Counterflow Flames}

\author{Claudia-F. L\'{o}pez-C\'{a}mara$^{1,*}$}
\cortext[mycorrespondingauthor]{Corresponding author}
\ead{clopezca@uci.edu}
\author{Albert Jord\`{a} Juan\'{o}s$^{2}$}
\author{William A. Sirignano$^{2}$}
\address{$^{1}$Civil and Environmental Engineering, University of California, Irvine, CA 92697, USA}
\address{$^{2}$Mechanical and Aerospace Engineering, University of California, Irvine, CA 92697, USA}

\begin{abstract}
This study presents methane-air counterflow simulations, in computationally efficient similar form, allowing combustible mixtures to flow from one or both directions in order to learn more about multi-branched propagating flame structures (e.g., a triple flame). These structures with both premixed and non-premixed flames are commonly seen in more practical combustion analyses. A range of realistic mass mixture fractions and asymmetric chemical rate laws are examined while avoiding the commonly forced unreal symmetric behavior with one-step second-order kinetics. Moreover, a survey of critical parameters is performed varying pressure and normal strain rate to define the flame structure and detect different characters. Three flames can co-exist if the strain rate is low enough and the pressure is high enough. However, at higher strain rate and/or lower pressure, only one or two flames might be obtained. Negative regions of heat release rate are observed and linked to potential endothermic reactions. With a rich premixed mixture at low strain rates and pressures, high exothermic reactions producing CO$_2$ and H$_2$O, and consuming CO and H$_2$ causes a heat-release-rate peak. Unexpected character of the lean and rich premixed flames is observed, leading to the conclusion that these flames are diffusion controlled.
\end{abstract}


\end{frontmatter}


\section{Introduction}
Flamelet models are attractive for modelling some subgrid behavior with reasonable fidelity for large-eddy simulations \cite{Nguyen2018,Nguyen_WAS2018}. Due to prohibitive computational costs, the typical turbulent combustor analysis with its inherently multiscale behavior cannot use direct numerical simulation to describe the flame behavior. While some success has been achieved with flamelet modelling in analyzing experimental combustors, there is need for improvement in the basic model. Current practice in the sub-grid modelling uses flamelets with only a single diffusion flame. Yet, we know from experience that multi-branched flames can occur. We need therefore advancement of the theory to address multi-branched flames undergoing strain over a wide range of pressure. Merging of the branches under high strain rates is also of interest in practical applications.\hfill \break
\indent Flamelet studies provide insights about common behaviors that can be observed in more convoluted turbulent flames. Flamelets subjected to high strain rates have been studied with findings of multi-branched flames and also merging of branches of triple flames \cite{Nguyen2018,Nguyen_WAS2018}. Triple flames are described as tri-brachial structures composed by a fuel-lean premixed flame, a fuel-rich premixed flame, and a diffusion flame derived from the unburned reactants \cite{Phillips,Dold,Buckmaster_1988}. These flames are crucial in several combustion applications such as ignition and extinction processes of non-premixed systems \cite{Reveillon}.\hfill \break

Rajamanickam et al. numerically investigated the influence of stoichiometry on strained triple flames in counterflow configuration \cite{RAJAMANICKAM}. They used a one-step chemical kinetic model with constant density and constant transport properties. Their main conclusion was that the classical tri-branchial structure might not be observed in some cases, especially at high equivalence ratio. However, that study and the majority of studies involving multi-branched flames are performed using global chemical kinetic models \cite{Dold,Hartley_1991,KIONI, Ruetsch_1995,JIMENEZ_2007}, which might not be sufficient to predict an accurate behavior of these flames as shown by previous authors \cite{Law_2015_reviewer1}. Use of unity reaction orders for triple flame studies with global kinetics are also common and result in unrealistically symmetric triple flames \cite{KIONI,Ruetsch_1995,AJJ_WAS}. They are also unable to predict flame-branch merging accurately.\hfill \break

The current study, however, considers all the possible scenarios referred to the mixture ratio of the reactants that could lead to a multi-branched flame. In particular, three cases are considered: (i) lean premixed mixture coming from one side and non-reactive mixture with fuel coming from the other side; (ii) rich premixed mixture from one side and non-reactive mixture with oxygen from the other side; and (iii) lean premixed mixture coming from one side and rich premixed mixture coming from the other side.\hfill \break

Therefore, this work advances further with examination of the flames that co-exist at different mixture ratios and pressures when a variable normal strain rate is applied in a transverse direction to the flame. In contrast to previous studies, the numerical simulations herein employ a detailed chemical kinetic model. The counterflow-burner configuration is used since it has features relevant to the downstream behavior of triple flames.

\section{Model and analysis}
\subsection{Chemical kinetic model and software}
Results are obtained using the opposed-flow flame module (OPPDIF) from Chemkin-Pro\textsuperscript{\textregistered} software, for all studied cases of counterflow flames in axisymmetric configuration. Individual specific heats and enthalpies are calculated from temperature polynomial fits. Chemkin-Pro\textsuperscript{\textregistered} offers two detailed formulations to estimate the transport properties: mixture-averaged and multicomponent. The latter is deemed superior and chosen herein. The ideal gas law is used as the equation of state. For more details on the formulation for thermo-physical and transport properties, see the Chemkin-Pro\textsuperscript{\textregistered} theory manual \cite{Chemkin_manual}. \hfill \break

Two parameters control the adaptive grid refinement in Chemkin-Pro\textsuperscript{\textregistered} based on the solution curvature and gradient. They are both set to 0.5. Another two parameters control the solution convergence. They are the absolute and relative tolerances, which are set to 10$^{-9}$ and 10$^{-4}$, respectively.\hfill \break 


The chemical kinetic model here is the most recent version of the San Diego Mechanism, which was last updated in 2018 to include two new reactions related to CHCHO production and consumption. The San Diego Mechanism includes a total of 270 reactions and 58 species \cite{San_Diego_mech}.

\subsection{Boundary conditions}
Boundary conditions are prescribed for the species mass fractions, temperature, pressure, and inlet velocities for each nozzle (Tables \ref{tab:Chemkin_Conditions} and \ref{tab:Chemkin_Cases}). The selected equivalence ratios ensure to deliver the outcomes for the interest regimes of this study, which are lean flame (Case 1), rich flame (Case 2) and both lean and rich flames (Case 3). The distance between the two nozzles is 2 cm. It is chosen based on the literature \cite{JORDA_2017}, to prevent flame proximity to the boundaries from affecting too heavily the solution -- i.e. zero gradients are desired at boundaries on species mass fraction and temperature curves -- while minimizing the computational cost.\hfill

\begin{table}[h]
	\centering
	\caption{Boundary conditions.}
	\label{tab:Chemkin_Conditions}
	\begin{tabular}{cccccl}
		\cline{2-5}
		& Inlet (left) & Inlet (right)     & Strain {($S$, [}s$^{-1}${])} & \\ \cline{1-5}
		& 0.1          & 0.1            & 10         &  \\
		Velocities {($V$, [}m/s{]}): 4 cases   & 0.5          & 0.5             & 50         &  \\
		& 1          & 1             & 100         &  \\
        & 1.5          & 1.5             & 150         &  \\
		Temperature {($T$, [}K{]}) & 298          & 298             & --    &  \\
		Pressure ({$P$, [}atm{]})  & 1 up to 20       & 1 up to 20  & -- & \\ \cline{1-5}
	\end{tabular}
	
\end{table}

The global strain rate $S$ is calculated following the expression $S$ = 2 $V_{left}/L$, where $V_{left}$ is the velocity at the left inlet nozzle, and $L$ is the distance between the two nozzles.

\begin{table}[ht!]
	\centering
	\caption{Species boundary conditions. Mass fractions and equivalence ratios ($\phi$).}
	\label{tab:Chemkin_Cases}
	\begin{tabular}{c|cc|cc|cc}
		\cline{2-7}
		& \multicolumn{2}{c|}{Case 1}    & \multicolumn{2}{c|}{Case 2}    & \multicolumn{2}{c}{Case 3}    \\\cline{1-7} 
		Nozzle:	& Left          & Right        & Left           & Right        & Left           & Right        \\ 
		Mixture Type:	& Fuel Lean          & One Reactant        & Fuel Rich           & One Reactant        & Fuel Lean           & Fuel Rich \\
		& $\phi$=0.34 & & $\phi$=5.70 & & $\phi$=0.34 & $\phi$=5.70
		\\\hline
		Methane (CH$_{4}$) & 0.02           & 0.25         & 0.25           & 0   & 0.02 & 0.25         \\
		Nitrogen (N$_{2}$)  & 0.7546         & 0.75         & 0.5775          & 0.77         & 0.7546         & 0.5775         \\
		Oxygen (O$_{2}$)  & 0.2254         & 0            & 0.1725         & 0.23         & 0.2254         & 0.1725         \\ \hline
	\end{tabular}
\end{table}

\section{Results and Discussion}
This section is divided in three sub-sections that correspond to each of the three studied cases according to Table \ref{tab:Chemkin_Cases}.\hfill \break

The non-orthodox variable $\xi$ is employed to separate the heat-release-rate peaks while normalizing the x-axis (Eq. \ref{eq:x-axis}). In Equation \ref{eq:x-axis}, $s.p$ stands for \textit{stagnation plane} and $\chi$ is a dummy variable, used only as integration limit, and corresponding to distance measured from the left nozzle $x$. In the plots, the stagnation plane location is marked by a green dashed vertical line and placed at zero. This should help visualize the position of the flames with respect to the stagnation plane and the mixing-layer thickness (denoted as $\delta$ in Eq. \ref{eq:MLT}). The range of $\xi$ magnitudes in the horizontal axis of each plot is unity, making them comparable visually even though their stagnation planes are shifted.


\begin{equation}
\label{eq:x-axis}
\xi=\frac{\int_0^\chi Tdx}{\int_0^L Tdx}- \frac{\int_0^{s.p} Tdx}{\int_0^L Tdx}
\end{equation}

\begin{equation}
\label{eq:MLT}
\delta\simeq\sqrt{{D_T}/{S}}
\end{equation}

The mixing-layer thickness is estimated using the expression in Equation \ref{eq:MLT}, where the thermal diffusivity $D_T$ is taken from the reacting mixture (Cases 1 and 2) or from the average of thermal diffusivities of both reacting mixtures (Case 3). RefProp \cite{LEMMON-RP91} is employed to calculate the properties of each mixture at the different pressures of study. Doing so, the Prandtl number in all cases is determined to be between 0.7 and 0.72. The definition of $\delta$ is arbitrary and results will show that significant mixing for both premixed and non-premixed flame structures can occur outside the indicated bounds.

	
	

For the following discussion, it is important to consider Tables \ref{tab:Adiabatic_flame_T} and \ref{tab:Adiabatic_flame_T_flame_front_Case1}, which show the computed adiabatic flame temperatures at stoichiometric mixture ratio, as well as for the relevant equivalence ratios for this study.\hfill \break

\begin{table}[ht!]
	\centering
	\caption{Adiabatic flame temperatures [K] for relevant mixtures at various pressures with initial temperature of 298 K. Upstream equivalence ratios ($\phi$).}
	\label{tab:Adiabatic_flame_T}
\begin{tabular}{ccccc}
\hline
\multicolumn{ 1}{c}{$P$ [atm]} &  {\bf Stoichiometric mixture} &{\bf Case 1} & {\bf Case 2} \\

\multicolumn{ 1}{c}{} &$\phi$=1 &  $\phi$=0.34 &    $\phi$=5.70 \\
\hline
         1 &  2224.54    & 1152.33 &      884.9 \\
\hline
        10 & 2267.12   &   1152.33 &     991.45 \\
\hline
        20 &  2276.6   &  1152.33 &    1026.86 \\
\hline
\end{tabular} 
\end{table}

\begin{table}[ht!]
	\centering
	\caption{Case 1. Local equivalence ratios at the premixed flame front ($\phi$) and their corresponding adiabatic flame temperatures ($T_{\text{adiabatic}}$ [K]) calculated with initial temperature of 298 K. Temperatures at the premixed flame heat-release-rate peak ($T_{\text{premixed flame}}$ [K]).}
	\label{tab:Adiabatic_flame_T_flame_front_Case1}
\begin{tabular}{c|cc|c|c|c|c}
\cline{2-7}
                       & \multicolumn{6}{c}{\textbf{Case 1}}                                                   \\ \hline
$P$ {[}atm{]}          & \multicolumn{2}{c|}{1}              & \multicolumn{2}{c|}{10} & \multicolumn{2}{c}{20} \\ \hline
$S$ {[}s$^{-1}${]}     & \multicolumn{1}{c|}{10}    & 50    & 10         & 50         & 10         & 50         \\ \hline
$\phi$    & \multicolumn{1}{c|}{0.17}  & 0.22  & 0.21       & 0.15       & 0.20       & 0.17       \\ \hline
T$_{\text{adiabatic}}$ & \multicolumn{1}{c|}{758.3} & 882.7 & 851.6      & 712.1      & 820.1      & 760.8      \\ \hline
T$_{\text{premixed flame}}$ & \multicolumn{1}{c|}{1363.9} & 1351.3 & 1525.8      & 1601.6      & 1521.5      & 1653.1      \\ \hline
\end{tabular}
\end{table}

\begin{table}[ht!]
	\centering
	\caption{Case 2. Local equivalence ratios at the premixed flame front ($\phi$) and their corresponding adiabatic flame temperatures ($T_{\text{adiabatic}}$ [K]) calculated with initial temperature of 298 K. Temperatures at the premixed flame heat-release-rate peak ($T_{\text{premixed flame}}$ [K]).}
	\label{tab:Adiabatic_flame_T_flame_front_Case2}
\begin{tabular}{c|cccccc}
\cline{2-7}
                       & \multicolumn{6}{c}{\textbf{Case 2}}                                                                                                                         \\ \hline
$P$ {[}atm{]}          & \multicolumn{2}{c|}{1}                                    & \multicolumn{2}{c|}{10}                                   & \multicolumn{2}{c}{20}              \\ \hline
$S$ {[}s$^{-1}${]}     & \multicolumn{1}{c|}{10}    & \multicolumn{1}{c|}{50}     & \multicolumn{1}{c|}{10}     & \multicolumn{1}{c|}{50}     & \multicolumn{1}{c|}{10}     & 50     \\ \hline
$\phi$    & \multicolumn{1}{c|}{3.23}  & \multicolumn{1}{c|}{3.07}   & \multicolumn{1}{c|}{3.58}   & \multicolumn{1}{c|}{3.11}   & \multicolumn{1}{c|}{4.14}   & 3.34   \\ \hline
T$_{\text{adiabatic}}$ & \multicolumn{1}{c|}{980.5} & \multicolumn{1}{c|}{1009.6} & \multicolumn{1}{c|}{1065.3} & \multicolumn{1}{c|}{1102.1} & \multicolumn{1}{c|}{1077.5} & 1121.2 \\ \hline
T$_{\text{premixed flame}}$ & \multicolumn{1}{c|}{1577.0} & \multicolumn{1}{c|}{1726.2} & \multicolumn{1}{c|}{1528.3}      & \multicolumn{1}{c|}{1654.7}      & \multicolumn{1}{c|}{1168.0}      & 1619.6     \\ \hline
\end{tabular}
\end{table}

\subsection{Case 1}
A fuel-lean mixture is injected from the left nozzle while diluted fuel enters the domain from the right side. Two flames are expected a priori: one premixed flame burning all the fuel from the left and one diffusion flame burning the leftover oxidizer from the left with the fuel from the right. Results are shown in Figure \ref{fig:SDMech_298K_case1_species}. Pressure and strain rate effects are described in the next two Subsections \ref{Subsection:case1_increaseP} and \ref{Subsection:case1_increaseS}.

\subsubsection{Pressure Effects}
\label{Subsection:case1_increaseP}
\begin{figure}[h!]
	\centering
	\includegraphics[width=.45\textwidth]{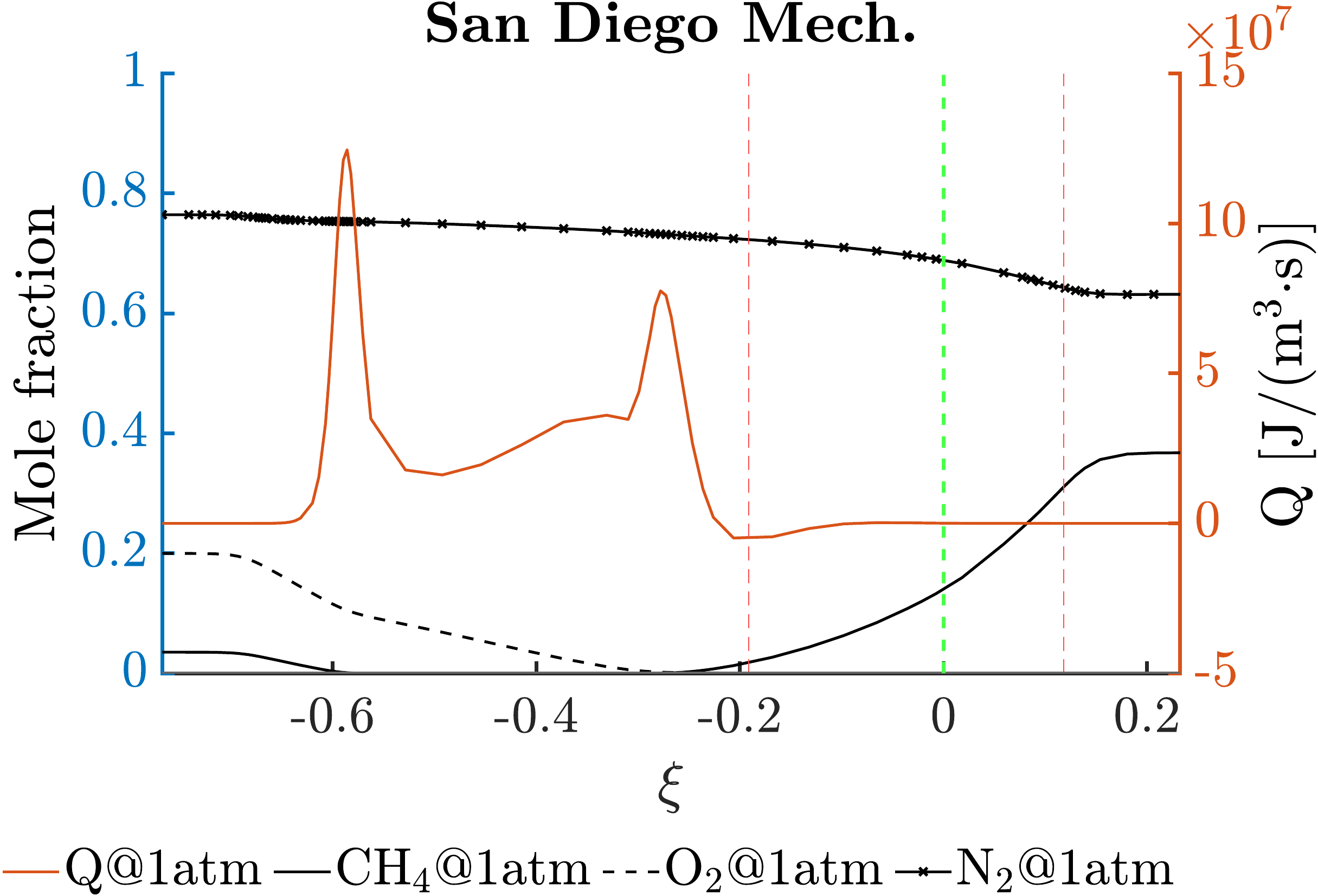}\quad
	\includegraphics[width=.45\textwidth]{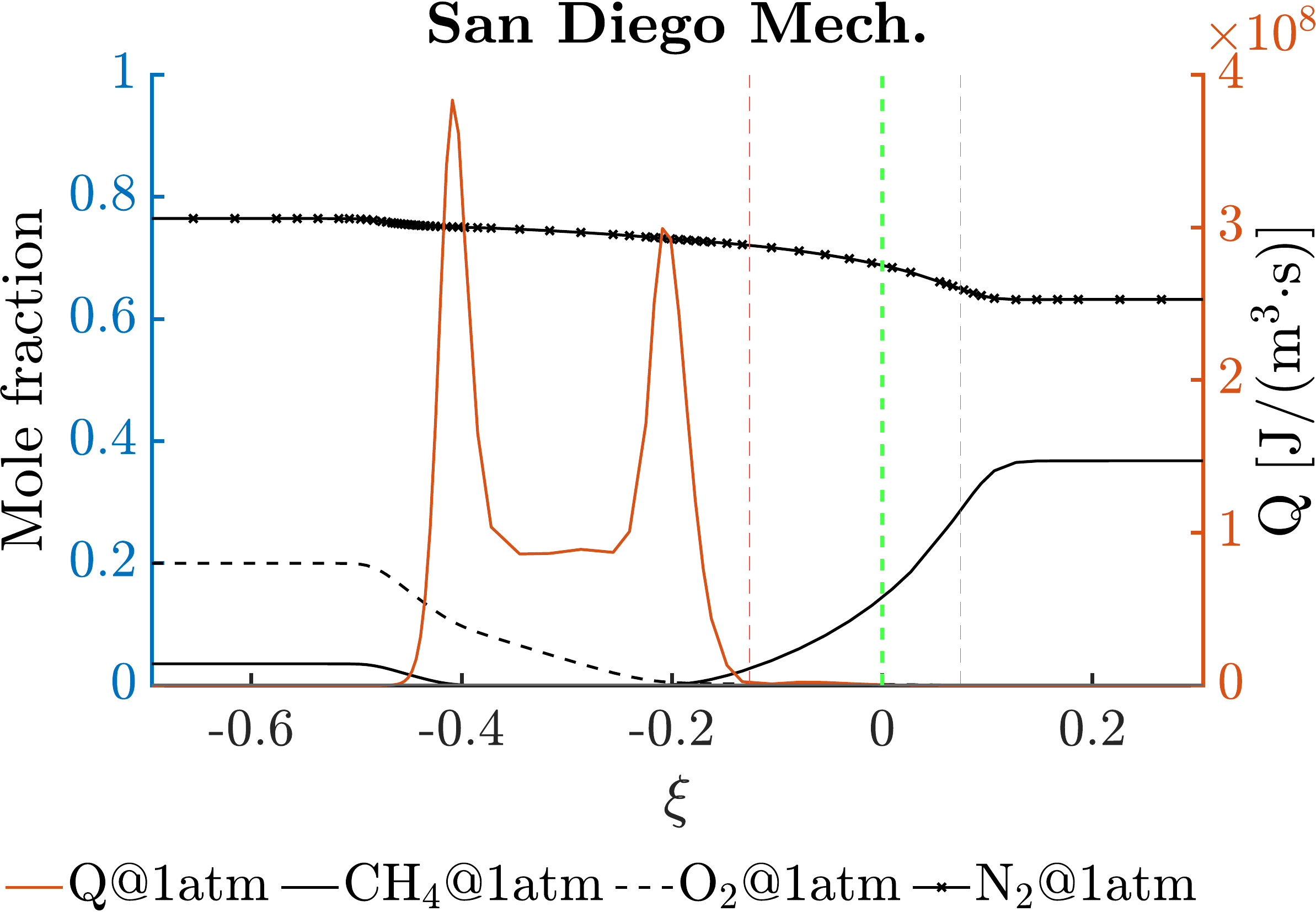}
	
	\medskip
	
	\includegraphics[width=.45\textwidth]{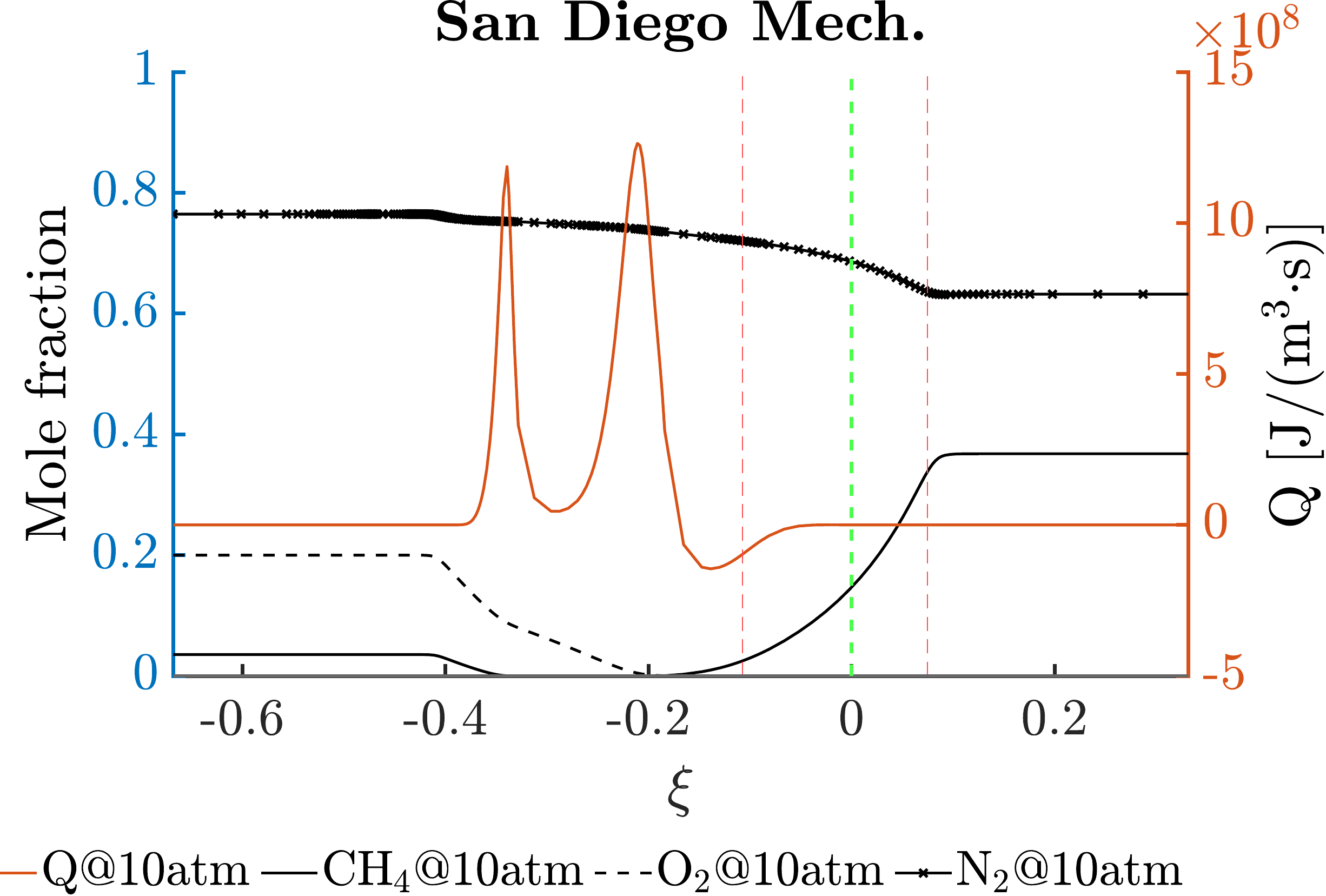}\quad
	\includegraphics[width=.45\textwidth]{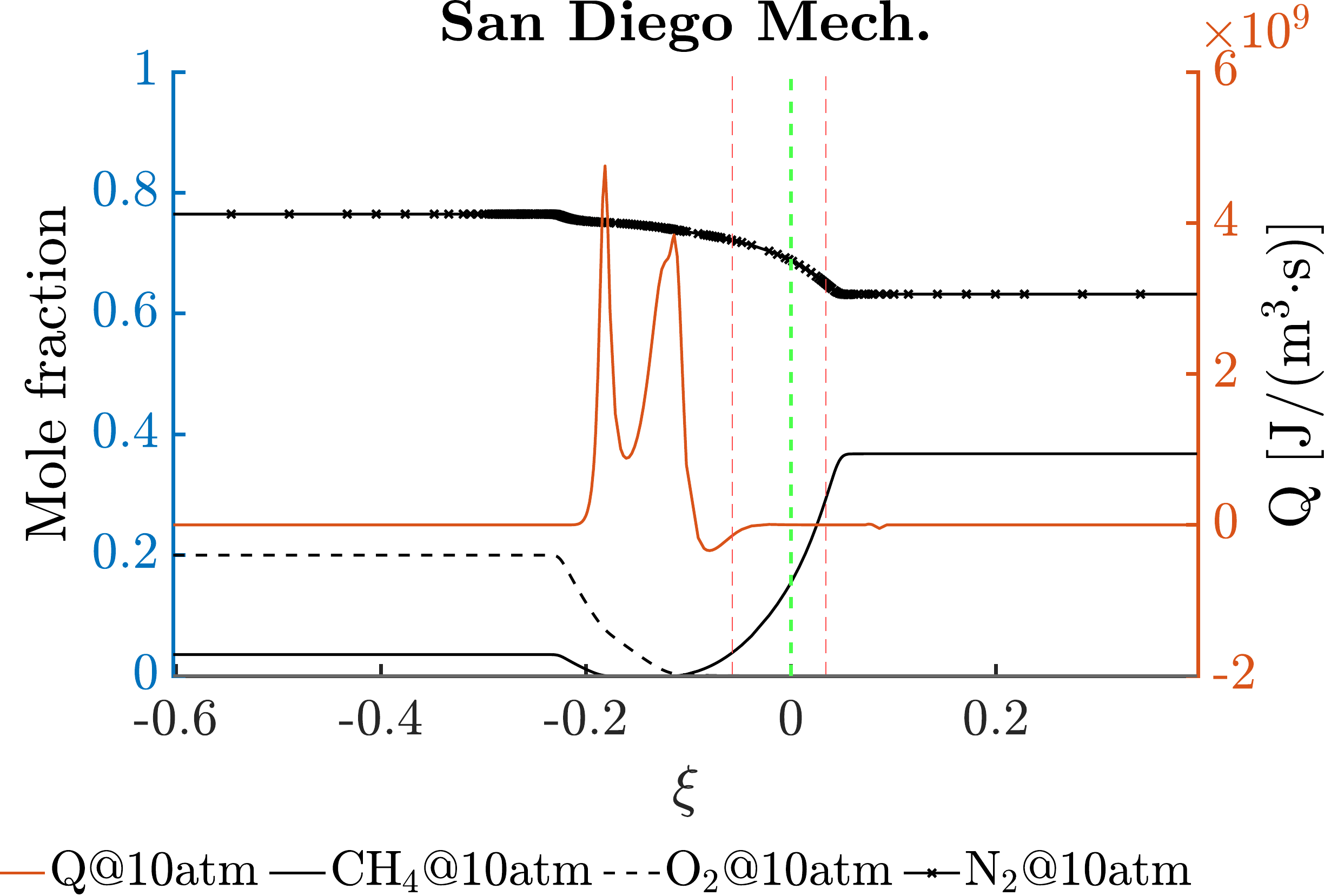}
	
	\medskip
	
	\includegraphics[width=.45\textwidth]{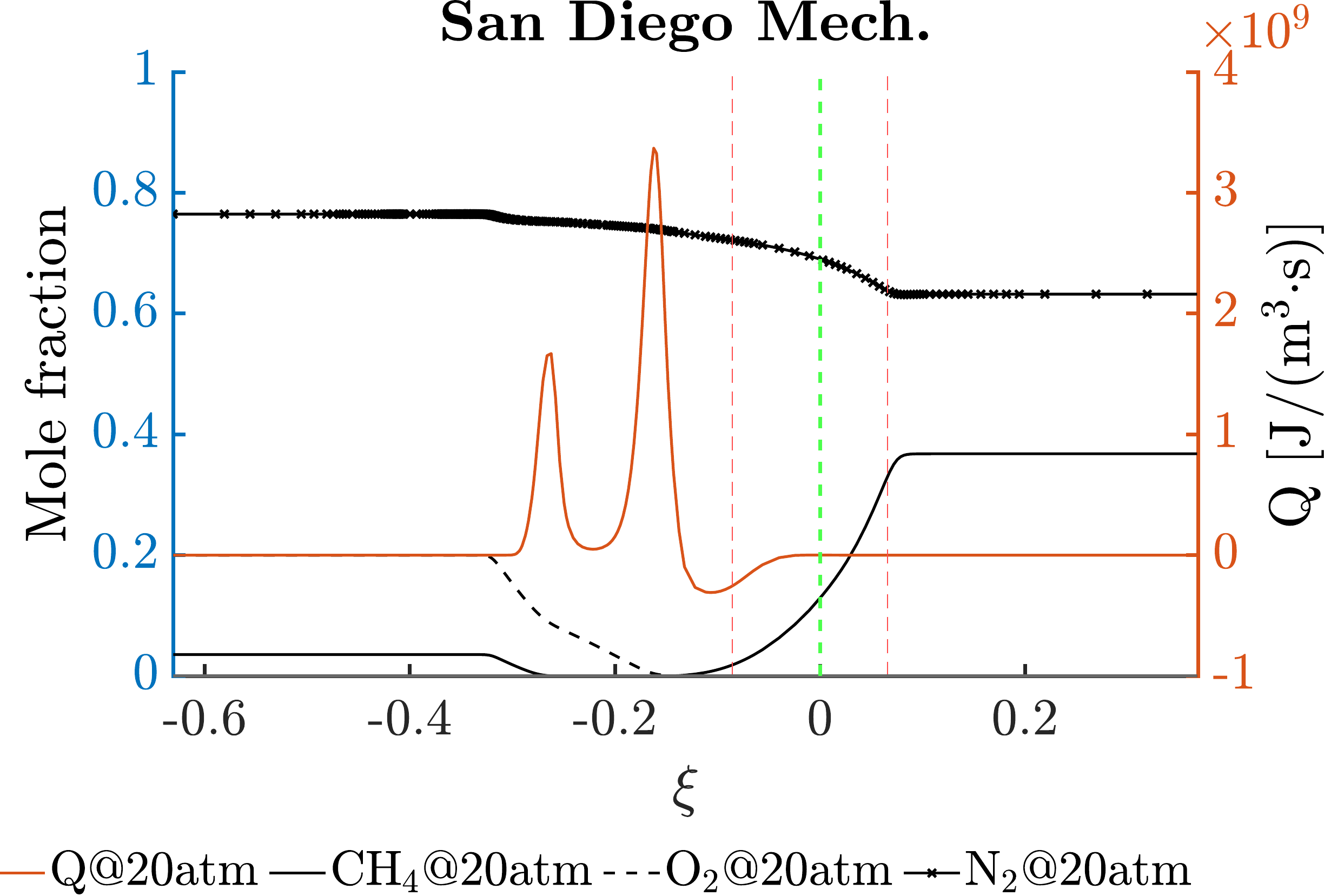}\quad
	\includegraphics[width=.45\textwidth]{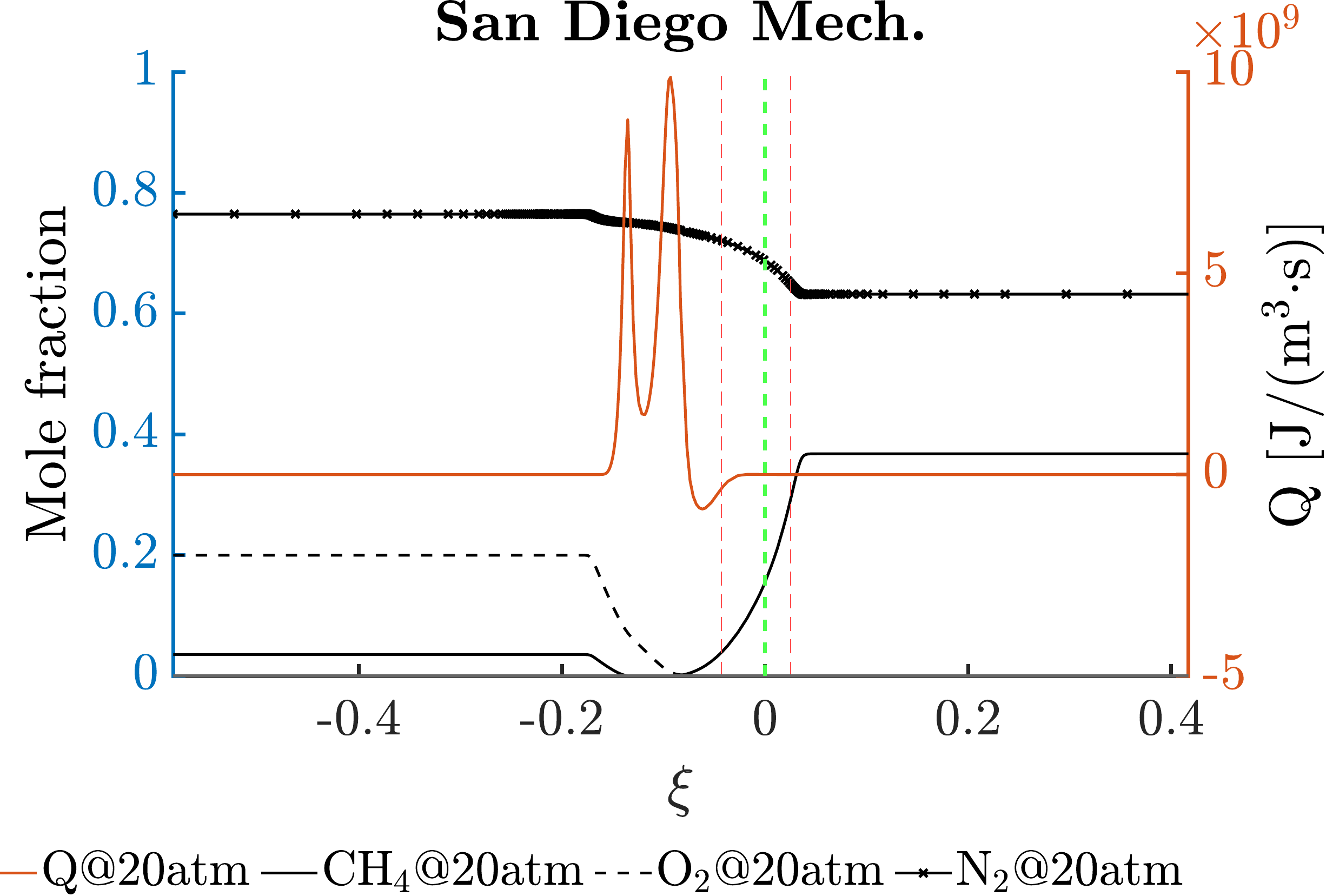}
	
	\caption{Comparison between two different strain rates for Case 1 at $1\:\text{atm}$, $10\:\text{atm}$ and $20\:\text{atm}$. $S = 10\text{\:s}^{-1}$ (left) and $50\text{\:s}^{-1}$ (right). San Diego Mechanism. Mole fractions of CH$_4$, O$_2$ and N$_2$ (black) and heat release rate (orange). Stagnation plane location (green) and the estimated mixing-layer edge (red). See the online version for color references.}
	\label{fig:SDMech_298K_case1_species}
\end{figure}

The heat release rate is taken as a marker of flame presence in the figures. The dominant flame, in terms of the peak value of heat release rate in Figure \ref{fig:SDMech_298K_case1_species}, is the premixed flame on the left rather than the diffusion flame at low pressures (upper plot). However, this relative importance switches at higher pressures for both strain rates studied.\hfill \break

From the same figure, comparing the different heat release rates at constant strain rate, it is seen that the heat-release-rate peaks are more distinct, even though the flames are getting closer. As pressure increases, the heat release peaks are higher and physically closer together. This behavior is expected since the combustion rate is enhanced when increasing pressure and it is in agreement with recent analysis using one-step Westbrook-Dryer kinetics for propane \cite{WAS_WSCI2019}.\hfill \break

\begin{figure}[h!]
	\centering

	\includegraphics[width=.45\textwidth]{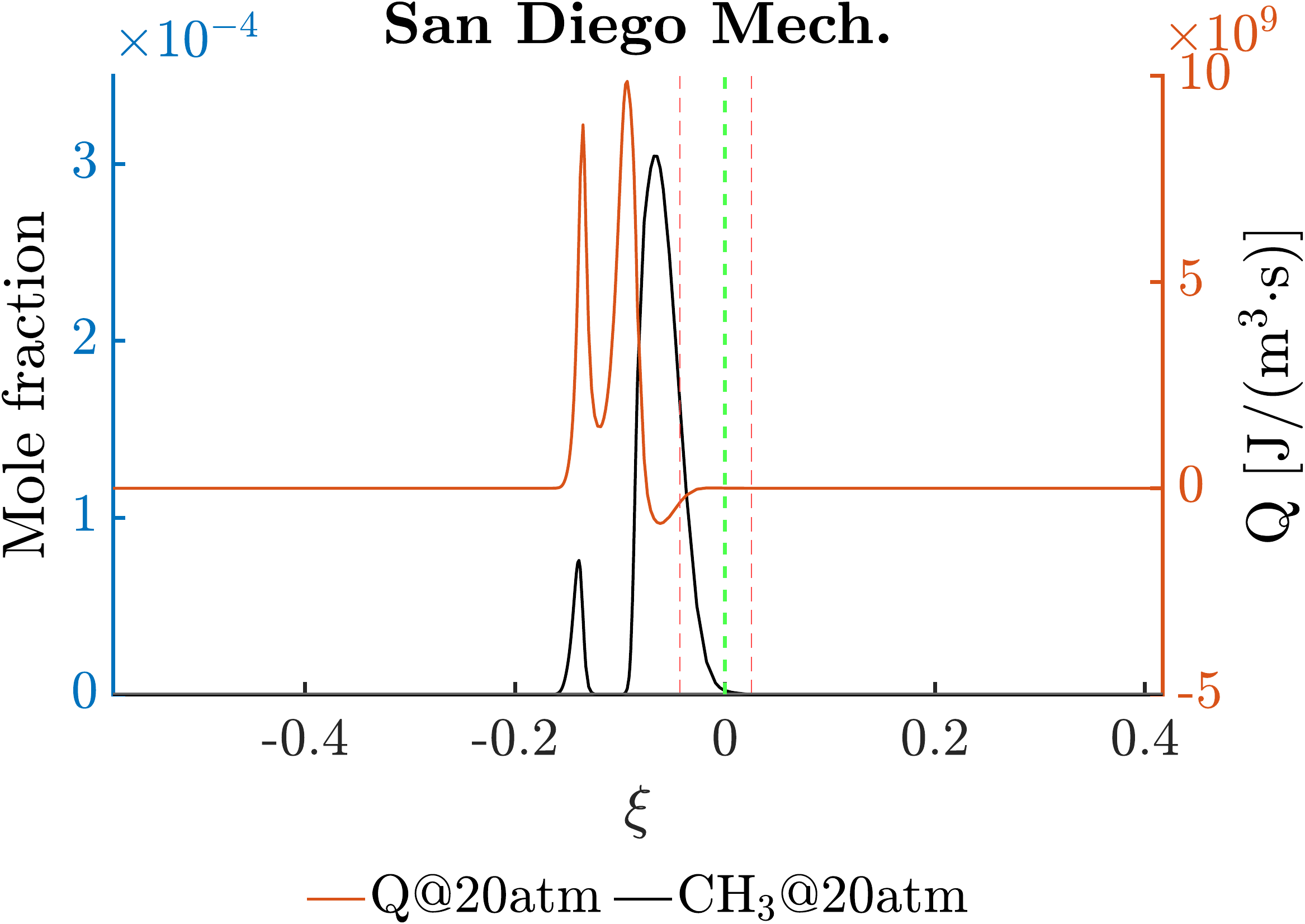}\quad
	\includegraphics[width=.45\textwidth]{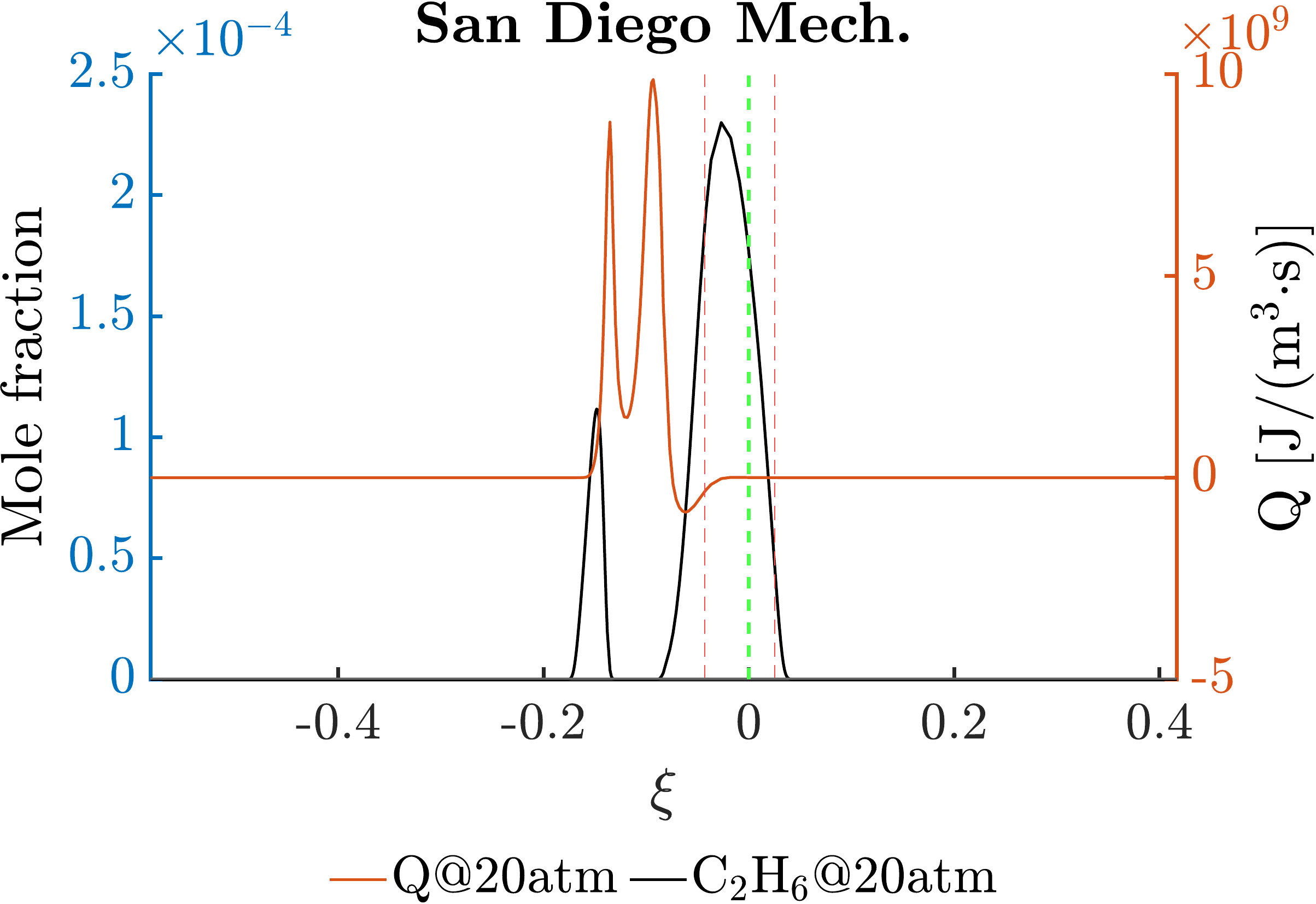}\quad
	\includegraphics[width=.45\textwidth]{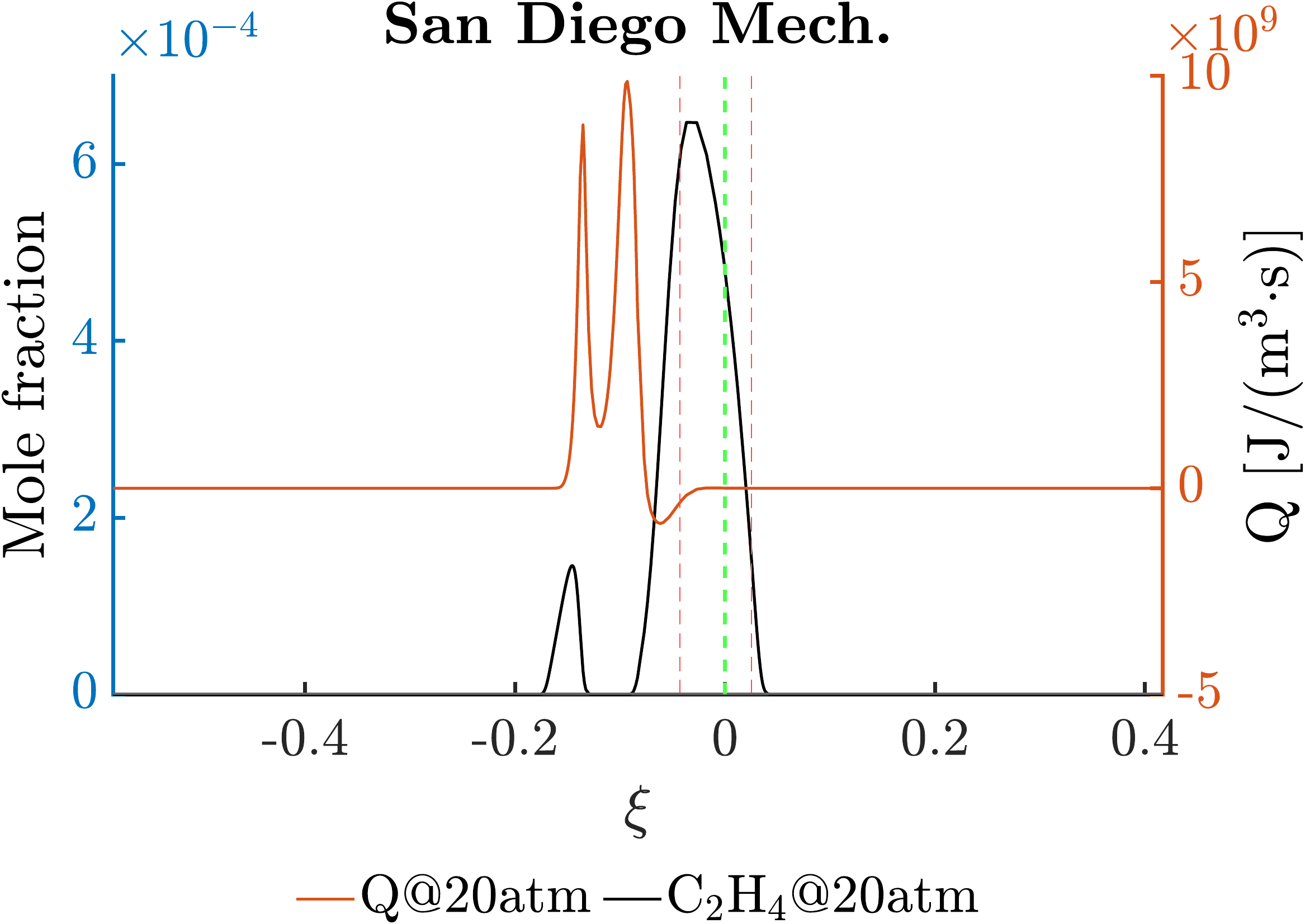}
	
	\caption{Result for Case 1 with $S = 50\text{\:s}^{-1}$ at 20 atm. San Diego Mechanism. Mole fractions of CH$_3$, C$_2$H$_6$ and C$_2$H$_4$ (black). Heat release rate (orange). Stagnation plane location (green) and the estimated mixing-layer edge (red). See the online version for color references.}
\label{fig:SDMech_case1_S25_P20_speciesednothermic}
\end{figure}

\begin{figure}[h!]
	\centering
	
	
	\includegraphics [width=.45\textwidth]{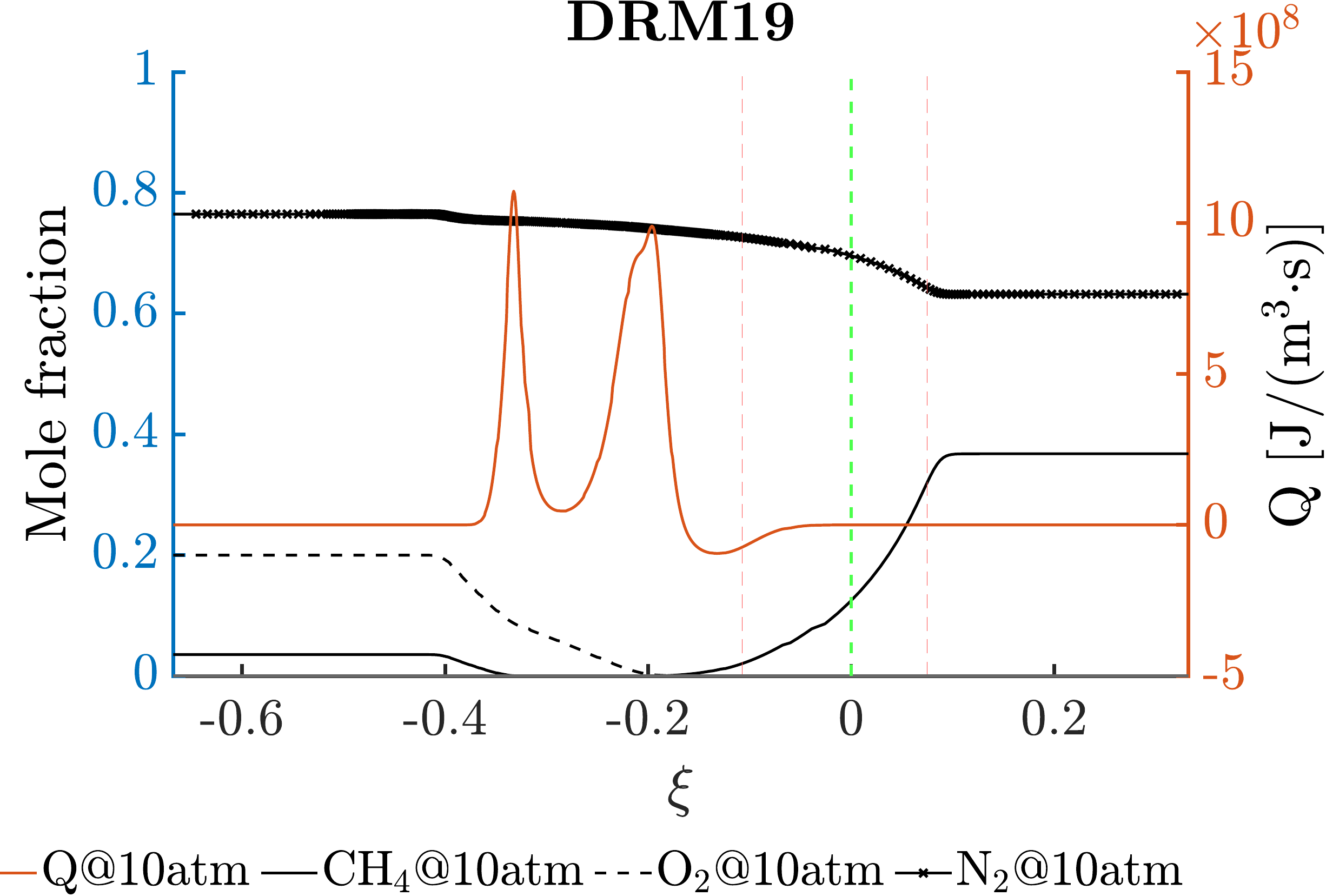}\quad
	\includegraphics [width=.45\textwidth]{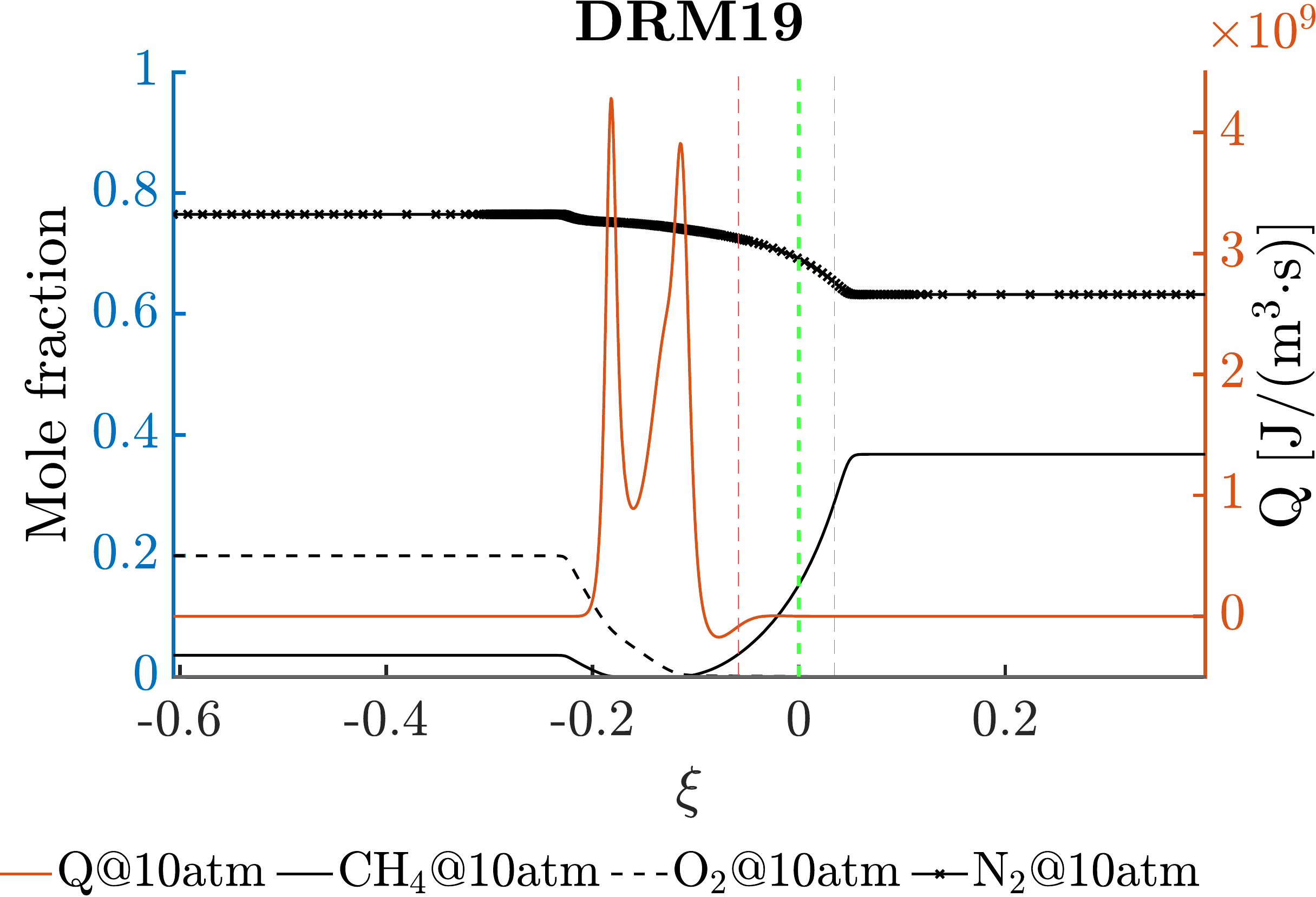}
	
\caption{Comparison between two different strain rates for Case 1 at $10\:\text{atm}$. $S = 10\text{\:s}^{-1}$ (left) and $50\text{\:s}^{-1}$ (right). DRM19 model. Mole fractions of CH$_4$, O$_2$ and N$_2$ (black) and heat release rate (orange). Stagnation plane location (green) and the estimated mixing-layer edge (red). See the online version for color references.}
		
	\label{fig:DRM19_298K_case1_species}
\end{figure}

Note that an endothermic region is found in most cases. A plausible hypothesis based on the review of that region (Figure \ref{fig:SDMech_case1_S25_P20_speciesednothermic}) is that, under certain conditions, methane converts to methyl (CH$_3$), which might recombine to form ethane (C$_2$H$_6$) and then, ethane dehydrogenates to ethylene (C$_2$H$_4$). This is a highly endothermic process \cite{pyrolysis_larkins} that is only considered in detailed chemical kinetic models and explains the endothermic region found in these simulations. Similar trends and a characteristic endothermic region were observed when simulating this case using the DRM19 chemical kinetic model (84 reactions, 19 species \cite{DRM19}). The results shown using that model (Figure \ref{fig:DRM19_298K_case1_species}) at lower pressures at these strain rates qualitatively support the features that were obtained using the San Diego Mechanism. Therefore, this endothermic region is not restricted to the San Diego mechanism and it is expected to appear when detailed chemistry is employed.\hfill \break

\begin{table}[ht!]
	\centering
	\caption{Characteristics of the heat-release-rate ($Q$) peaks for Cases 1 and 2. $T$ stands for temperature [K], $P$ for pressure [atm], and $S$ to strain rate [$\text{\:s}^{-1}$].}
	\label{tab:Characteristics_flames_Case1_Case2}
\begin{tabular}{c|c|c|c|c|c}
\hline
      Case & $P$ & $S$ & $\xi$ at $Q$ peak & Character of $Q$ peak & $T$ at $Q$ peak \\
\hline
\multicolumn{ 1}{c|}{} & \multicolumn{1}{|c|}{1} & \multicolumn{ 1}{|c|}{10} &     -0.586 & Lean premixed flame with diffusive character &    1363.90 \\
\cline{4-6}
\multicolumn{ 1}{c|}{} & \multicolumn{ 1}{|c|}{} & \multicolumn{ 1}{|c|}{} &     -0.277 & Diffusion flame &    1945.35 \\
\cline{3-6}
\multicolumn{ 1}{c|}{} & \multicolumn{ 1}{|c|}{} & \multicolumn{ 1}{|c|}{50} &     -0.409 & Lean premixed flame with diffusive character &    1351.28 \\
\cline{4-6}
\multicolumn{ 1}{c|}{} & \multicolumn{ 1}{|c|}{} & \multicolumn{ 1}{|c|}{} &     -0.209 & Diffusion flame &    1896.72 \\
\cline{2-6}
\multicolumn{ 1}{c|}{1} & \multicolumn{ 1}{|c|}{10} & \multicolumn{ 1}{|c|}{10} &     -0.339 & Lean premixed flame with diffusive character &    1525.77 \\
\cline{4-6}
\multicolumn{ 1}{c|}{} & \multicolumn{ 1}{|c|}{} & \multicolumn{ 1}{|c|}{} &     -0.210 & Diffusion flame &    2100.02 \\
\cline{3-6}
\multicolumn{ 1}{c|}{} & \multicolumn{ 1}{|c|}{} & \multicolumn{ 1}{|c|}{50} &     -0.181 & Lean premixed flame with diffusive character &    1601.63 \\
\cline{4-6}
\multicolumn{ 1}{c|}{} & \multicolumn{ 1}{|c|}{} & \multicolumn{ 1}{|c|}{} &     -0.114 & Diffusion flame &    2042.00 \\
\cline{2-6}
\multicolumn{ 1}{c|}{} & \multicolumn{ 1}{|c|}{20} & \multicolumn{ 1}{|c|}{10} &     -0.264 & Lean premixed flame with diffusive character &    1521.45 \\
\cline{4-6}
\multicolumn{ 1}{c|}{} & \multicolumn{ 1}{|c|}{} & \multicolumn{ 1}{|c|}{} &     -0.162 & Diffusion flame &    2124.70 \\
\cline{3-6}
\multicolumn{ 1}{c|}{} & \multicolumn{ 1}{|c|}{} & \multicolumn{ 1}{|c|}{50} &     -0.135 & Lean premixed flame with diffusive character &    1653.12 \\
\cline{4-6}
\multicolumn{ 1}{c|}{} & \multicolumn{ 1}{|c|}{} & \multicolumn{ 1}{|c|}{} &     -0.093 & Diffusion flame &    2073.30 \\
\hline
\multicolumn{ 1}{c|}{} & \multicolumn{ 1}{|c|}{1} & \multicolumn{ 1}{|c|}{10} &     -0.440 & Rich premixed flame with diffusive character &    1576.99 \\
\cline{4-6}
\multicolumn{ 1}{c|}{} & \multicolumn{ 1}{|c|}{} & \multicolumn{ 1}{|c|}{} &     -0.240 & Diffusion flame &    2010.19 \\
\cline{4-6}
\multicolumn{ 1}{c|}{} & \multicolumn{ 1}{|c|}{} & \multicolumn{ 1}{|c|}{} &     -0.155 & Exothermic reactions (\textit{no flame}) &    2009.53 \\
\cline{3-6}
\multicolumn{ 1}{c|}{} & \multicolumn{ 1}{|c|}{} & \multicolumn{ 1}{|c|}{50} &      0.039 & Rich premixed flame with diffusive character &    1726.23 \\
\cline{4-6}
\multicolumn{ 1}{c|}{} & \multicolumn{ 1}{|c|}{} & \multicolumn{ 1}{|c|}{} &      0.150 & Diffusion flame &    2040.70 \\
\cline{4-6}
\multicolumn{ 1}{c|}{} & \multicolumn{ 1}{|c|}{} & \multicolumn{ 1}{|c|}{} &      0.232 & Exothermic reactions (\textit{no flame}) &    1953.67 \\
\cline{2-6}
\multicolumn{ 1}{c|}{2} & \multicolumn{ 1}{|c|}{10} & \multicolumn{ 1}{|c|}{10} &     -0.310 & Rich premixed flame with diffusive character&    1528.33 \\
\cline{4-6}
\multicolumn{ 1}{c|}{} & \multicolumn{ 1}{|c|}{} & \multicolumn{ 1}{|c|}{} &     -0.106 &  Diffusion flame &    2217.03 \\
\cline{3-6}
\multicolumn{ 1}{c|}{} & \multicolumn{ 1}{|c|}{} & \multicolumn{ 1}{|c|}{50} &      0.005 & Rich premixed flame with diffusive character &    1654.69 \\
\cline{4-6}
\multicolumn{ 1}{c|}{} & \multicolumn{ 1}{|c|}{} & \multicolumn{ 1}{|c|}{} &      0.107 &  Diffusion flame &    2259.32 \\
\cline{2-6}
\multicolumn{ 1}{c|}{} & \multicolumn{ 1}{|c|}{20} & \multicolumn{ 1}{|c|}{10} &     -0.305 & Rich premixed flame with diffusive character &    1168.04 \\
\cline{4-6}
\multicolumn{ 1}{c|}{} & \multicolumn{ 1}{|c|}{} & \multicolumn{ 1}{|c|}{} &     -0.078 &  Diffusion flame &    2253.03 \\
\cline{3-6}
\multicolumn{ 1}{c|}{} & \multicolumn{ 1}{|c|}{} & \multicolumn{ 1}{|c|}{50} &     -0.002 & Rich premixed flame with diffusive character &    1619.59 \\
\cline{4-6}
\multicolumn{ 1}{c|}{} & \multicolumn{ 1}{|c|}{} & \multicolumn{ 1}{|c|}{} &      0.085 & Diffusion flame &    2308.72 \\
\hline
\end{tabular}  

\end{table}

\subsubsection{Strain Rate Effects}
\label{Subsection:case1_increaseS}
In all pressure and strain rate conditions, two flames can be observed. Both flames are placed on the left of the stagnation plane and the flame on the left becoming dominant over the diffusion flame at low pressures. Our results indicate that increasing strain rate at constant pressure favors flame merging, especially at low pressures. Therefore, lower strain rates allow easier distinction between existing flame peaks. \hfill \break

\begin{figure}[h!]
	\centering
	
	\includegraphics[width=.75\textwidth]{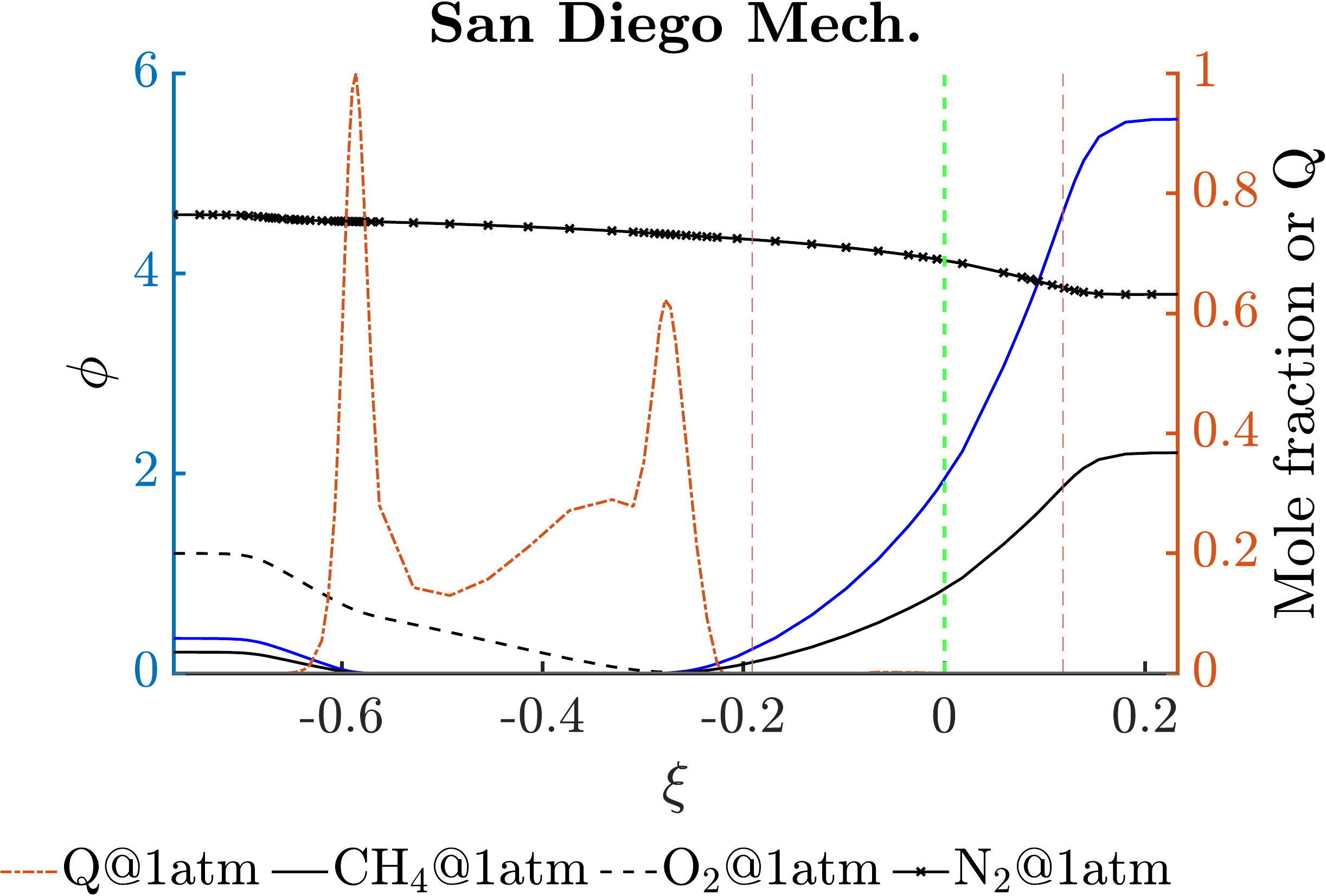}
	
	\caption{Case 1 at $S = 10\text{\:s}^{-1}$ and $1\:\text{atm}$. San Diego Mechanism. Mole fractions of CH$_4$, N$_2$ and O$_2$ (black), normalized heat release rate (orange), and local equivalence ratio (blue). Stagnation plane location (green) and the estimated mixing-layer edge (red). See the online version for color references.}
	\label{fig:SDMech_298K_case1_phi}
\end{figure}

Analysing the character of these flames in depth, it is concluded that the flame expected to be premixed (left) does not follow the classical premixed characteristics where flame peak temperature is lower than adiabatic flame temperature, and velocity remains constant at constant pressure when increasing strain rate. Rather, the temperature of this flame is over the adiabatic flame temperature expected for the upstream equivalence ratio ($\phi$ = 0.34, see Tables \ref{tab:Adiabatic_flame_T} and \ref{tab:Adiabatic_flame_T_flame_front_Case1}). This is an indication that the premixed flame is diffusion controlled. The possibility that this high flame temperature is due to a variation on the local equivalence ratio with respect to its upstream value has also been explored. Figure \ref{fig:SDMech_298K_case1_phi} shows equivalence ratio as a function of $\xi$ at $1\:\text{atm}$ and $S = 10\text{\:s}^{-1}$. The heat release rate has been normalized by its maximum for graphical purposes. The local equivalence ratio at the premixed flame front is $\phi$ = 0.17, with a corresponding adiabatic flame temperature of 758.3 K, which is still lower than the premixed flame temperature reported in Table \ref{tab:Adiabatic_flame_T_flame_front_Case1}. Similar results are found for all the pressures and strain rates conditions considered in Case 1 (see Table \ref{tab:Adiabatic_flame_T_flame_front_Case1}). Hence, the conclusion that the premixed flame behaves as a diffusion flame still holds.\hfill \break


\begin{figure}[h!]
	\centering
	\includegraphics[width=.45\textwidth]{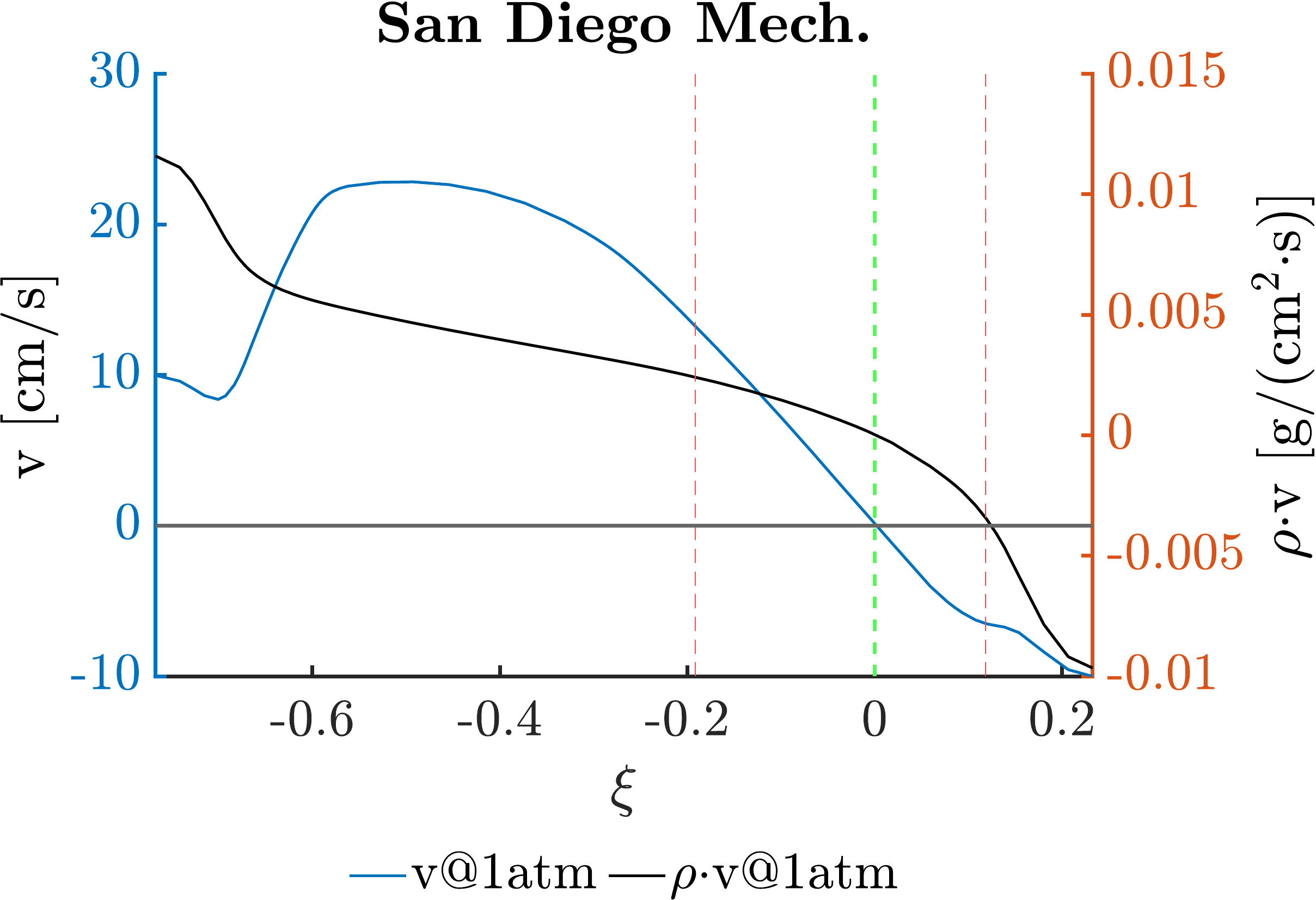}\quad
	\includegraphics[width=.45\textwidth]{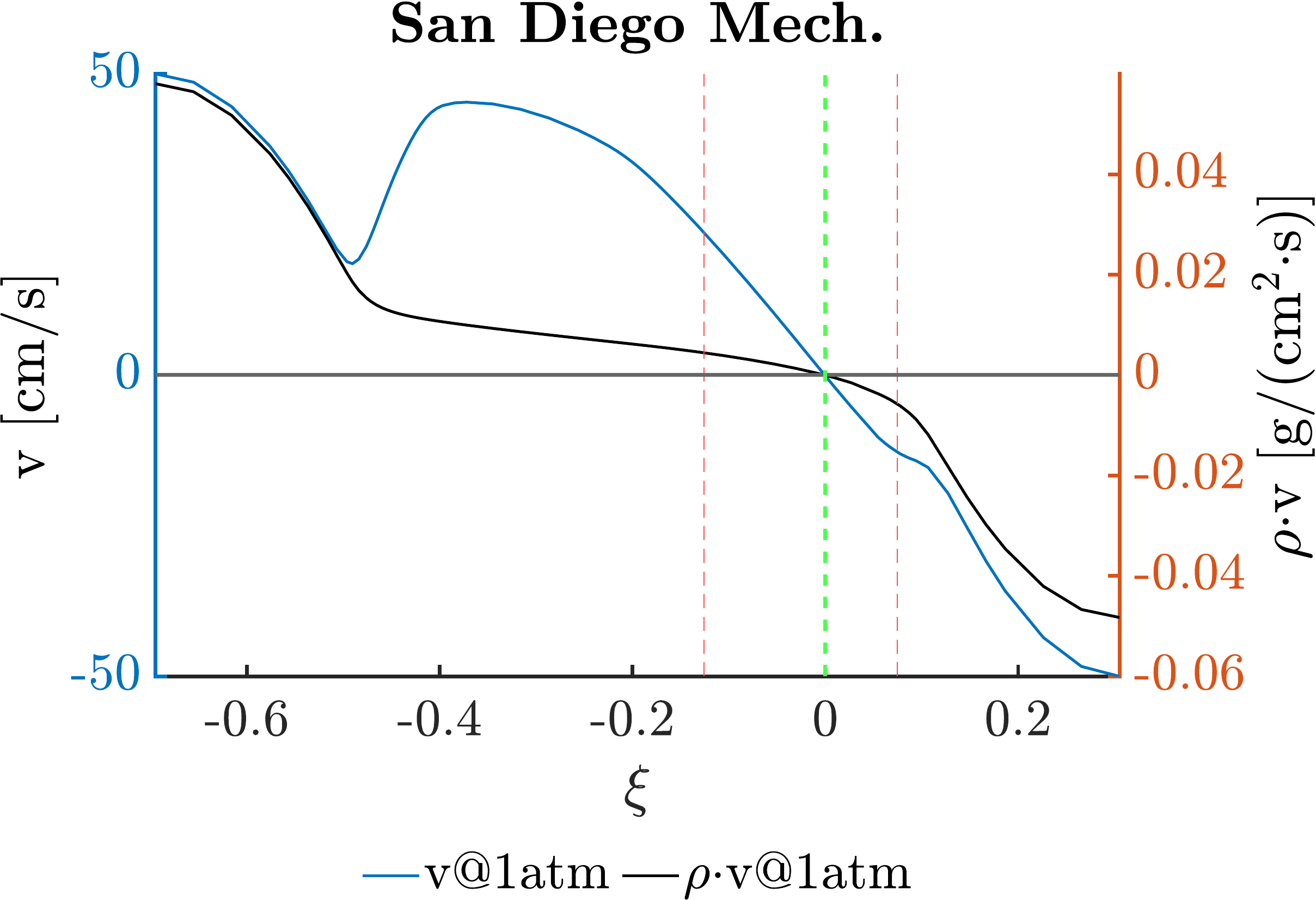}
	
	\medskip
	
	\includegraphics[width=.45\textwidth]{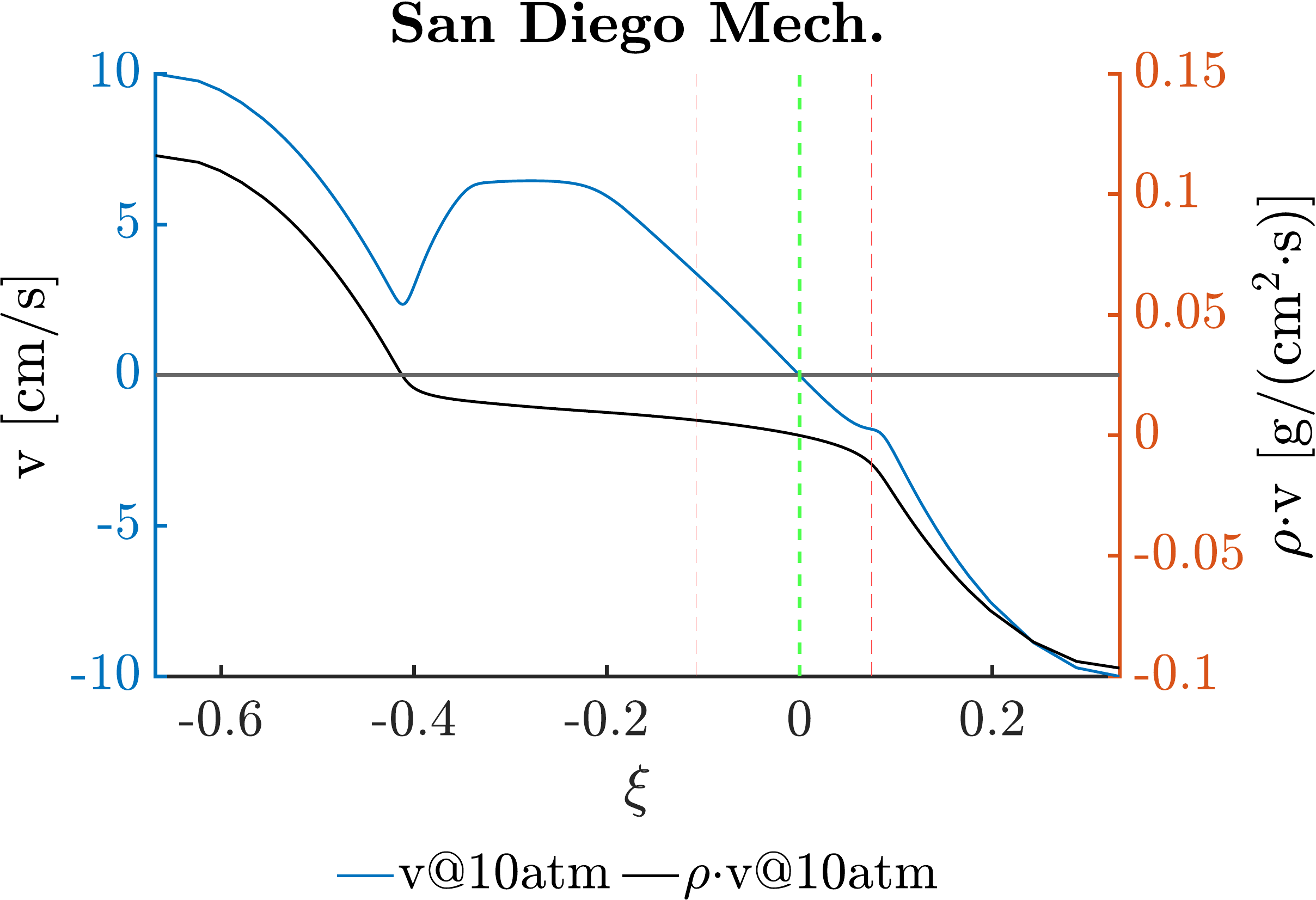}\quad
	\includegraphics[width=.45\textwidth]{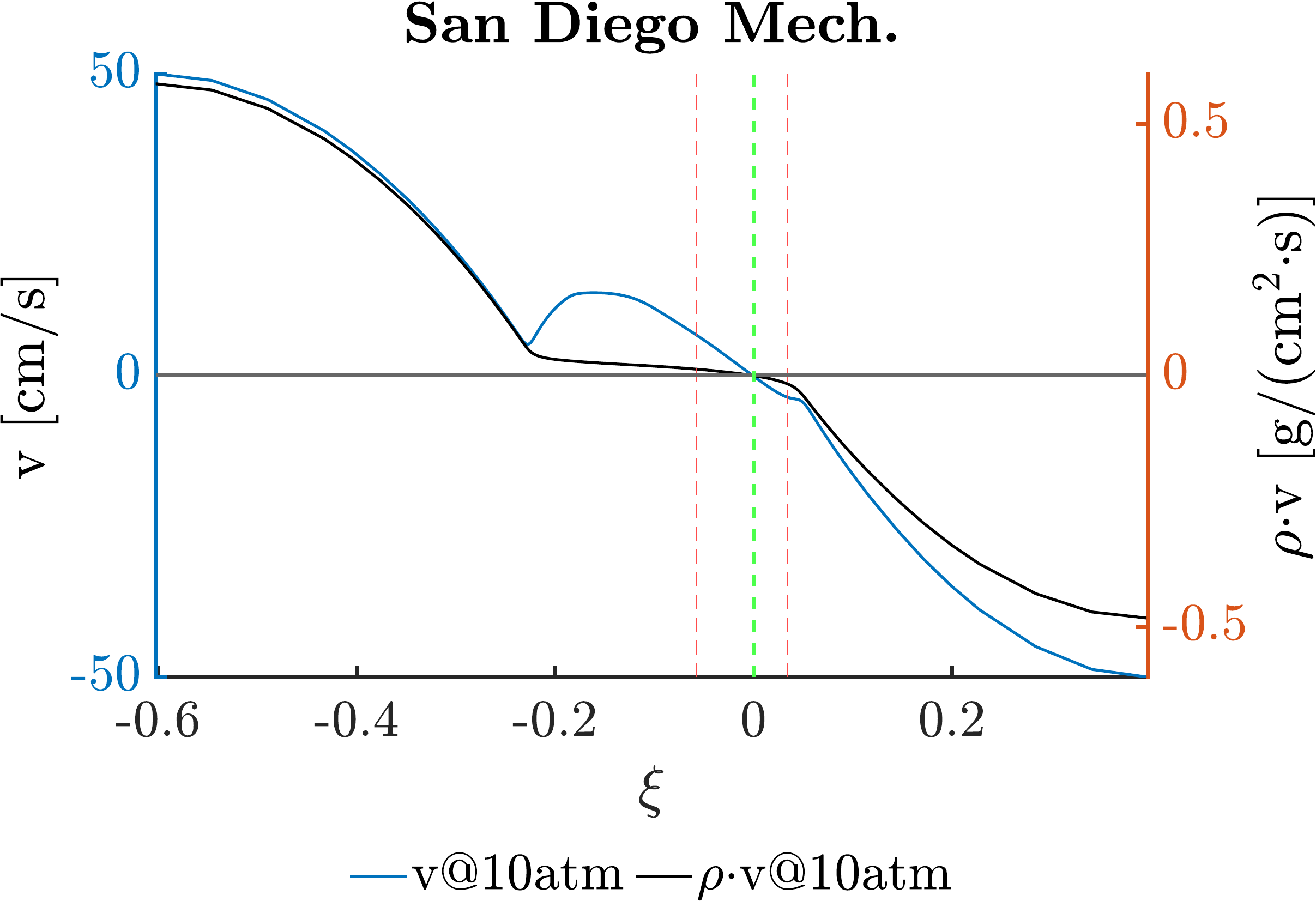}
	
	\medskip
	
	\includegraphics[width=.45\textwidth]{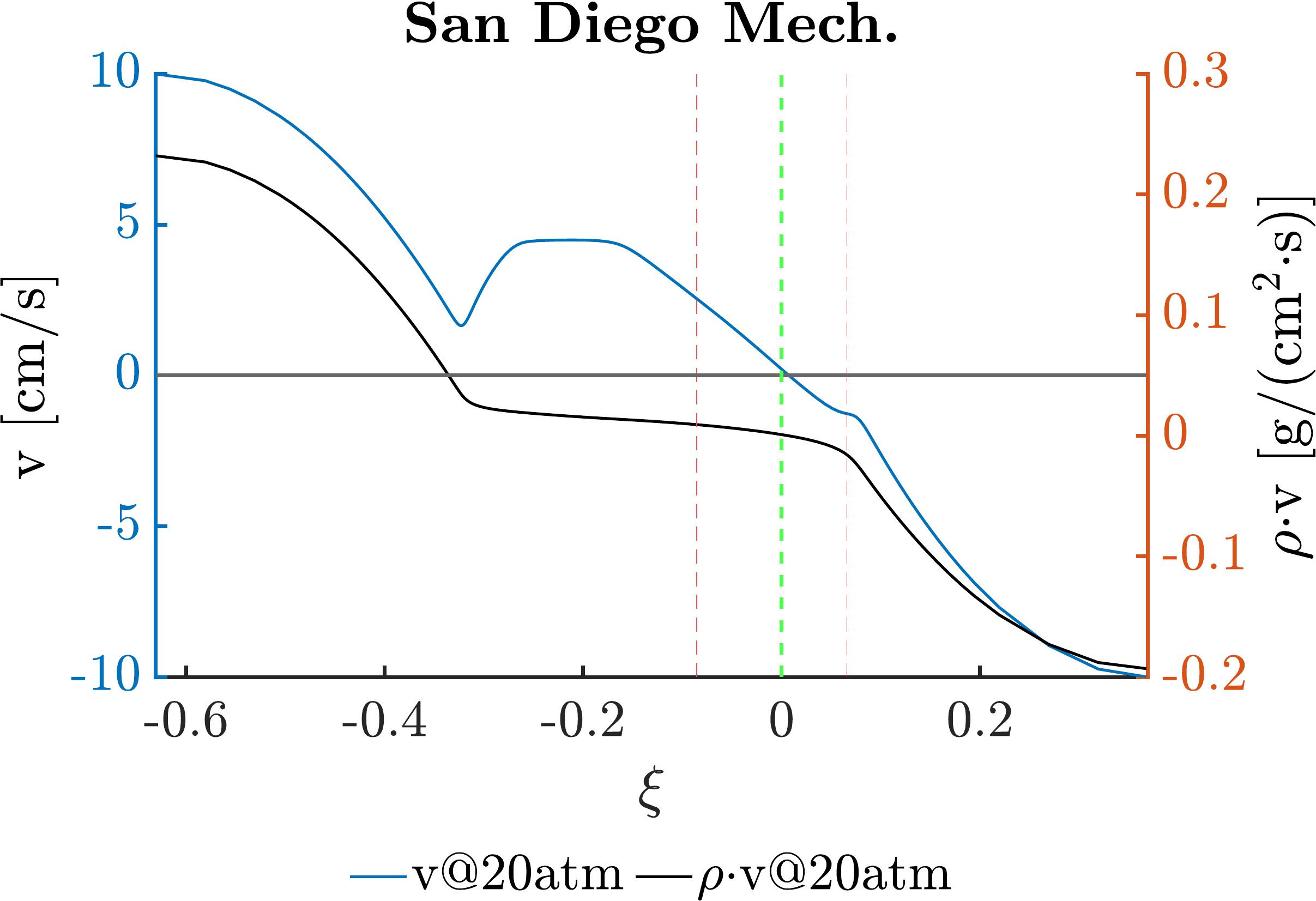}\quad
	\includegraphics[width=.45\textwidth]{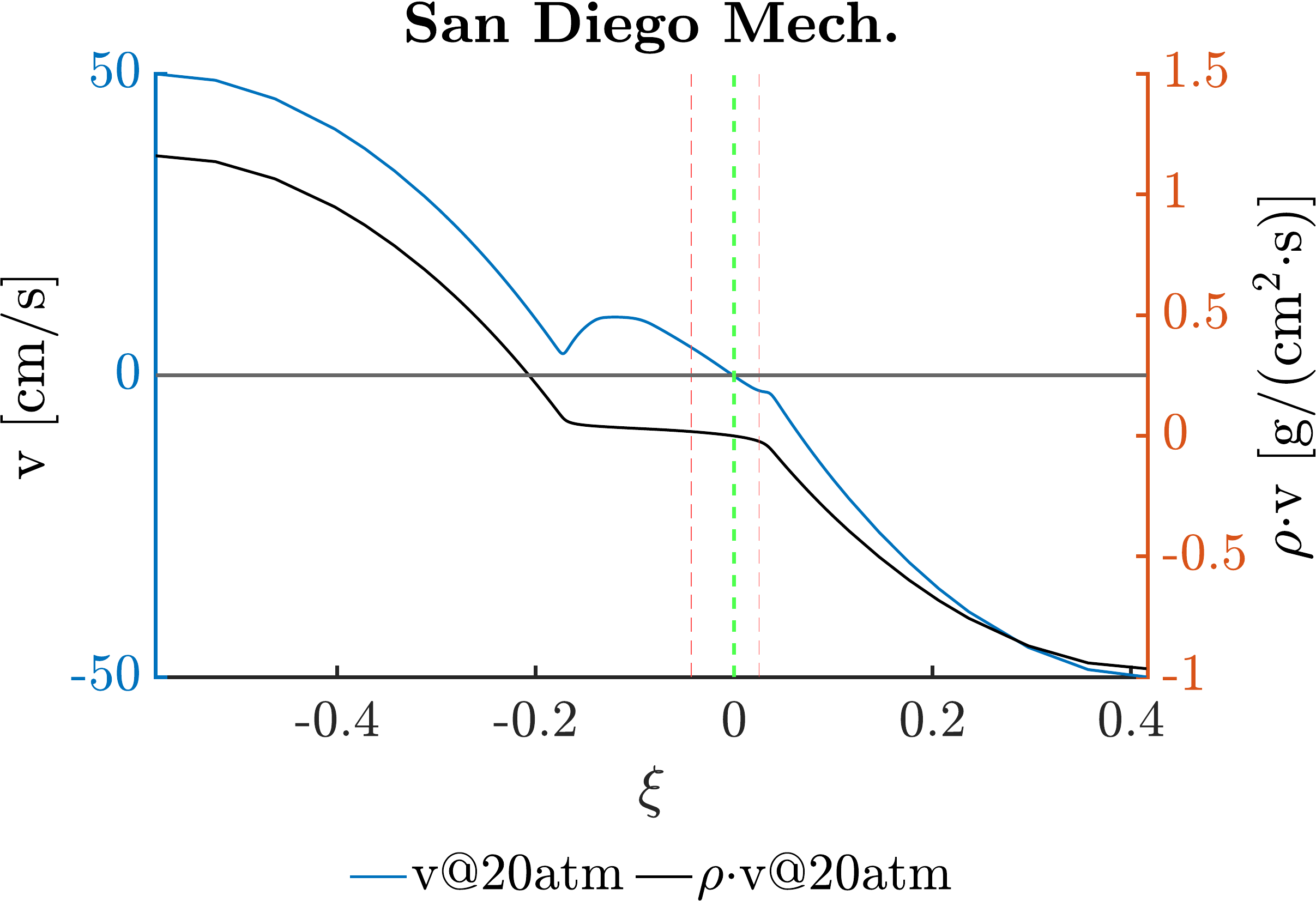}
	
\caption{Comparison between two different strain rates for Case 1 at $1\:\text{atm}$, $10\:\text{atm}$ and $20\:\text{atm}$. $S = 10\text{\:s}^{-1}$ (left) and $50\text{\:s}^{-1}$ (right). Velocity ($v$) and density ($\rho$) times velocity are plotted. San Diego Mechanism. Stagnation plane location (green) and the estimated mixing-layer edge (red). See the online version for color references.}
	\label{fig:SDMech_298K_case1_densityvelocity}
\end{figure}

Additionally, Figure \ref{fig:SDMech_298K_case1_densityvelocity} and Table \ref{tab:SDMech_densityvelocity} show that at constant pressure, the velocity at the premixed flame front does not remain constant when varying the strain rate, but rather increases with it. These observations indicate that as strain rate increases, the distance between flames decreases, leading to steeper gradients in the axial direction for the reactants mass-fraction curves and temperature profile. Consequently, diffusion rates are enhanced. In other words, increasing the strain rate enhances reactants diffusion mass transport towards the diffusion flame, producing more heat, which in turn is diffused in the negative x-direction at a faster rate towards the premixed flame. The latter also produces more heat - highlighted by its peak temperature being higher than its corresponding adiabatic flame temperature- and is able to sustain higher velocities than it would if it behaved like a typical premixed flame. \hfill \break

\begin{table}[ht!]
	\centering
\caption{Flame front velocities [cm/s] at different pressure ($P$) and strain rate ($S$) conditions for Cases 1 and 2.}
\label{tab:SDMech_densityvelocity}
\begin{tabular}{c|c|c|c}
\centering
&\backslashbox{$P$ [atm]}{$S$ [s$^{-1}$]}&{10}&{50}\\\cline{1-4}
& 1 &17.0&33.01\\\cline{2-4}
Case 1 & 10 &4.76&10.16\\\cline{2-4}
& 20 &3.18&7.147\\\hline
& 1 &3.98&9.02\\\cline{2-4}
Case 2 & 10 &1.90&3.89\\\cline{2-4}
& 20 &1.69&3.03\\\hline
\end{tabular}
\end{table}

\subsection{Case 2}
In this case, a fuel-rich mixture is injected from the left nozzle while air enters the domain from the right side. Here, two flames are expected: a premixed flame burning the fuel stoichiometrically from the left and a diffusion flame burning the leftover fuel from the left with the air from the right. However, due to the same reasons discussed in Case 1, the premixed fuel-lean flame is also diffusion controlled. Its peak temperature is also higher than the corresponding adiabatic flame temperature at the in-flowing mixture ratio, and the velocity at the flame front increases with strain rate at constant pressure. Solutions for this case are portrayed in Figure \ref{fig:SDMech_298K_case2_species} and more details follow in the next Subsections \ref{Subsection:case2_increaseP} and \ref{Subsection:case2_increaseS}.

\subsubsection{Pressure Effects} 
\label{Subsection:case2_increaseP}
\begin{figure}[h!]
	\centering
	\includegraphics [width=.45\textwidth]{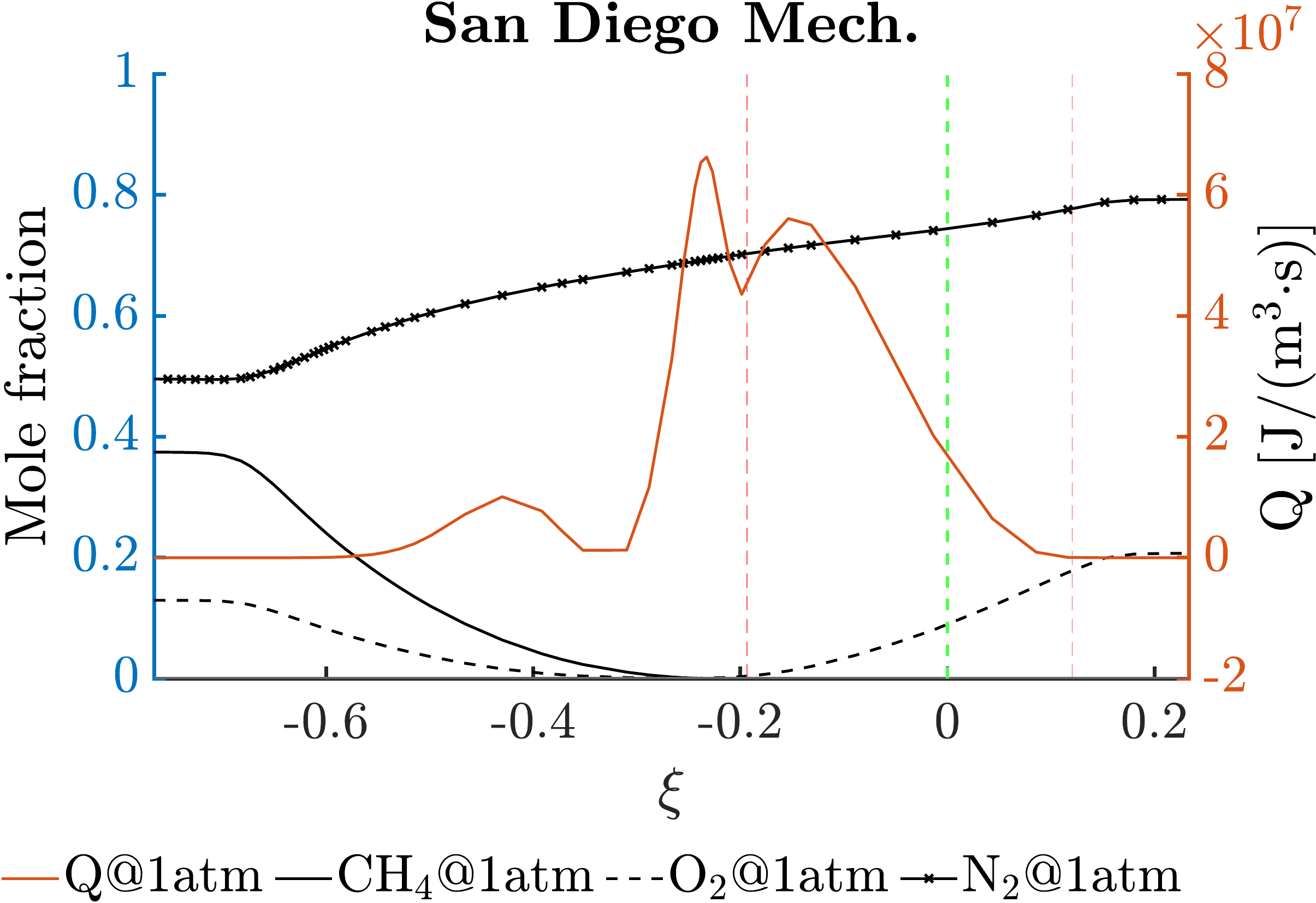}\quad
	\includegraphics [width=.45\textwidth]{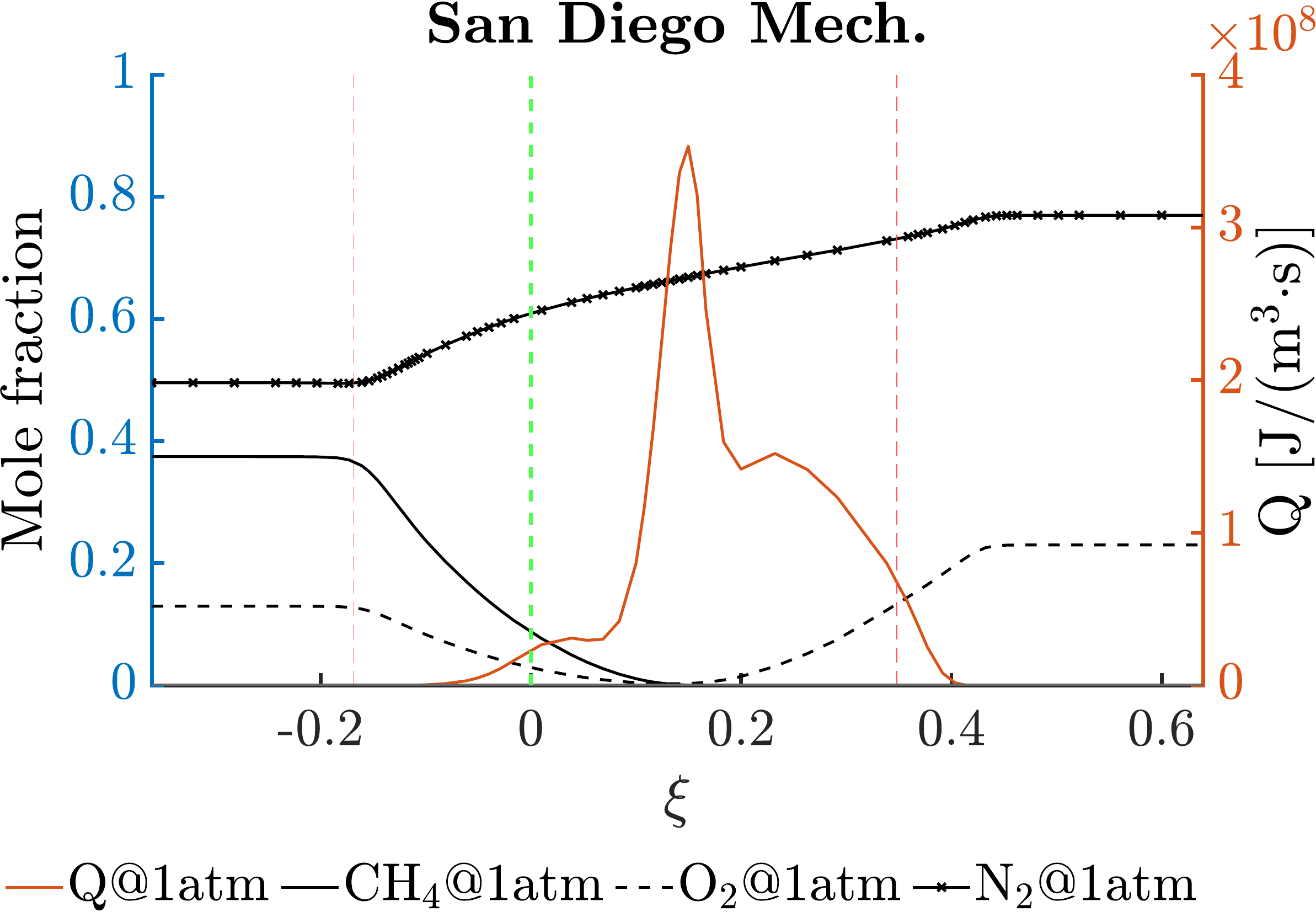}
	
	\medskip
	
	\includegraphics [width=.45\textwidth]{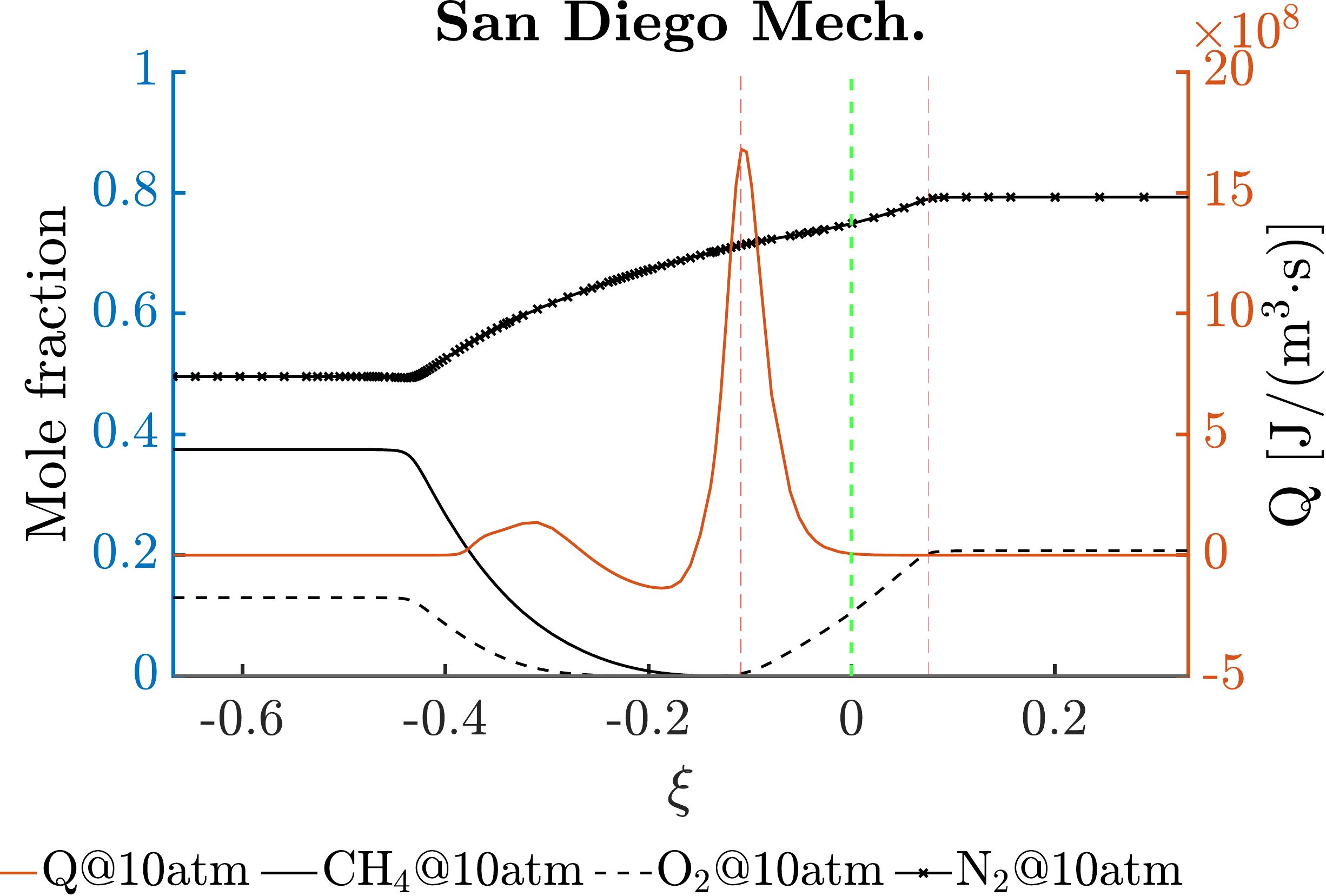}\quad
	\includegraphics [width=.45\textwidth]{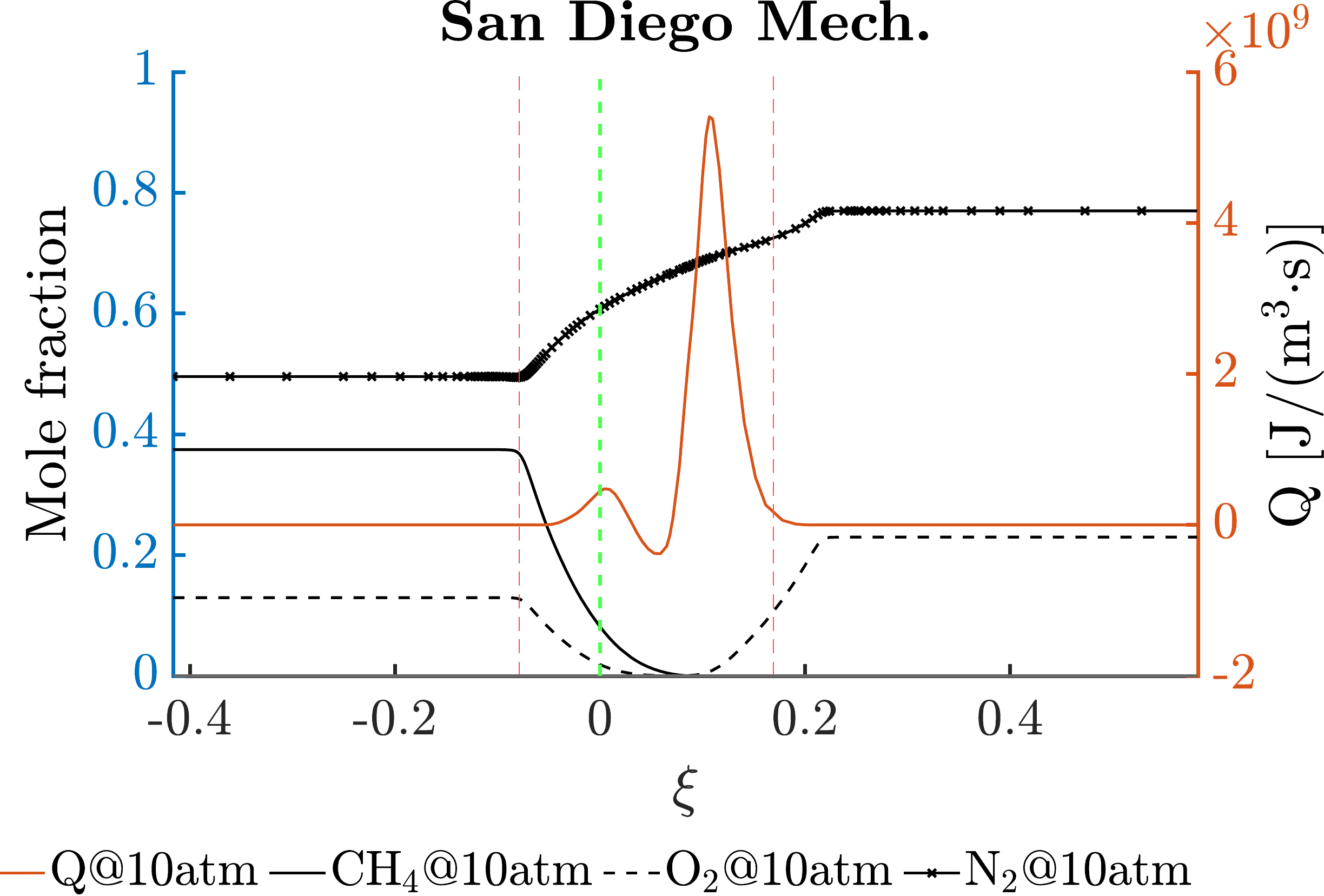}
	
	\medskip
	
	\includegraphics [width=.45\textwidth]{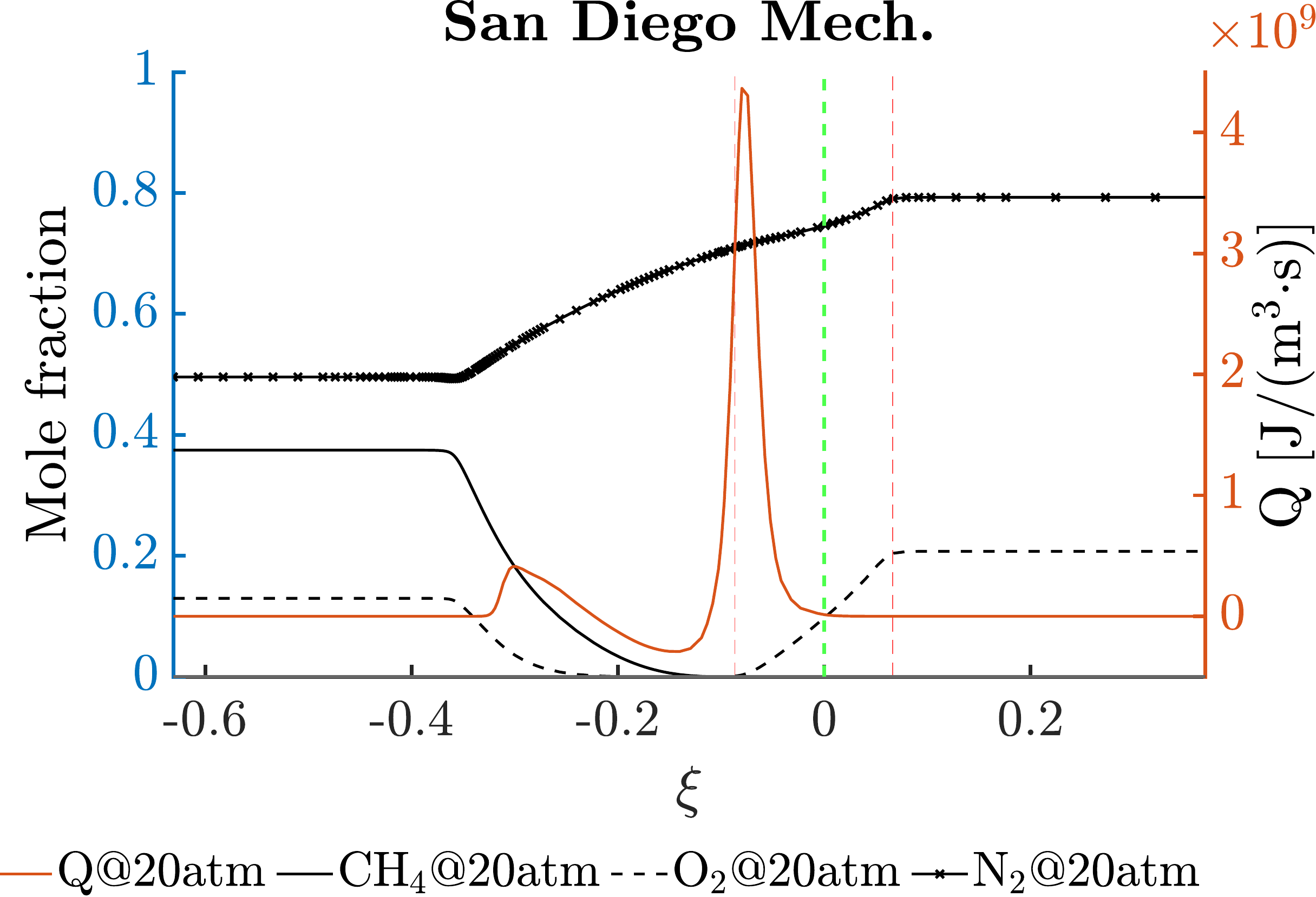}\quad
	\includegraphics [width=.45\textwidth]{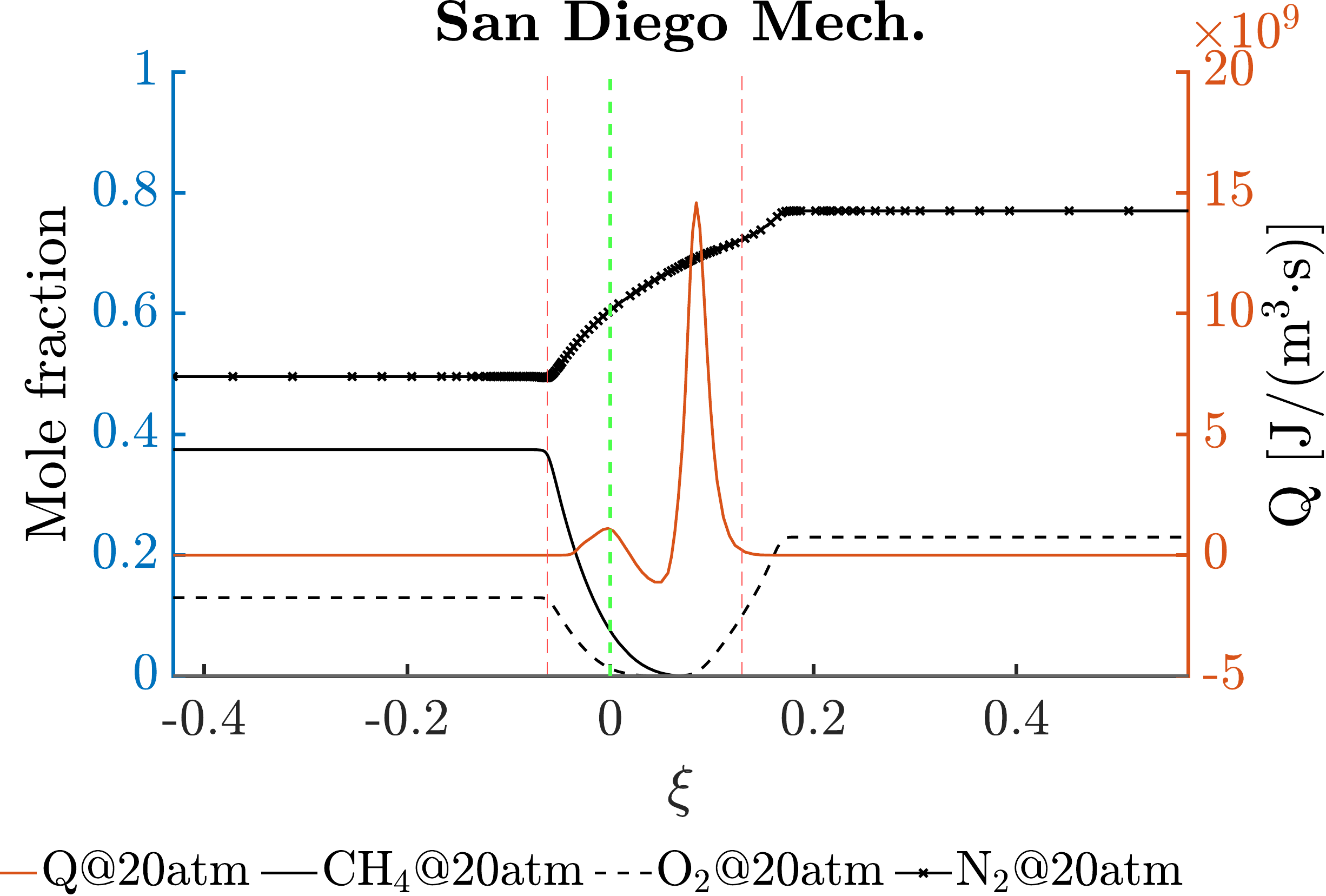}
	
	\caption{Comparison between two different strain rates for Case 2 at $1\:\text{atm}$, $10\:\text{atm}$ and $20\:\text{atm}$. $S = 10\text{\:s}^{-1}$ (left) and $50\text{\:s}^{-1}$ (right). San Diego Mechanism. Mole fractions of CH$_4$, O$_2$ and N$_2$ (black) and heat release rate (orange). Stagnation plane location (green) and the estimated mixing-layer edge (red). See the online version for color references.}
	
	\label{fig:SDMech_298K_case2_species}
\end{figure}

As in the previous case, Figure \ref{fig:SDMech_298K_case2_species} shows that increasing pressure enhances the combustion; therefore, higher heat release rates are obtained. The diffusion flame (right heat-release-rate peak) is dominant for all pressures and strain rates. \hfill \break

\begin{figure}[h!]
	\centering
	
	\includegraphics[width=.45\textwidth]{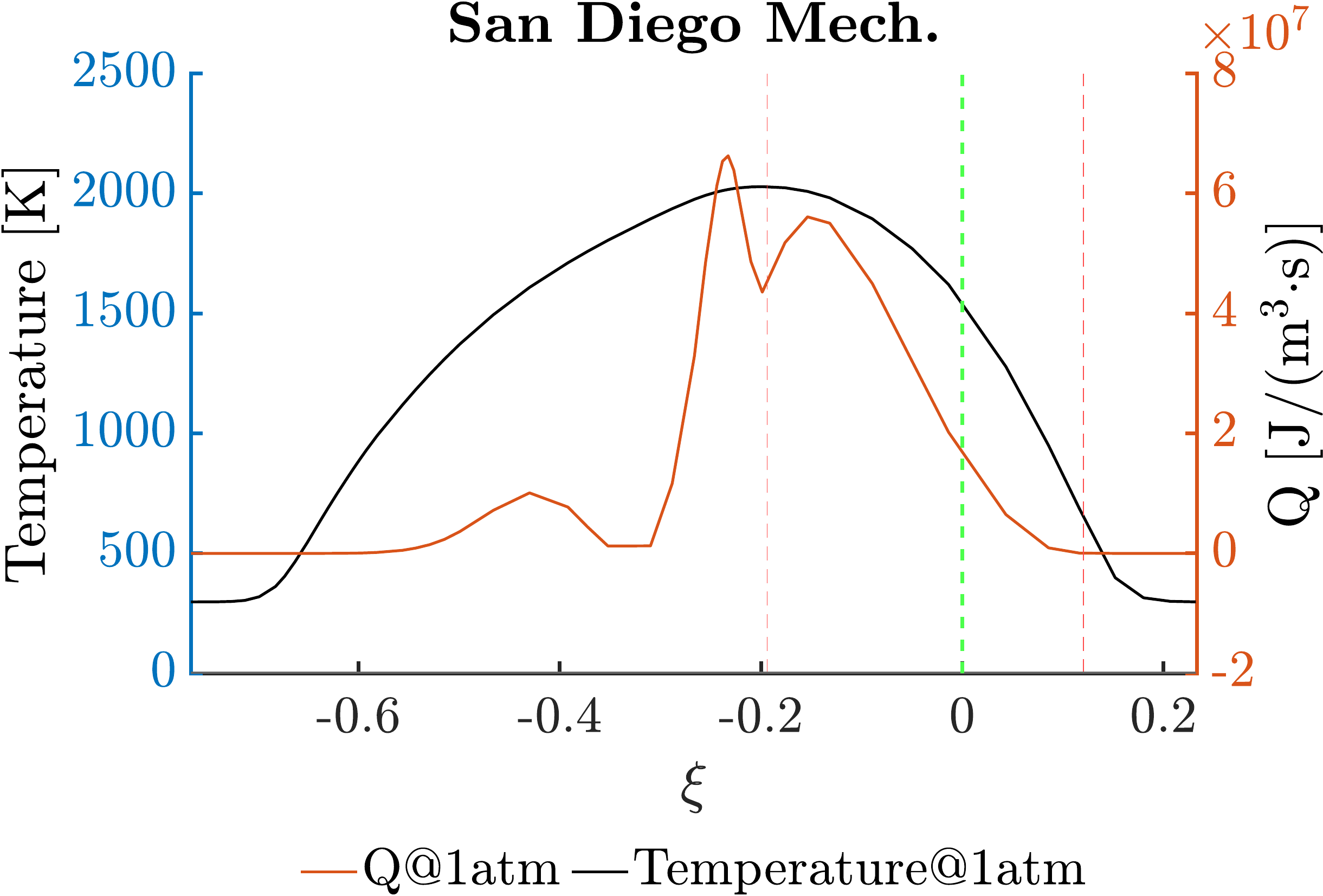}
	
	\caption{Case 2 at $S = 10\text{\:s}^{-1}$ and $1\:\text{atm}$. San Diego Mechanism. Temperature (black) and heat release rate (orange). Stagnation plane location (green) and the estimated mixing-layer edge (red). See the online version for color references.}
	\label{fig:SDMech_case2_S5_P1_temperature}
\end{figure}

At low pressure ($1\:\text{atm}$) and especially at high strain rate, merged flames are observed which become distinct and branch into two co-existing flames when increasing pressure and maintaining constant strain rate. Figure \ref{fig:SDMech_298K_case2_species} also shows that depending on the pressure and strain, the diffusion flame (right) possesses two local peaks in heat release rate: one at $\xi$ = - 0.240 and the other at $\xi$ = - 0.155 ($1\:\text{atm}$ and $10\text{\:s}^{-1}$), while the peak at $\xi$ = - 0.440 corresponds to the premixed fuel-rich flame (see Table \ref{tab:Characteristics_flames_Case1_Case2}). There is only one peak in temperature (see Figure \ref{fig:SDMech_case2_S5_P1_temperature}). In the local absence of methane, the heat-release-rate peak at $\xi$ = - 0.155 is mostly produced by reactions \ref{reac:Production_CO2} and \ref{reac:Consumption_H2}, which are highly exothermic (i.e. negative reaction enthalpy $\Delta H_{\ce{R}}$) and displaced to the formation of products. 

\begin{chequation}
\begin{align}
\ce{CO + OH <=>> CO2 + H $\quad \Delta H_{\ce{R}} = -320.443 \;$ kJ $\;$ mol^{-1}} \label{reac:Production_CO2}\\
\ce{H2 + OH <=>> H2O + H $\quad \Delta H_{\ce{R}} = -279.328 \;$ kJ $\;$ mol^{-1} } \label{reac:Consumption_H2}
\end{align}
\end{chequation}

\begin{figure}[h!]
	\centering
	
	\includegraphics[width=.45\textwidth]{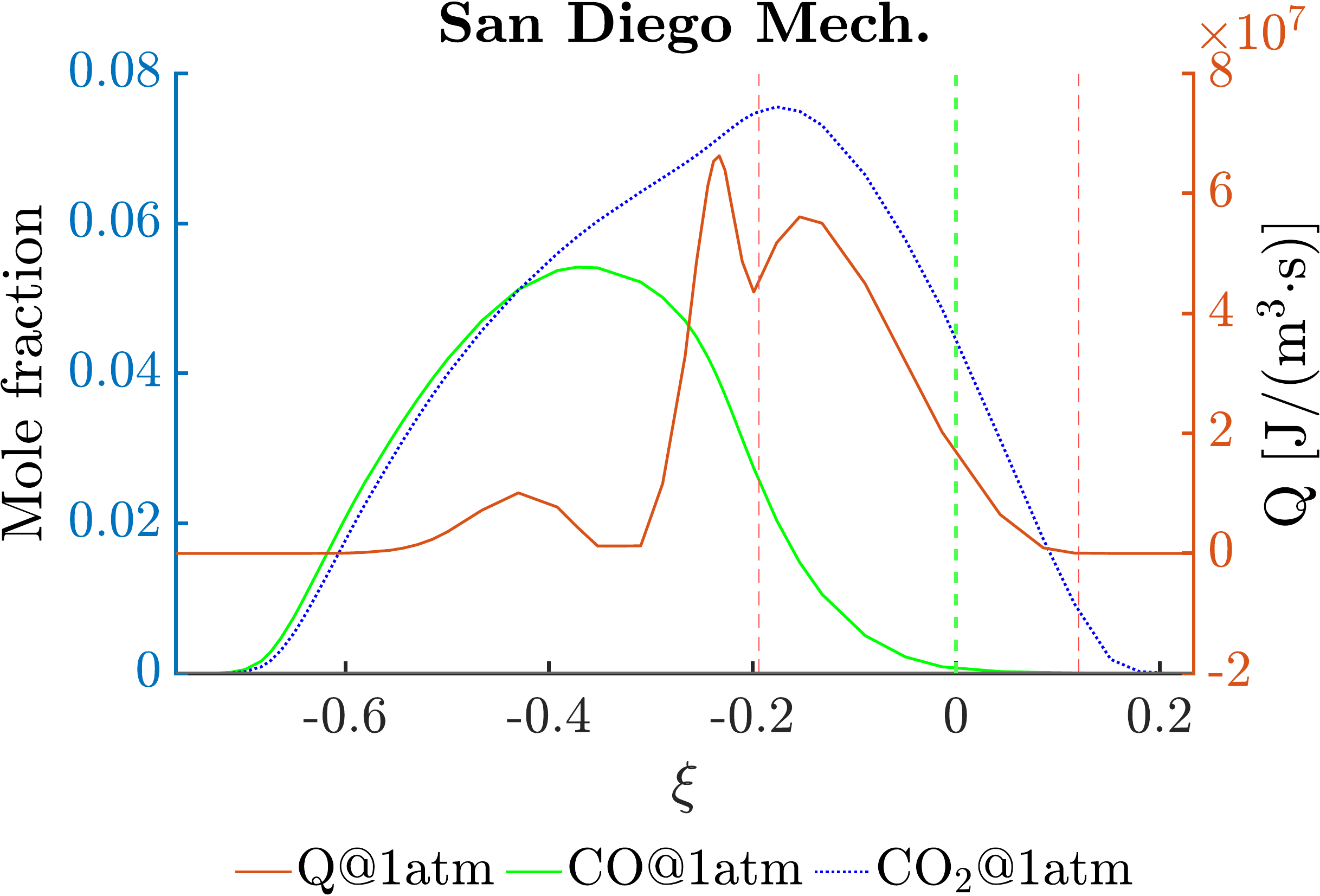}\quad
	\includegraphics [width=.45\textwidth]{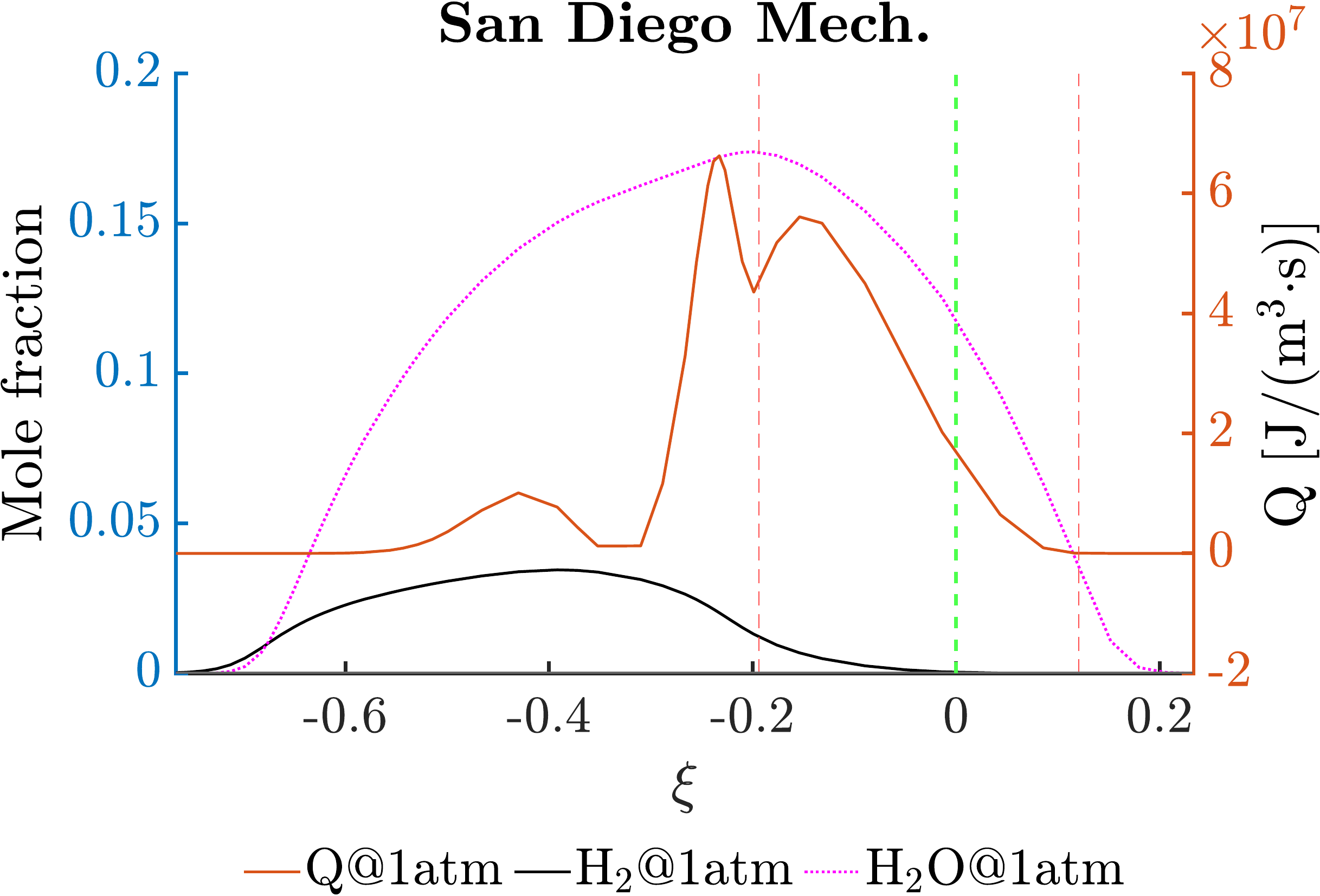}
	
	\caption{Case 2 at $S = 10\text{\:s}^{-1}$ and $1\:\text{atm}$. San Diego Mechanism. Mole fractions of CO (green), CO$_2$ (blue), H$_2$ (black), and H$_{2}$O (magenta) and heat release rate (orange). Stagnation plane location (green) and the estimated mixing-layer edge (red). See the online version for color references.}
	\label{fig:SDMech_case2_S5_P1_speciesdetail}
\end{figure}

\begin{figure}
\centering
\begin{subfigure}[b]{0.55\textwidth}
   \caption{}
	\includegraphics[width=.50\textwidth]{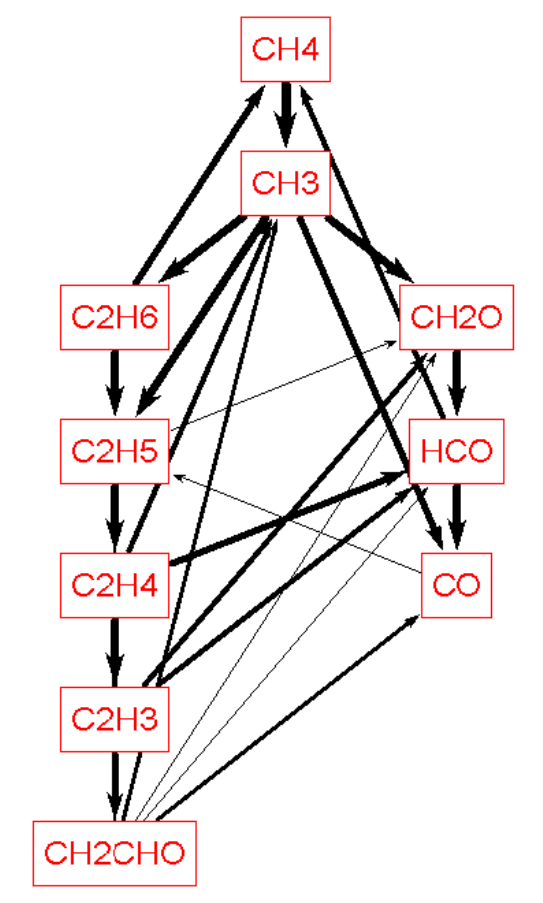}
   \label{fig:SDMech_case2_S5_P1_pathways_a_C} 
\end{subfigure}

\begin{subfigure}[b]{0.7\textwidth}
   \caption{}
	\includegraphics[width=.78\textwidth]{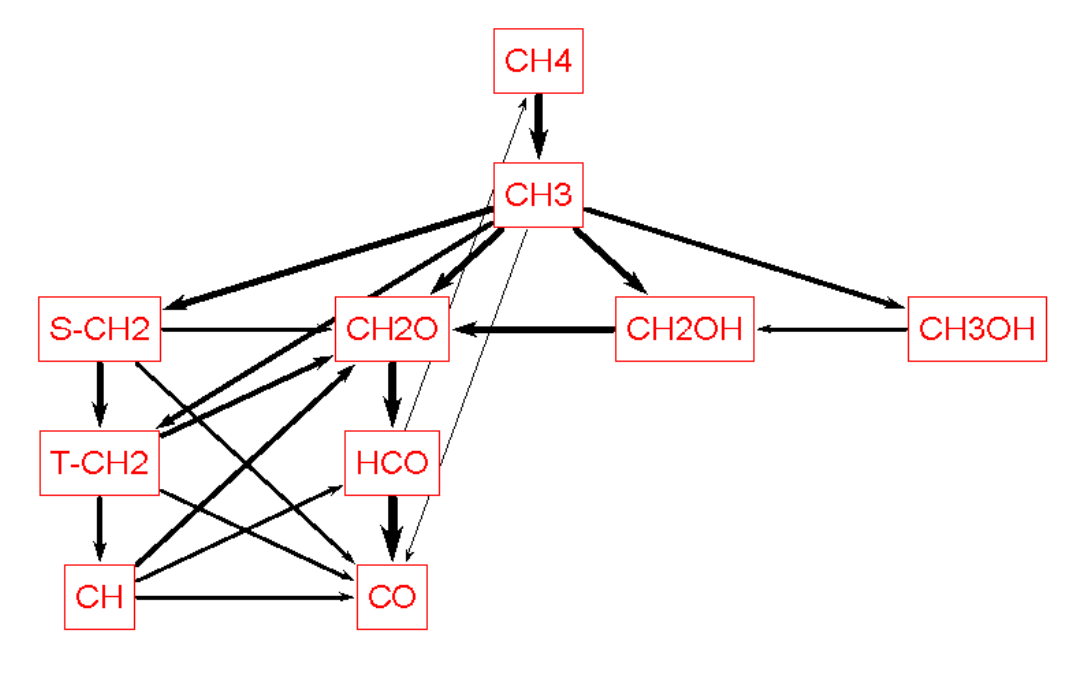}
   \label{fig:SDMech_case2_S5_P1_pathways_b_C}
\end{subfigure}

\begin{subfigure}[b]{0.8\textwidth}
   \caption{}
    \includegraphics[width=.52\textwidth]{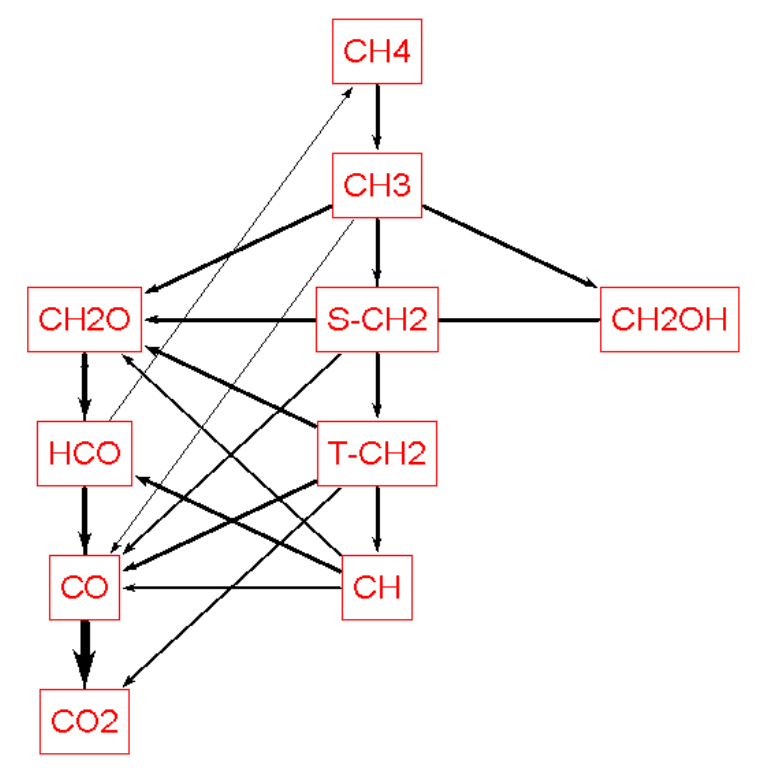}
   \label{fig:SDMech_case2_S5_P1_pathways_c_C}
\end{subfigure}

\caption{Main reaction pathways for species containing carbon at the three locations where reaction rate has a local maximum, (a) $\xi$ = -0.440 (b) $\xi$ = -0.240 (c) $\xi$ = -0.155. Case 2 with $S = 10\text{\:s}^{-1}$ at $1\:\text{atm}$. San Diego Mechanism. Thickness of the arrows represent the importance of the reaction pathway. Generated with ANSYS Chemkin-Pro\textsuperscript{\textregistered} Reaction Path Analyzer (RPA) \cite{Chemkin_RPA_manual}.}
\label{fig:SDMech_case2_S5_P1_pathways_C}
\end{figure}

\begin{figure}
\centering
\begin{subfigure}[b]{0.55\textwidth}
   \caption{}
	\includegraphics[width=1.2\textwidth]{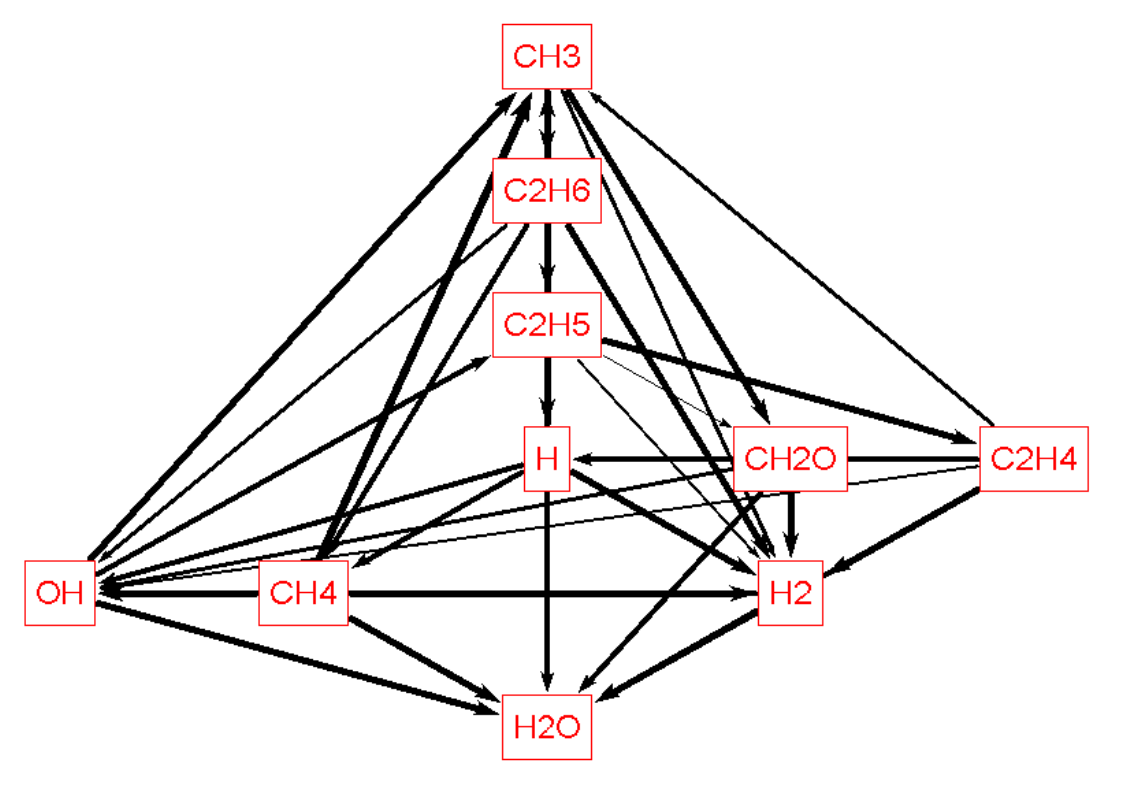}
   \label{fig:SDMech_case2_S5_P1_pathways_a_H} 
\end{subfigure}

\begin{subfigure}[b]{0.7\textwidth}
   \caption{}
	\includegraphics[width=.90\textwidth]{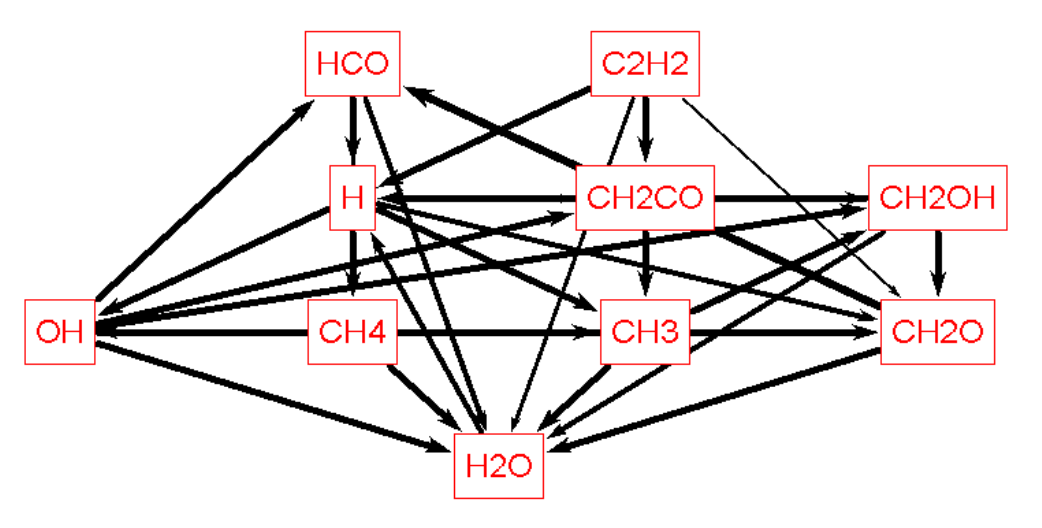}
   \label{fig:SDMech_case2_S5_P1_pathways_b_H}
\end{subfigure}

\begin{subfigure}[b]{0.6\textwidth}
   \caption{}
	\includegraphics[width=.90\textwidth]{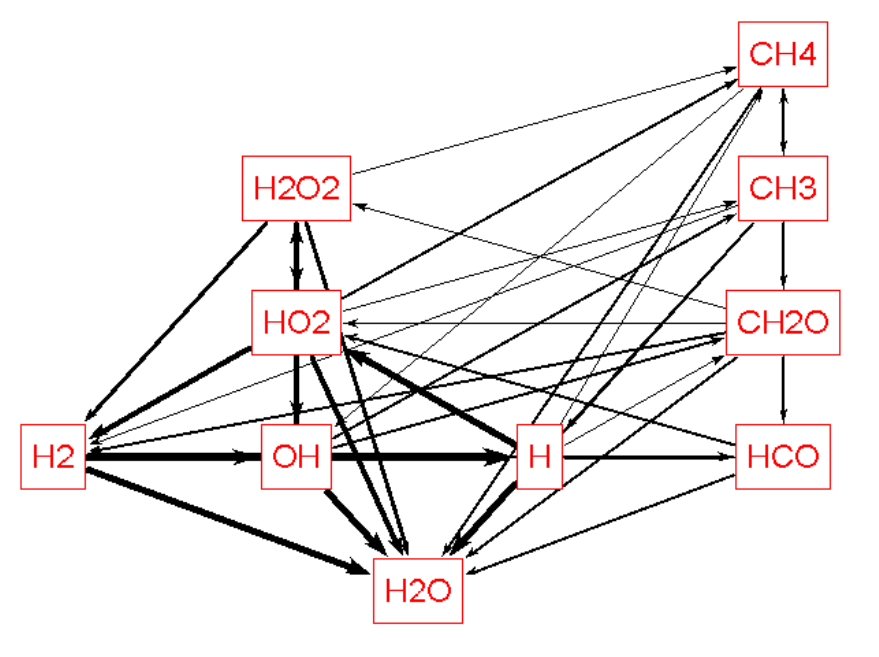}
   \label{fig:SDMech_case2_S5_P1_pathways_c_H}
\end{subfigure}

\caption{Main reaction pathways for species containing hydrogen at the three locations where reaction rate has a local maximum, (a) $\xi$ = -0.440 (b) $\xi$ = -0.240 (c) $\xi$ = -0.155. Case 2 with $S = 10\text{\:s}^{-1}$ at $1\:\text{atm}$. San Diego Mechanism. Thickness of the arrows represent the importance of the reaction pathway. Generated with ANSYS Chemkin-Pro\textsuperscript{\textregistered} Reaction Path Analyzer (RPA) \cite{Chemkin_RPA_manual}.}
\label{fig:SDMech_case2_S5_P1_pathways_H}
\end{figure}

\begin{figure}[h!]
	\centering
	
	\includegraphics[width=.75\textwidth]{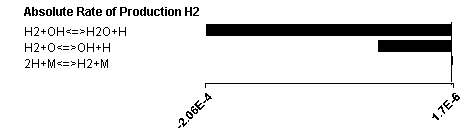}
	
	\caption{Contribution of each reaction on a linear scale to the absolute rate of production of H$_2$ at $\xi$ = -0.155. Case 2 with $S = 10\text{\:s}^{-1}$ at $1\:\text{atm}$. San Diego Mechanism. Rate values at the right and left of the vertical line represent the production and consumption of H$_2$ [mol/(cm$^3$ s)], respectively.}
	\label{fig:SDMech_case2_S5_P1_pathways_sensitivity_analysis}
\end{figure}

Figures \ref{fig:SDMech_case2_S5_P1_temperature} and \ref{fig:SDMech_case2_S5_P1_speciesdetail} show that the maximum temperature and concentration of CO$_2$ and H$_2$O occur at the valley between $\xi$ = - 0.240 and $\xi$ = - 0.155. However, concentrations of CO and H$_2$ peak in the region between the premixed rich flame and the diffusion flame. These two species are not fully depleted across the left peak of the diffusion flame, hence enabling reactions \ref{reac:Production_CO2} and \ref{reac:Consumption_H2} to be active in the right peak of the same flame. The importance of certain reaction pathways at the locations of the three reaction peaks are highlighted in Figures \ref{fig:SDMech_case2_S5_P1_pathways_C} and \ref{fig:SDMech_case2_S5_P1_pathways_H}. Bolder lines imply a more important pathway. Thus, it is expected that the pathways that relate to the premixed fuel-rich flame (Subfigure \ref{fig:SDMech_case2_S5_P1_pathways_a_C}, $\xi$ = - 0.440) are thicker than the ones presented for the diffusion flame (Subfigure \ref{fig:SDMech_case2_S5_P1_pathways_b_C}, $\xi$ = - 0.240) or when there is no flame (Subfigures \ref{fig:SDMech_case2_S5_P1_pathways_c_C} and \ref{fig:SDMech_case2_S5_P1_pathways_c_H}, $\xi$ = - 0.155). Moreover, Subfigures \ref{fig:SDMech_case2_S5_P1_pathways_c_C} and \ref{fig:SDMech_case2_S5_P1_pathways_c_H} further support that reactions \ref{reac:Production_CO2} and \ref{reac:Consumption_H2} are the cause for the right peak of the diffusion flame, since pathways that lead to reactions \ref{reac:Production_CO2} and \ref{reac:Consumption_H2} are represented by darker and thicker arrows. Figure 	\ref{fig:SDMech_case2_S5_P1_pathways_sensitivity_analysis} shows the contribution of the most important reactions on a linear scale to the absolute rate of production of H$_2$ at $\xi$ = -0.155, clarifying that the consumption of H$_2$ is mostly given by reaction \ref{reac:Consumption_H2}. The negative sign in the values at the left side of the vertical line indicates the reverse reaction rate while the positive values at the right side indicates that it is a forward reaction rate. The remainder of the reactions that are not shown in Figure \ref{fig:SDMech_case2_S5_P1_pathways_sensitivity_analysis} do not have a significant contribution to the production or consumption of H$_2$.\hfill \break

\subsubsection{Strain Rate Effects}
\label{Subsection:case2_increaseS}
Figure \ref{fig:SDMech_298K_case2_species} shows that lowering strain rate separates the different flame branches, while increasing it at constant pressure causes merger. The effect of strain rate is better observed at low pressure ($1\:\text{atm}$) and agrees with the observations from Sirignano \cite{WAS_WSCI2019}.\hfill \break

\begin{figure}[h!]
	\centering
	
	\includegraphics[width=.75\textwidth]{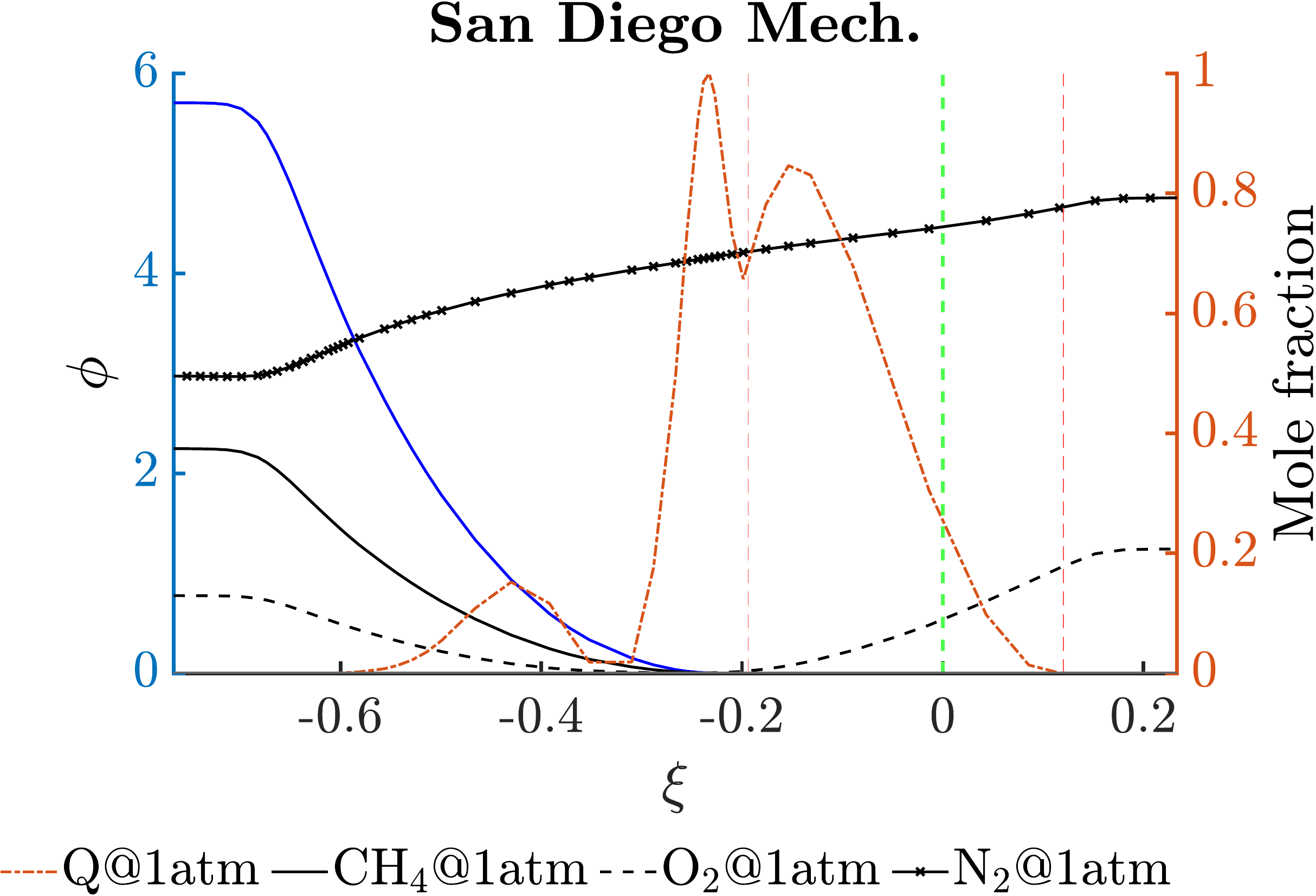}
	
	\caption{Case 2 at $S = 10\text{\:s}^{-1}$ and $1\:\text{atm}$. San Diego Mechanism. Mole fractions of CH$_4$, N$_2$ and O$_2$ (black), normalized heat release rate (orange), and local equivalence ratio (blue). Stagnation plane location (green) and the estimated mixing-layer edge (red). See the online version for color references.}
	\label{fig:SDMech_298K_case2_phi}
\end{figure}

Variations in strain rate have a major impact on the location and the character of the flames. In this Case 2, as in the previous case and contrary to expectation, the character of the premixed flame does not follow the classical premixed flame behavior for any of the conditions studied. The temperature of this flame is, again, higher than the adiabatic flame temperature expected for the equivalence ratio ($\phi$)= 5.70 (see Tables \ref{tab:Adiabatic_flame_T} and \ref{tab:Characteristics_flames_Case1_Case2}). However, similarly to Case 1, Figure \ref{fig:SDMech_298K_case2_phi} shows that the equivalence ratio varies locally due to a local mixture fraction change. Also in Case 2, the adiabatic flame temperature corresponding to the local equivalence ratio at the premixed flame front ($\phi$ = 3.23) is 980.5K for 1 atm, which is inferior to the flame temperature. The same trend is observed for all the pressures and strain rates studied for Case 2 (see Table \ref{tab:Adiabatic_flame_T_flame_front_Case2}).\hfill \break

\begin{figure}[h!]
	\centering
	\includegraphics[width=.45\textwidth]{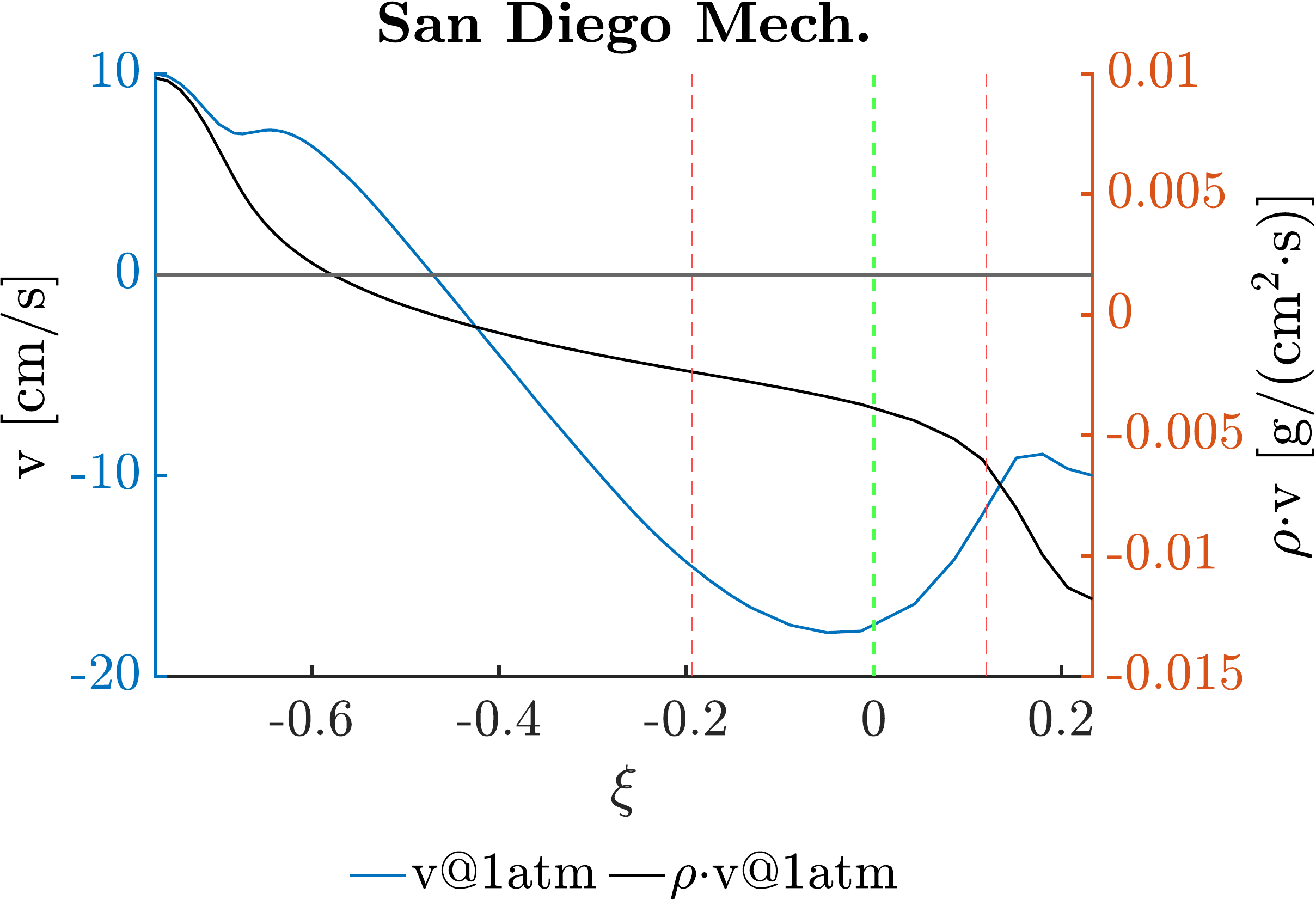}\quad
	\includegraphics[width=.45\textwidth]{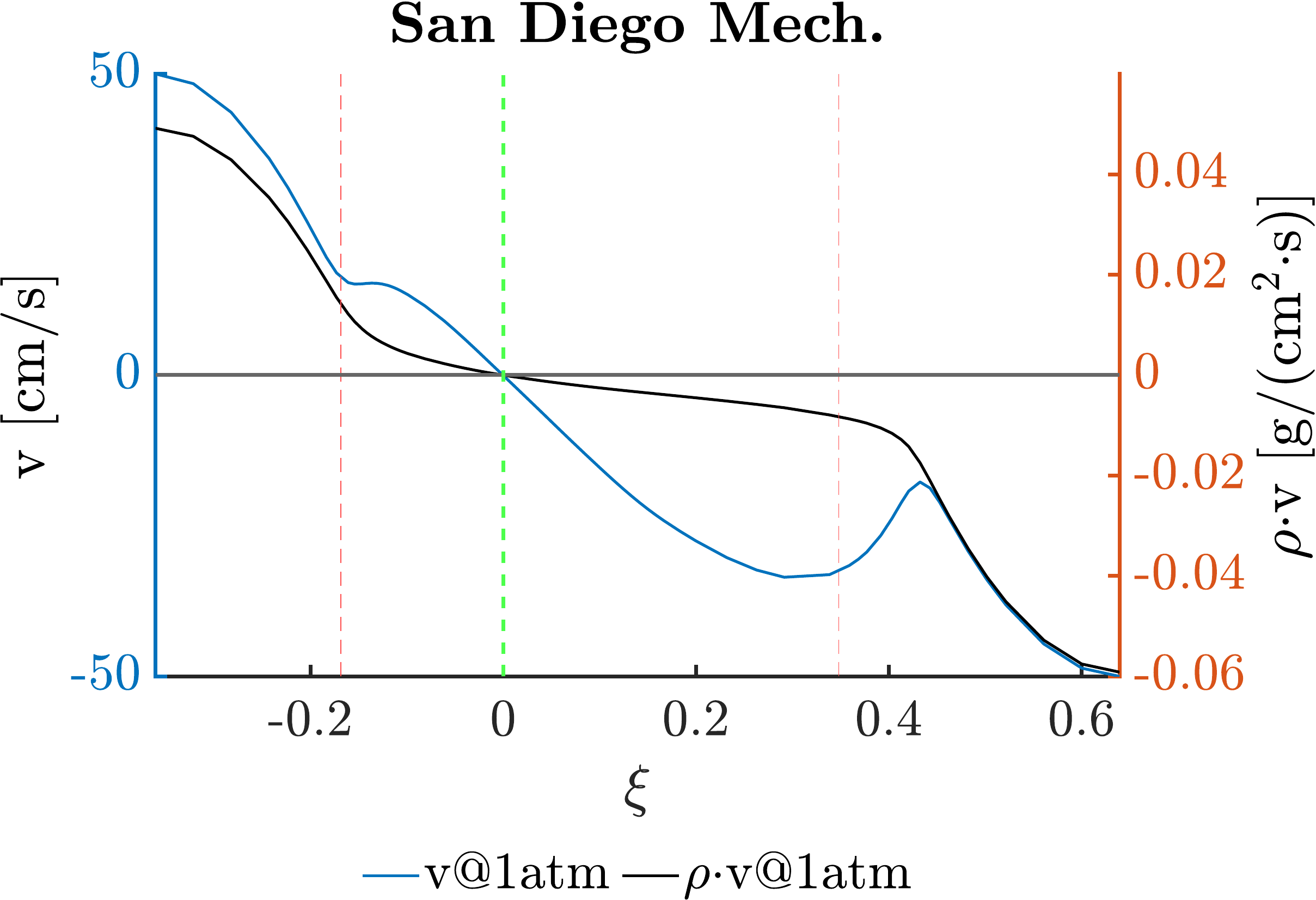}
	
	\medskip
	
	\includegraphics[width=.45\textwidth]{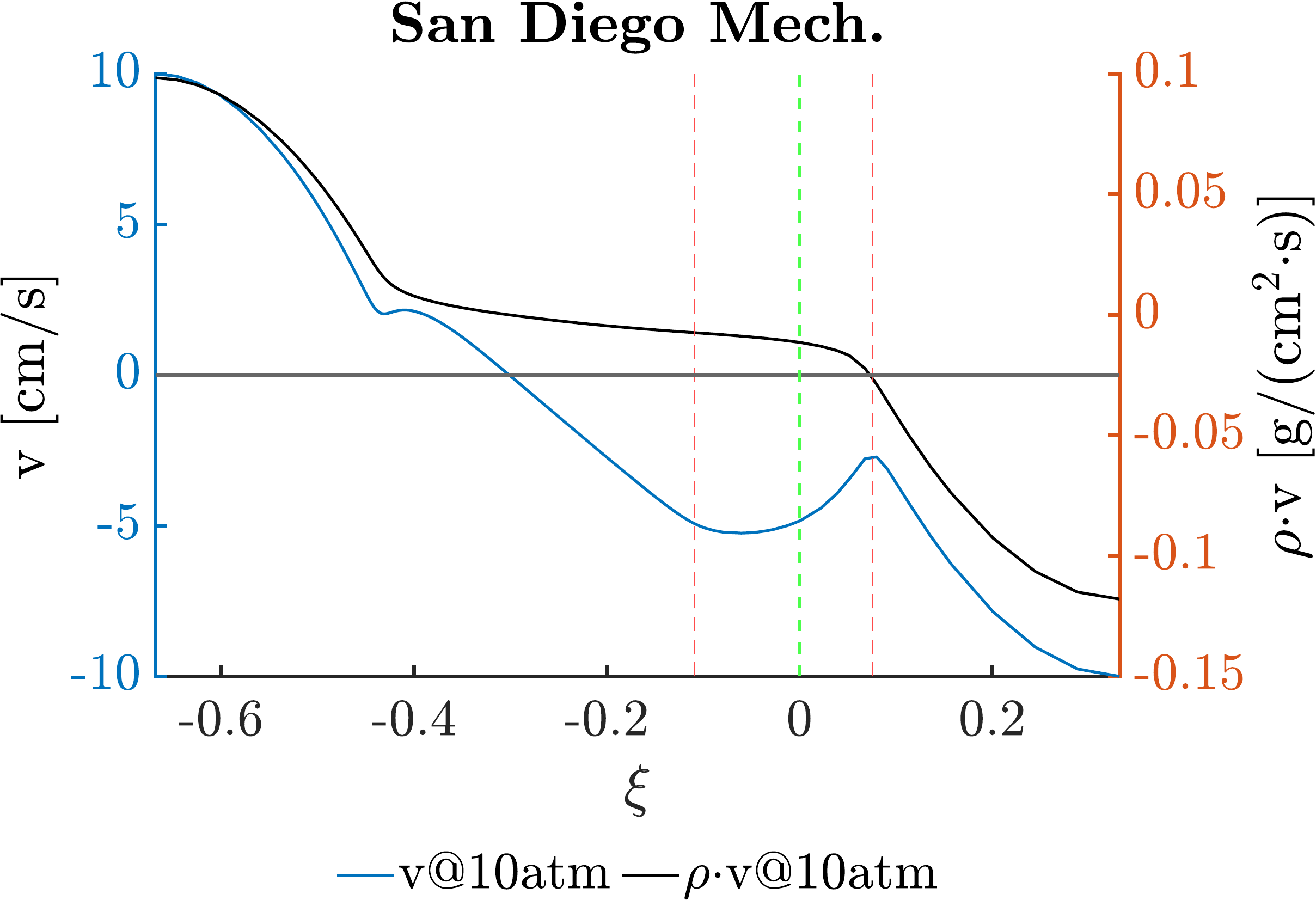}\quad
	\includegraphics[width=.45\textwidth]{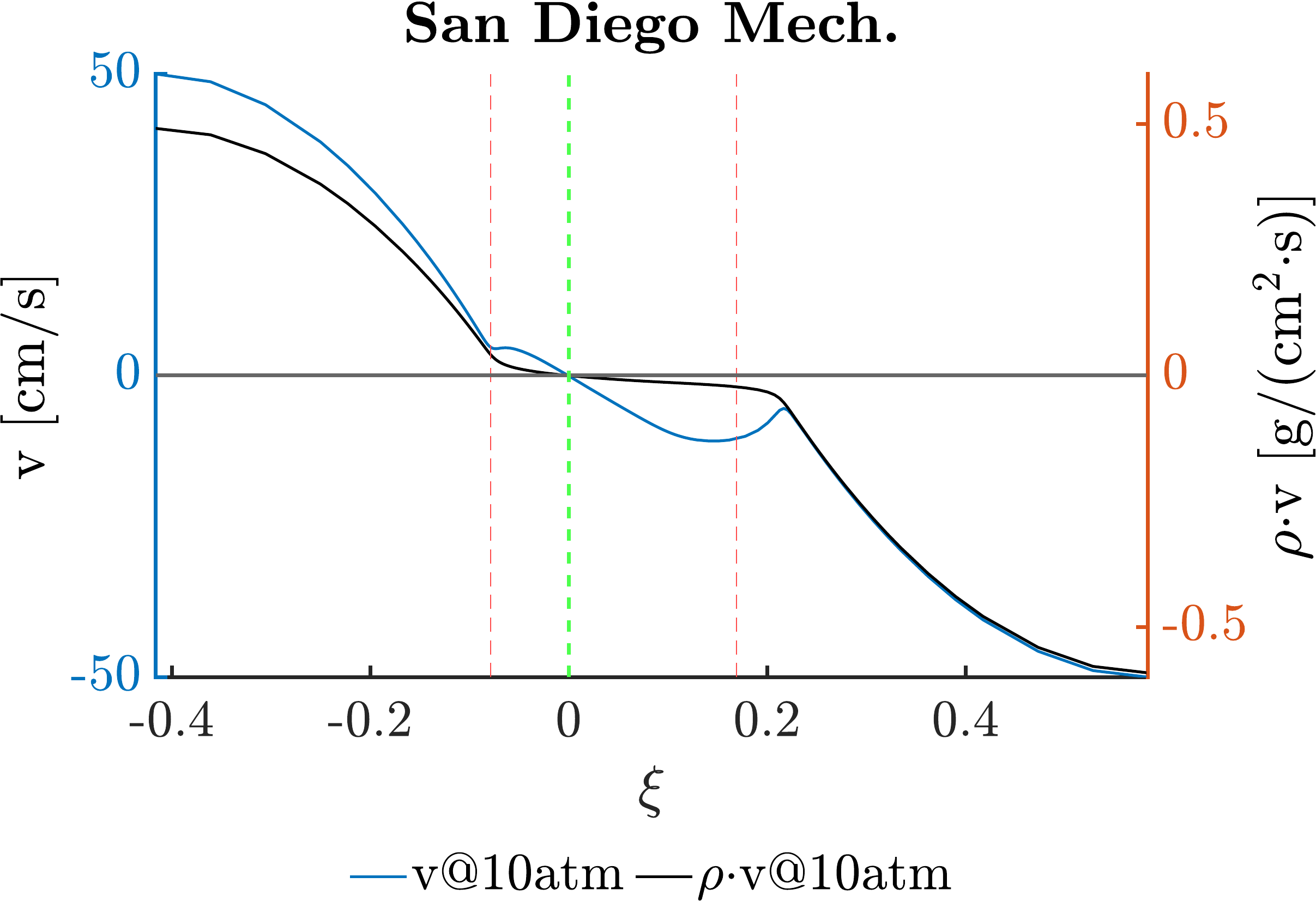}
	
	\medskip
	
	\includegraphics[width=.45\textwidth]{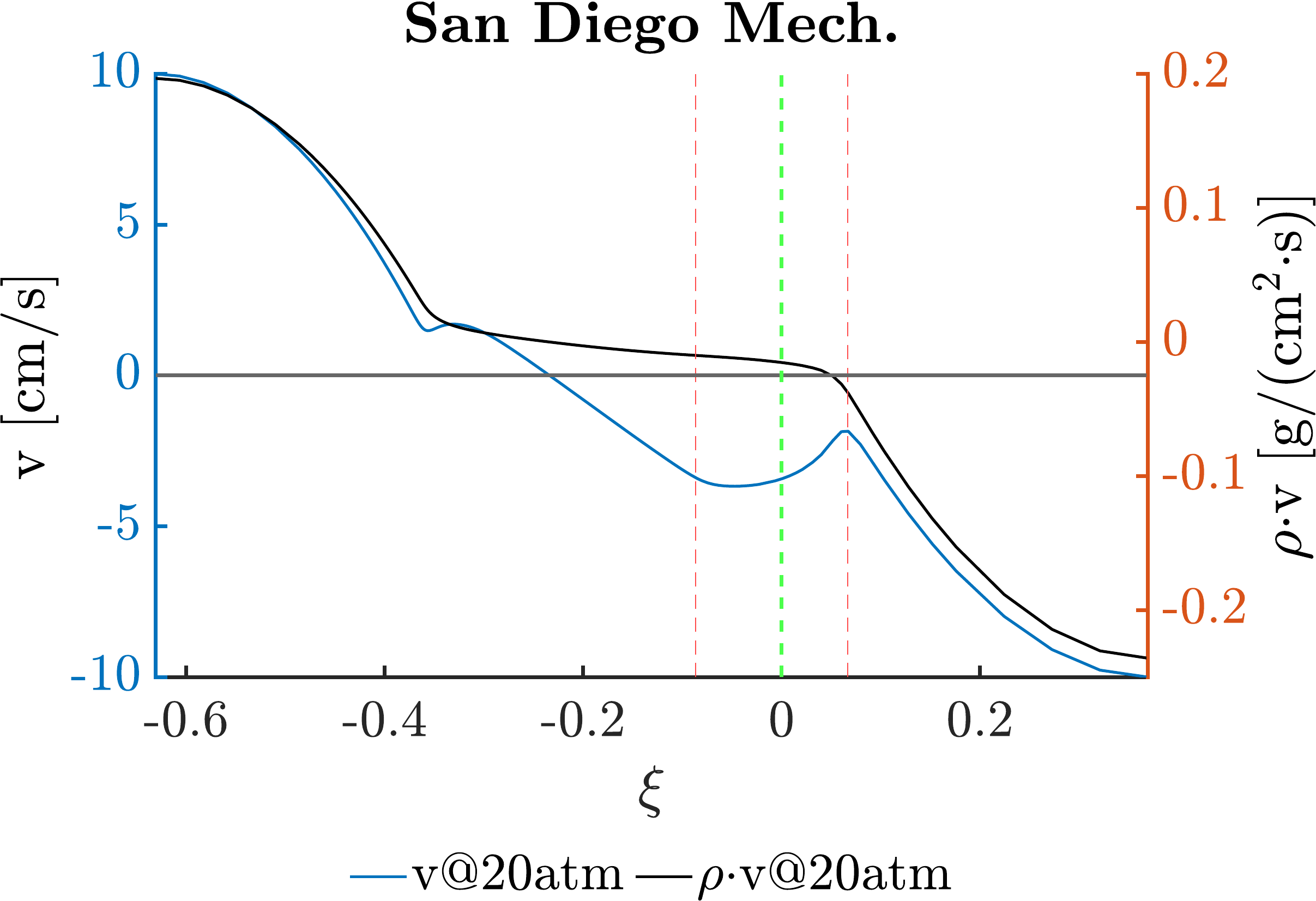}\quad
	\includegraphics[width=.45\textwidth]{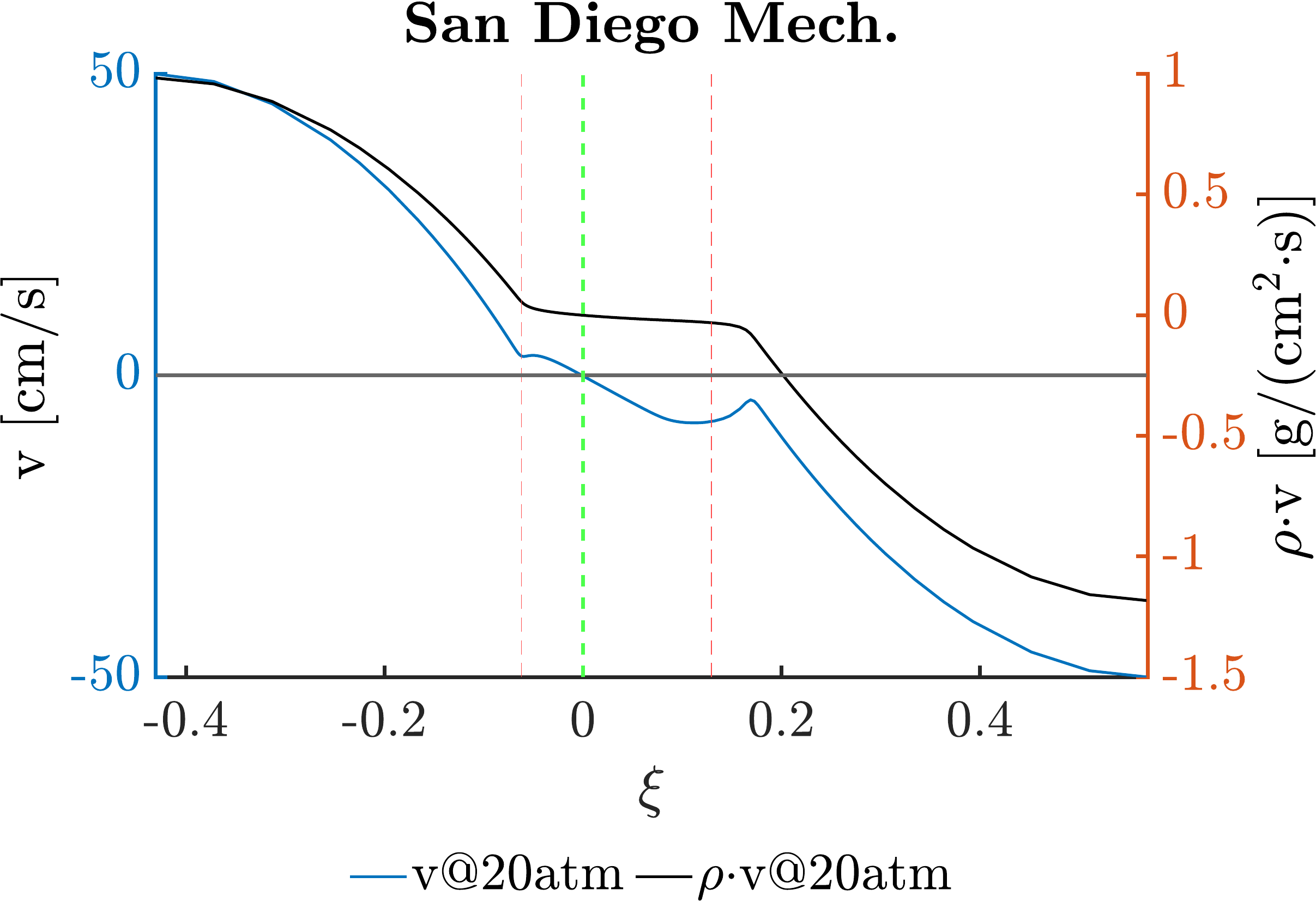}
	
		\caption{Comparison between two different strain rates for Case 2 at $1\:\text{atm}$, $10\:\text{atm}$ and $20\:\text{atm}$. $S = 10\text{\:s}^{-1}$ (left) and $50\text{\:s}^{-1}$ (right). Velocity ($v$) and density ($\rho$) times velocity are plotted. San Diego Mechanism. Stagnation plane location (green) and the estimated mixing-layer edge (red). See the online version for color references.}
	\label{fig:SDMech_298K_case2_densityvelocity}
\end{figure}

Figure \ref{fig:SDMech_298K_case2_densityvelocity} and Table \ref{tab:SDMech_densityvelocity} also show that, similarly to Case 1, the velocity does not remain constant when varying the strain rate. Analogous examination to that for Case 1 in Subsection \ref{Subsection:case1_increaseS} is followed. Consequently, similar to Case 1, the premixed flame is heat-diffusion controlled. Production of active radicals -- such as H, O, and OH -- goes along with the heat production. These radicals could be transported upstream by mass diffusion and also contribute to the temperature increment \cite{Gomez_reviewer1}. Nevertheless, literature has shown that heat diffusion alone is sufficient to cause that effect \cite{WAS_WSCI2019}. The observed diffusive character is strengthened for the higher strain rate ($50\text{\:s}^{-1}$) cases, where both flames are within the mixing-layer boundaries. For all pressures at high strain rate, the left flame falls on the stagnation plane while the right flame is placed on the right side. This implies once again that the premixed left flame has diffusion character even though it receives both reactants from a single side. \hfill \break 

At low strain rate ($10\:\text{s}^{-1}$), the purely diffusion flame (right flame) is found just where the mixing-layer ``edge" is. The edge point is based only on an order-of-magnitude estimate and the flame actually lies in the mixing layer.  All the remaining methane that passes through the premixed flame without burning is consumed in this flame. For all pressures at low strain rate, both flames are placed to the left side of the stagnation plane.  \hfill \break 


\subsection{Case 3}
A fuel-lean mixture is injected from the left nozzle while a fuel-rich mixture enters the domain from the right side. Three flames are expected in this case: one premixed flame burning the fuel stoichiometrically from the right, one premixed flame burning all the fuel from the left, and one diffusion flame in the middle burning the leftover fuel and air coming from the right and left sides, respectively.\hfill \break

For this more complex case, two extra strain-rate values have been studied ($100\:\text{s}^{-1}$ and $150\:\text{s}^{-1}$) and the corresponding pressure effects can be seen in Figures \ref{fig:SDMech_298K_case3_species} and \ref{fig:SDMech_298K_case3_S50_S75}.

\subsubsection{Pressure Effects}
\begin{figure}[ht!]
	\centering
	\includegraphics[width=.45\textwidth]{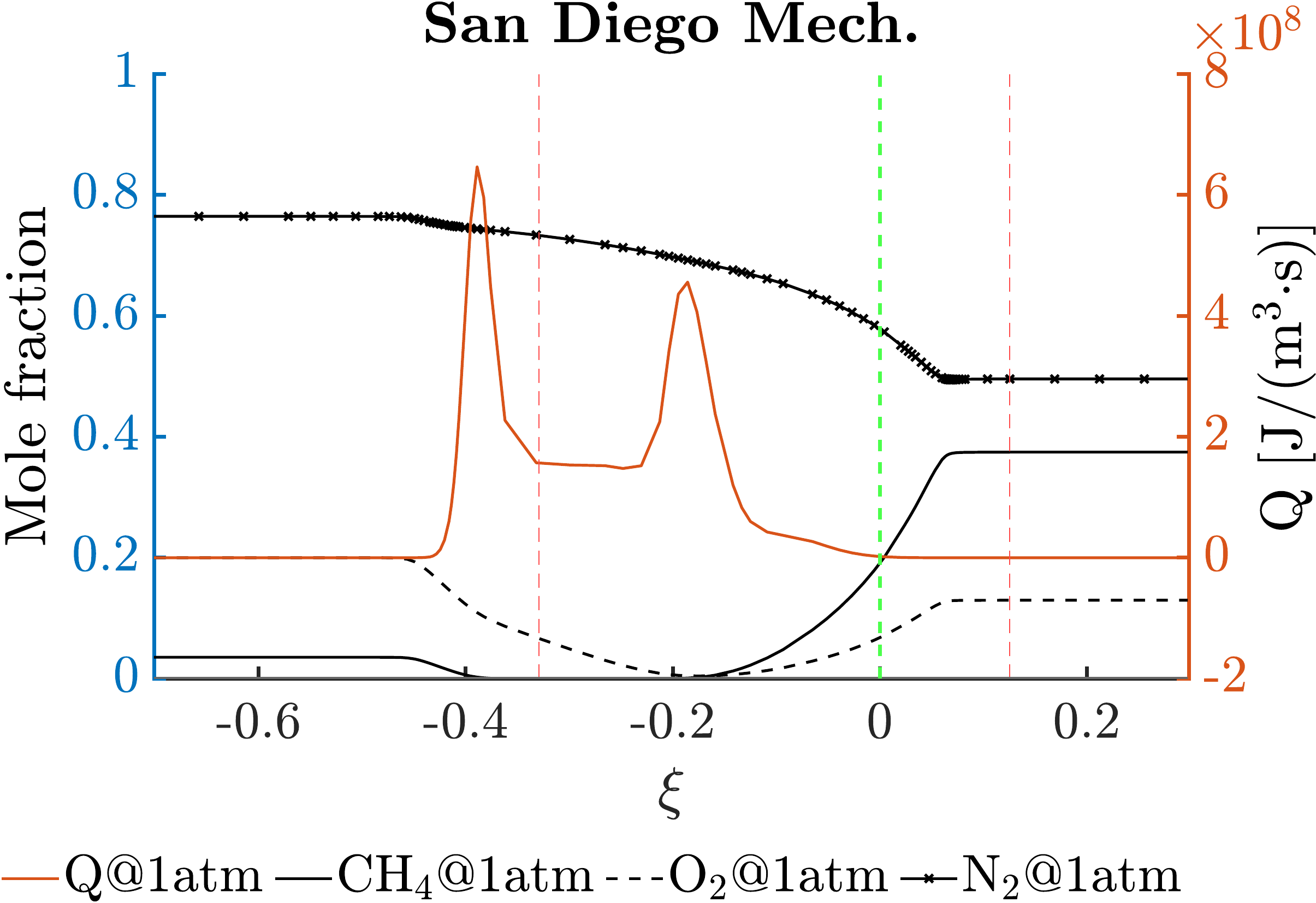}\quad
	\includegraphics[width=.45\textwidth]{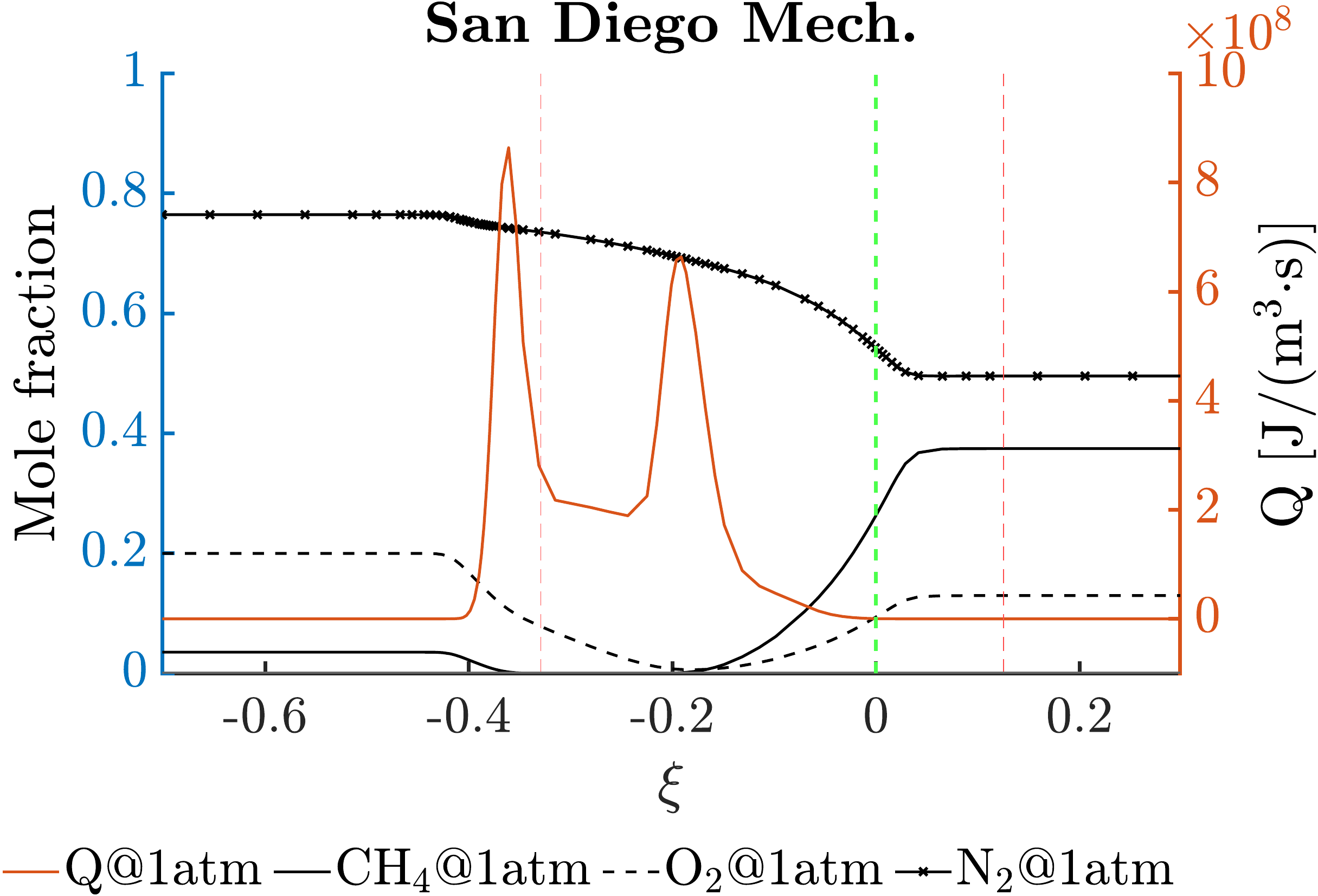}
	
	\medskip
	
	\includegraphics[width=.45\textwidth]{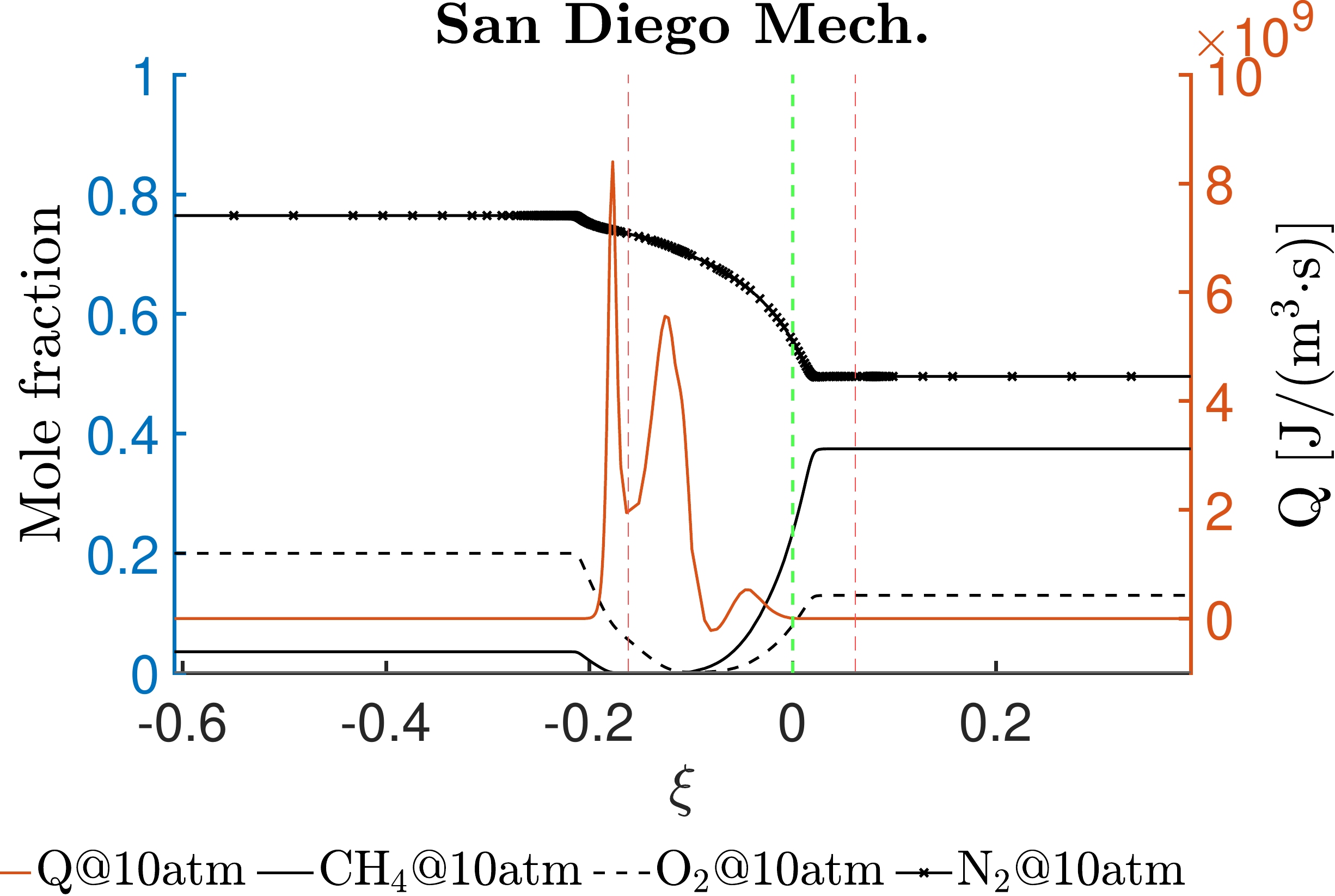}\quad
	\includegraphics[width=.45\textwidth]{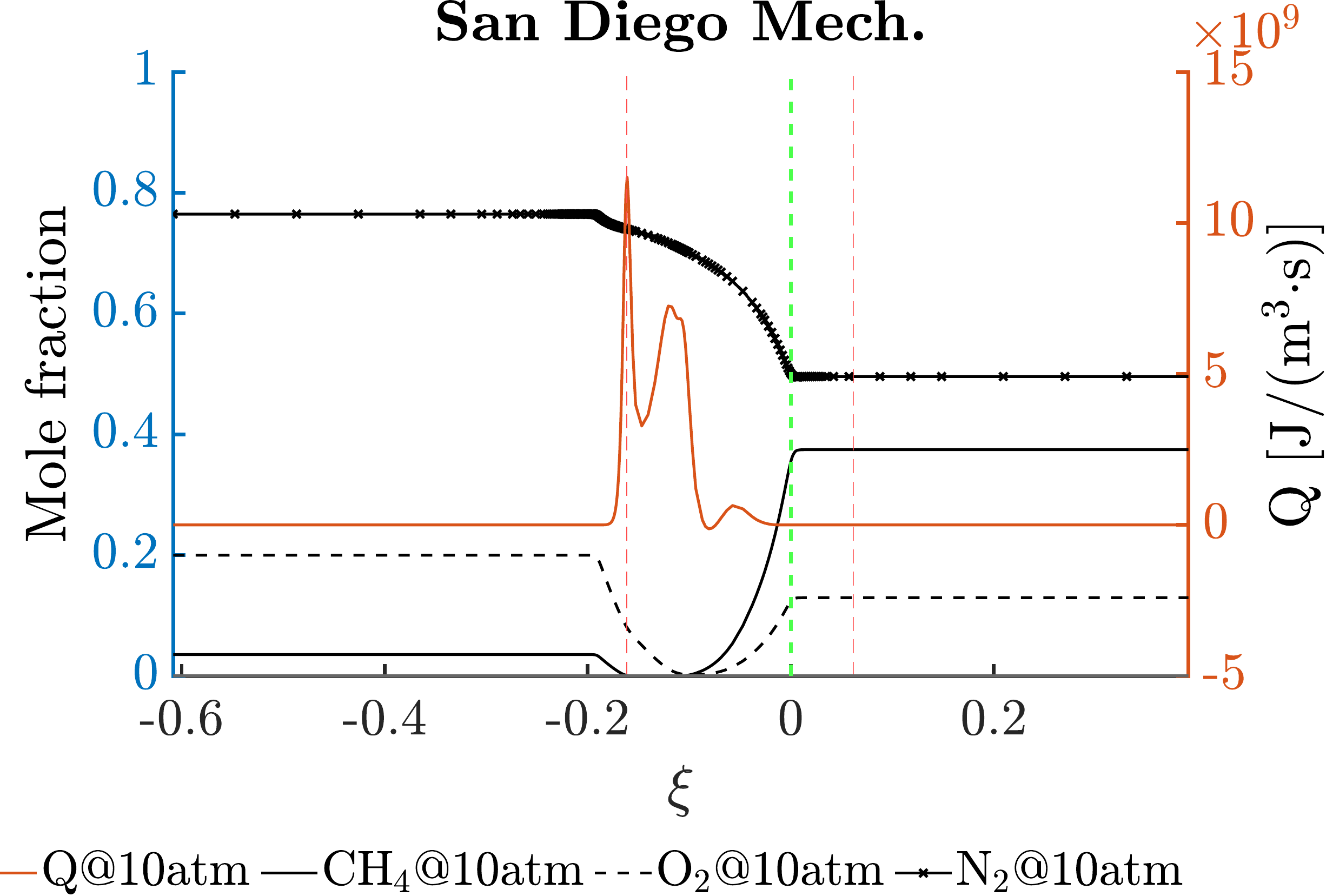}
	
	\medskip
	
	\includegraphics[width=.45\textwidth]{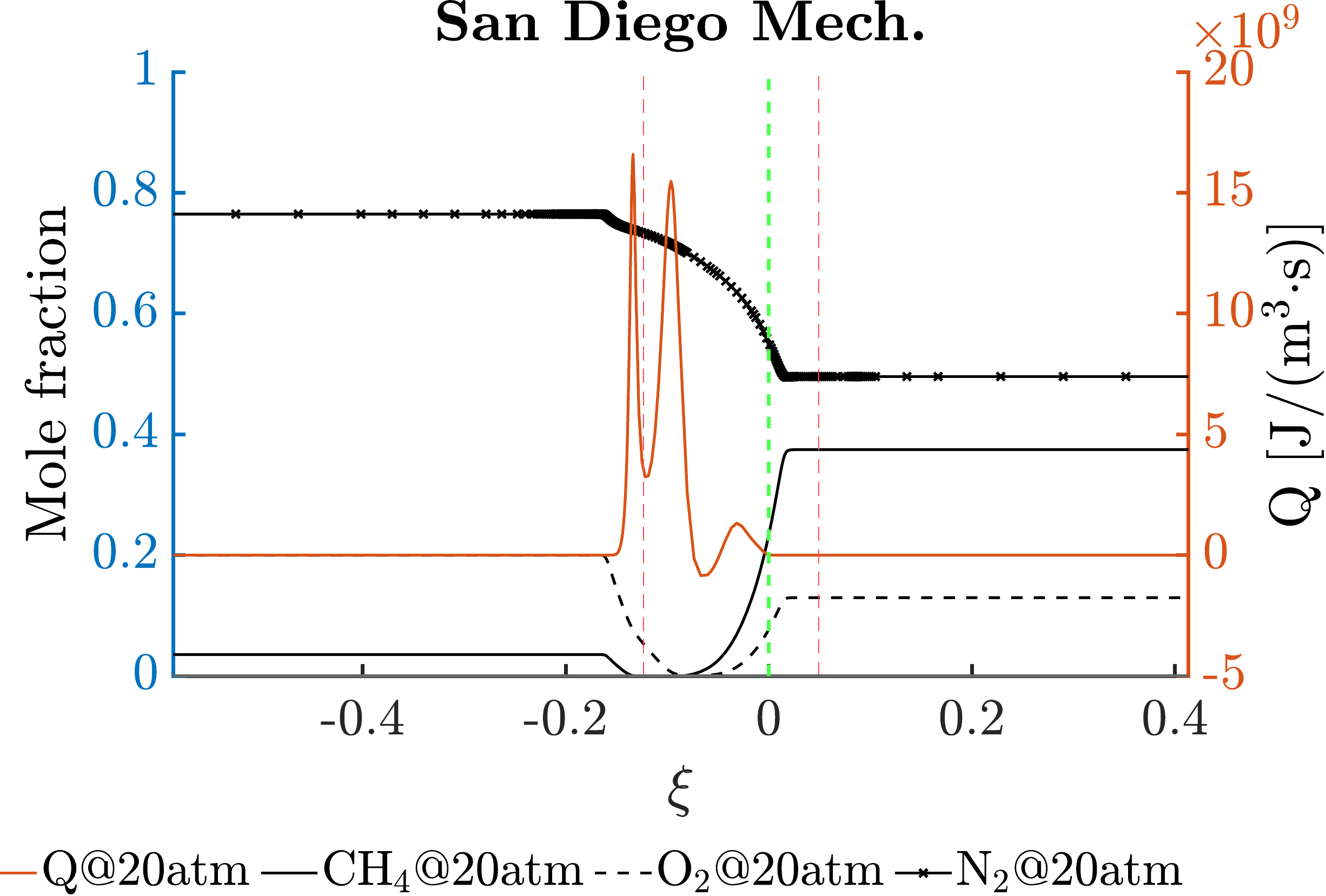}\quad
	\includegraphics[width=.45\textwidth]{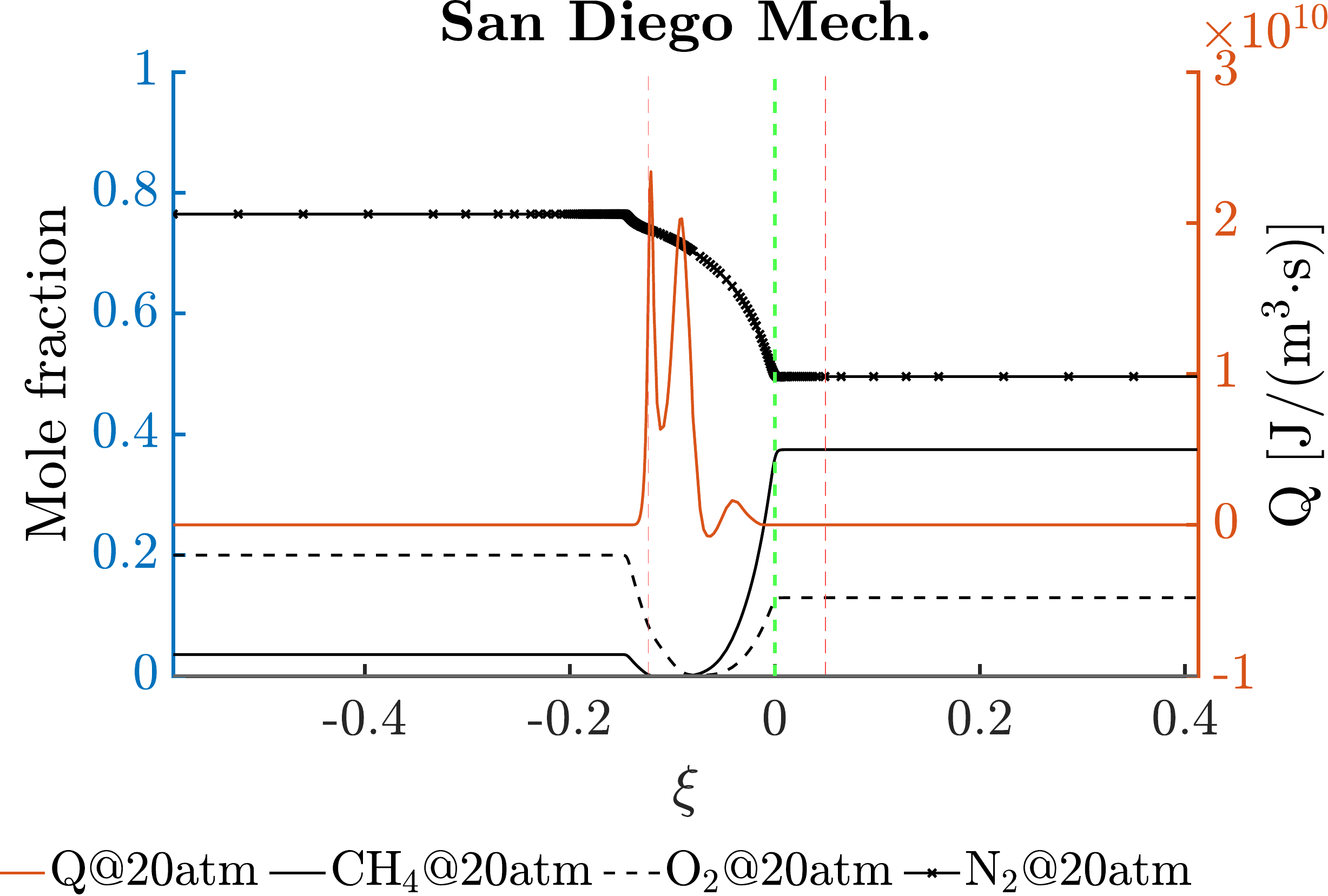}
	
		\caption{Comparison between two different strain rates for Case 3 at $1\:\text{atm}$, $10\:\text{atm}$ and $20\:\text{atm}$. $S = 100\text{\:s}^{-1}$ (left) and $150\text{\:s}^{-1}$ (right). San Diego Mechanism. Mole fractions of CH$_4$, O$_2$ and N$_2$ (black) and heat release rate (orange). Stagnation plane location (green) and the estimated mixing-layer edge (red). See the online version for color references.}
	\label{fig:SDMech_298K_case3_S50_S75}
\end{figure}

As described for Cases 1 and 2, flames become more distinct from each other as pressure is increased. This effect can be observed for all strain rates, but it manifests best at the highest strain rate. In the cases when $S = 100\text{\:s}^{-1}$ and $150\text{\:s}^{-1}$, only two flames are observed at low pressure, being a premixed fuel-lean diffusion character flame (left) and a diffusion flame (right). Notice that Figure \ref{fig:SDMech_298K_case3_S50_S75} clearly shows that the flame on the right has a diffusive character since reactants arrive from opposite sides of the domain and get consumed where the heat-release-rate peaks. Therefore, it is concluded that the expected premixed fuel-rich flame is merged with the diffusion flame. \hfill \break

\begin{figure}[ht!]
	\centering
	\includegraphics[width=.45\textwidth]{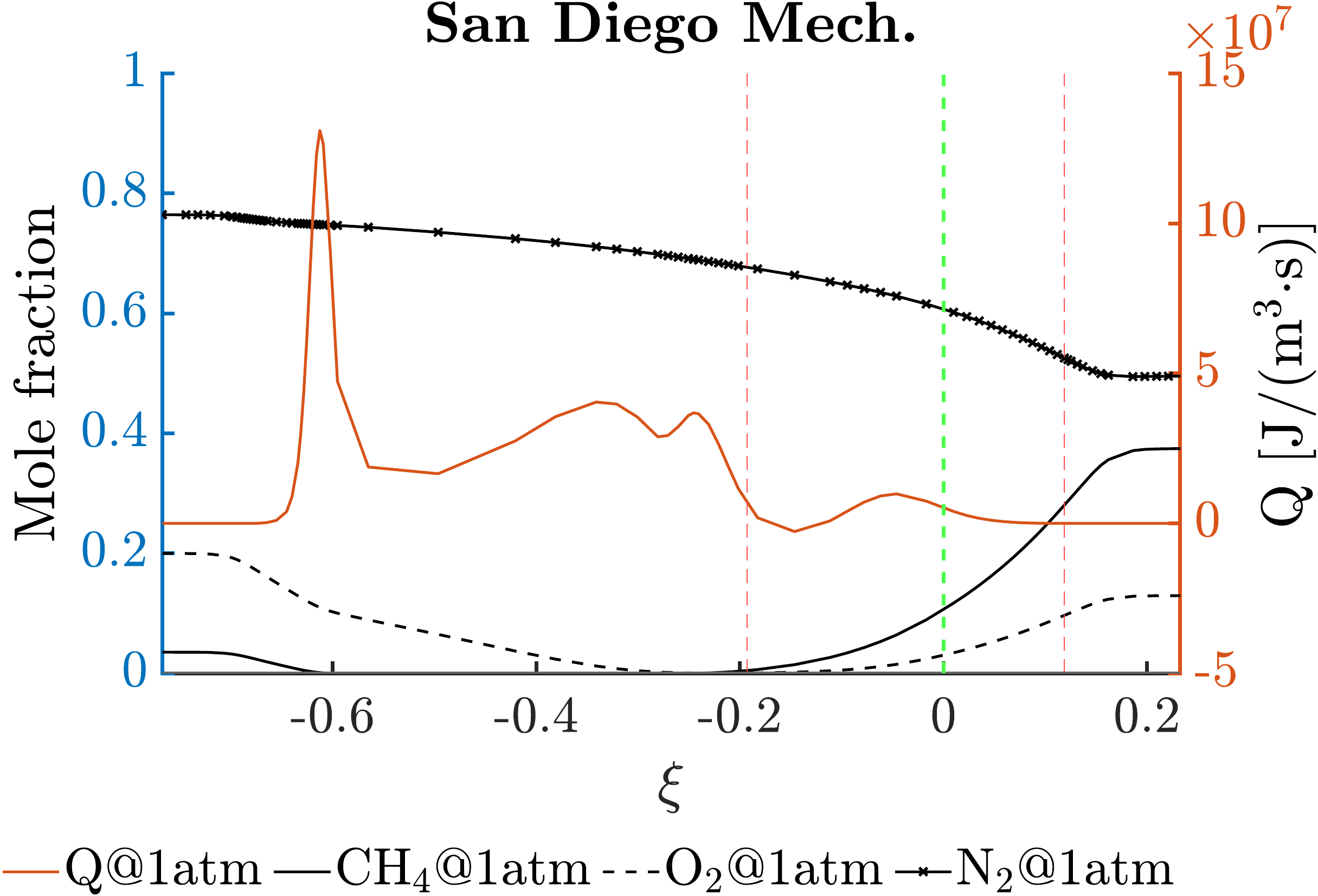}\quad
	\includegraphics[width=.45\textwidth]{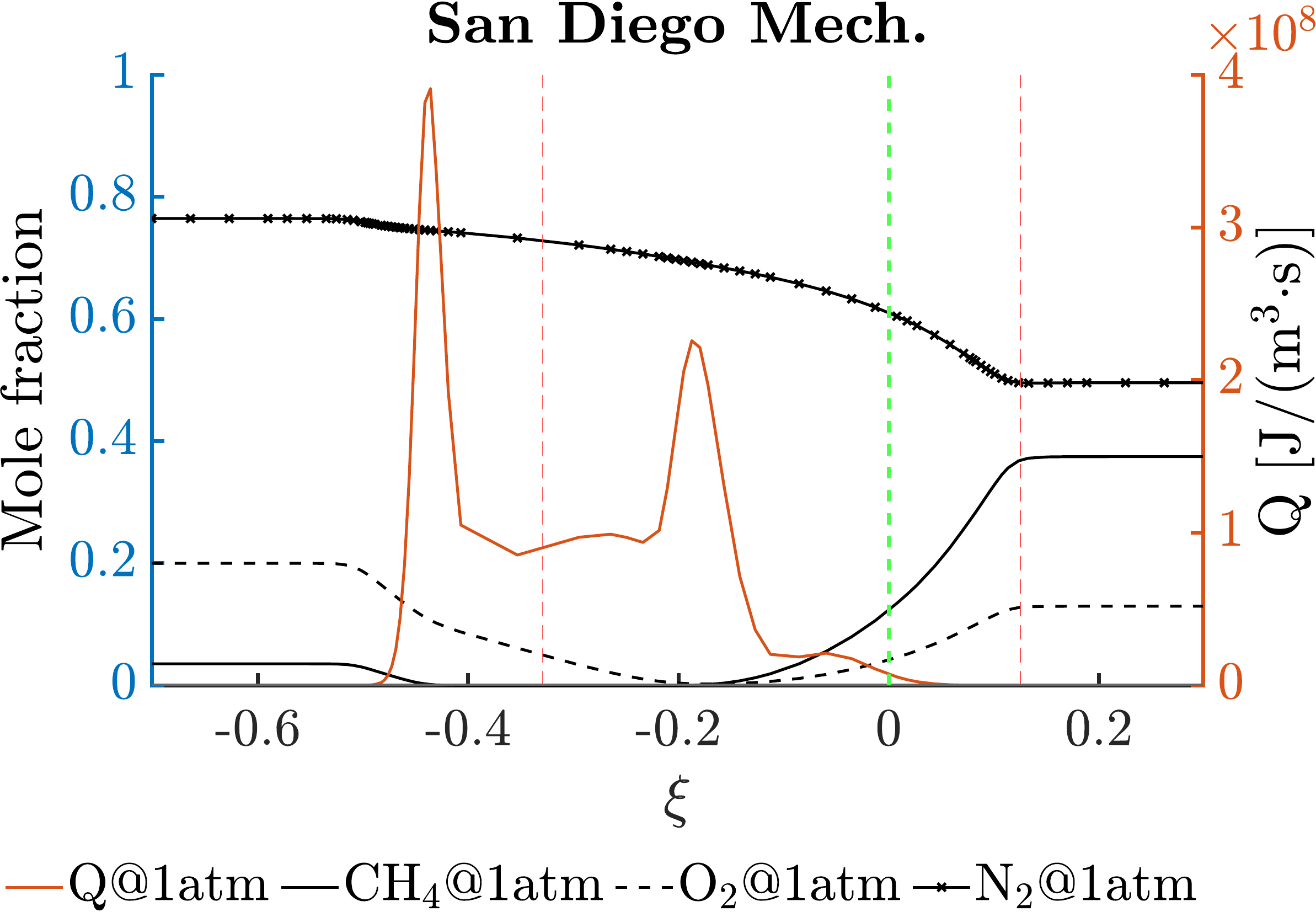}
	
	\medskip
	
	\includegraphics[width=.45\textwidth]{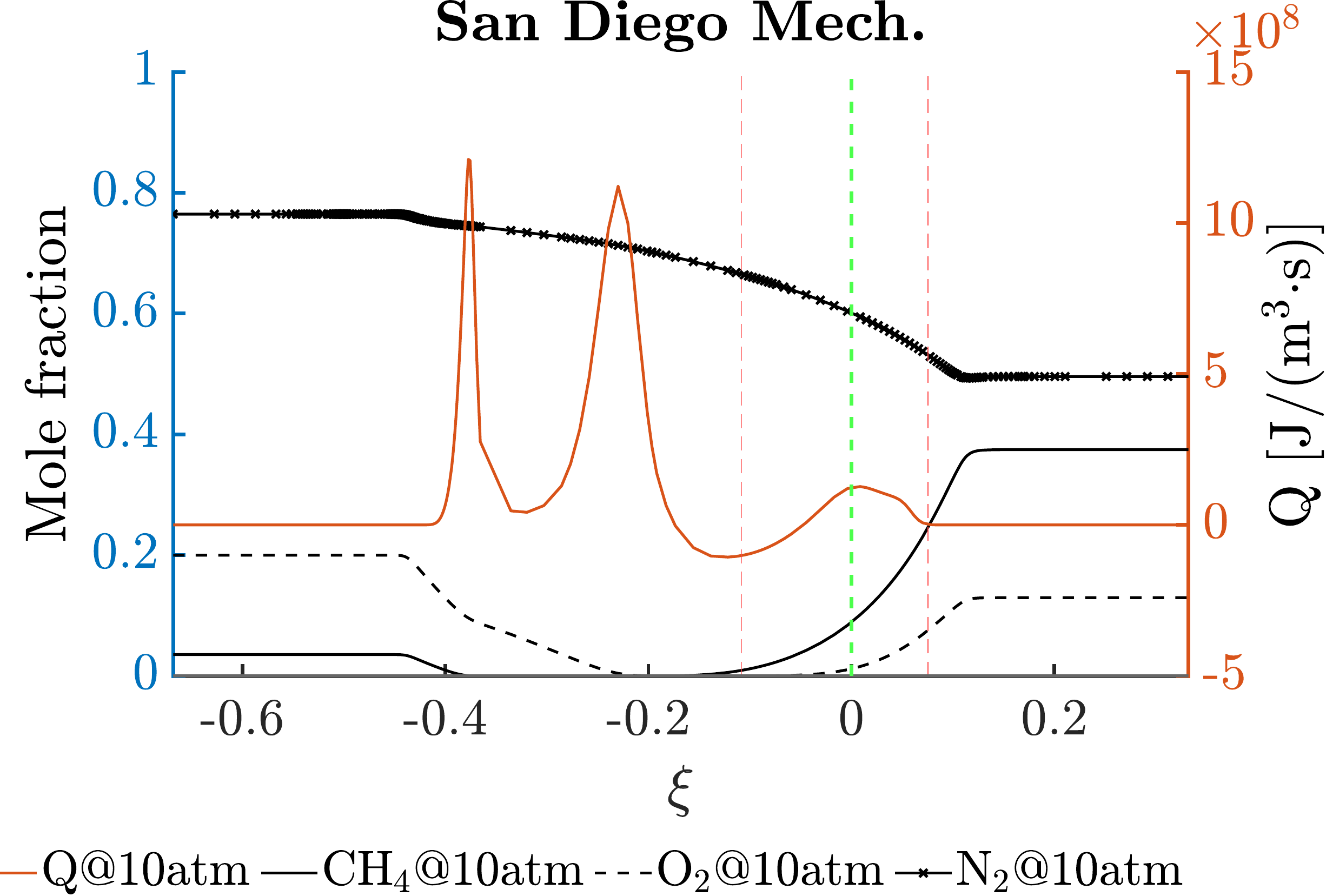}\quad
	\includegraphics[width=.45\textwidth]{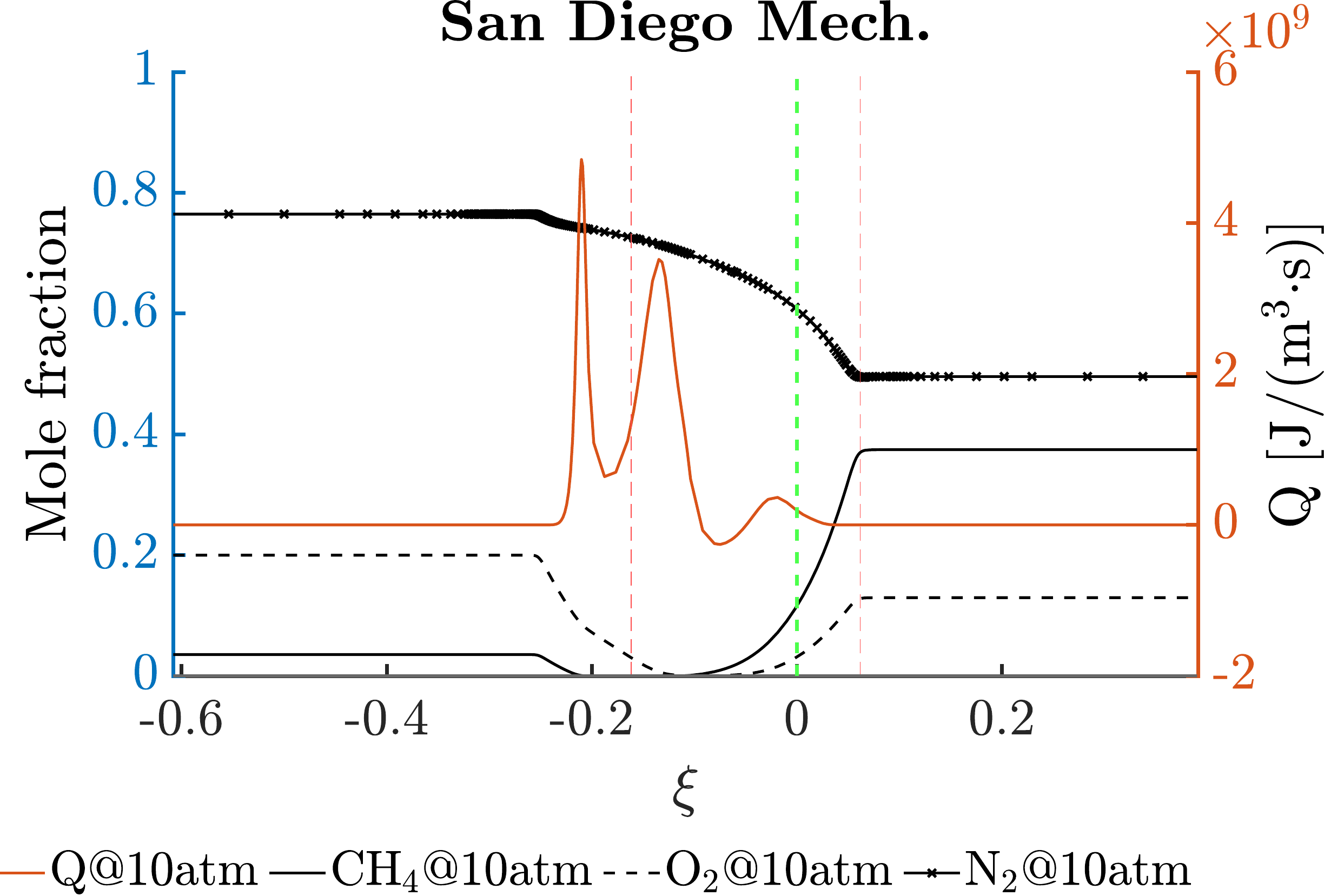}
	
	\medskip
	
	\includegraphics[width=.45\textwidth]{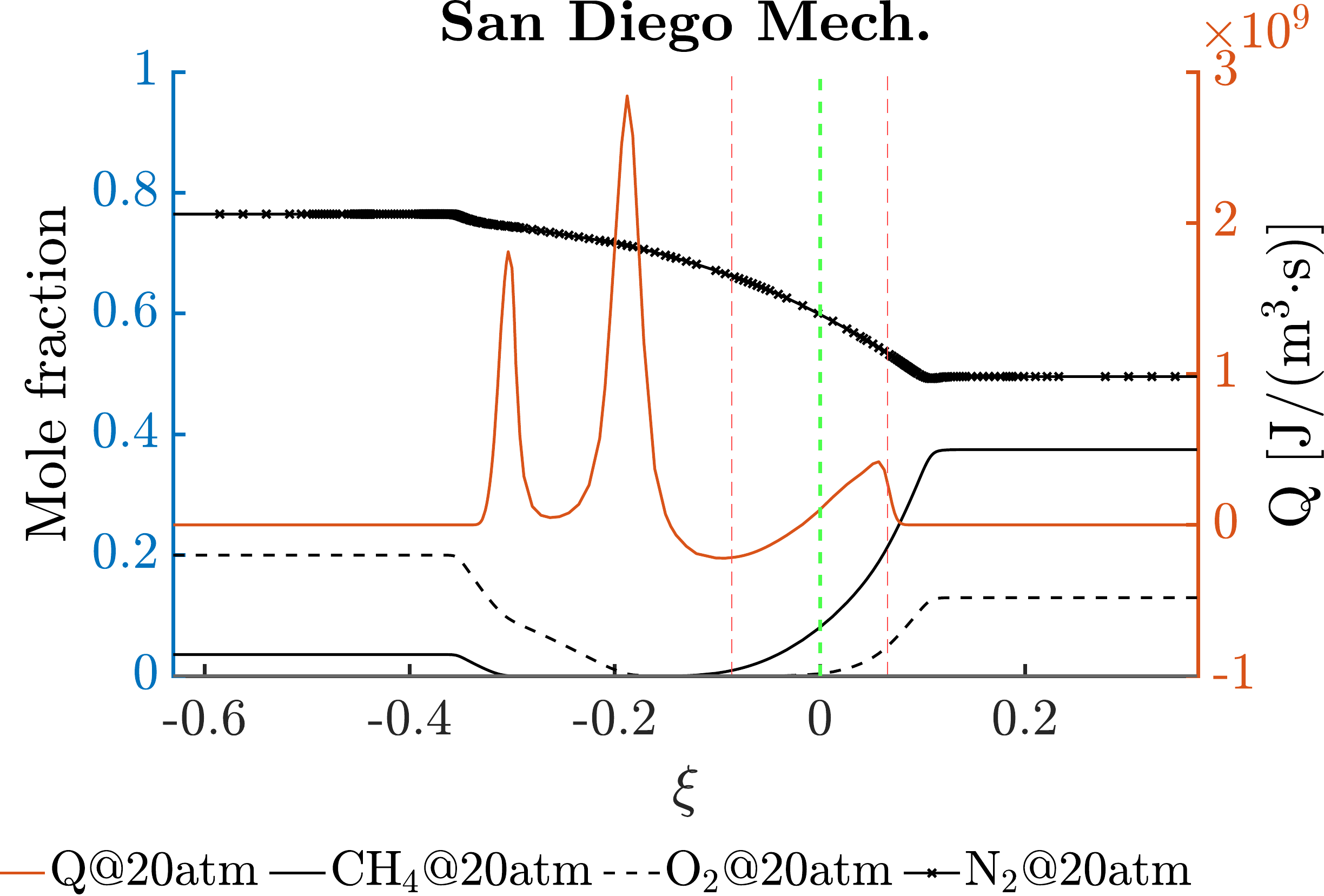}\quad
	\includegraphics[width=.45\textwidth]{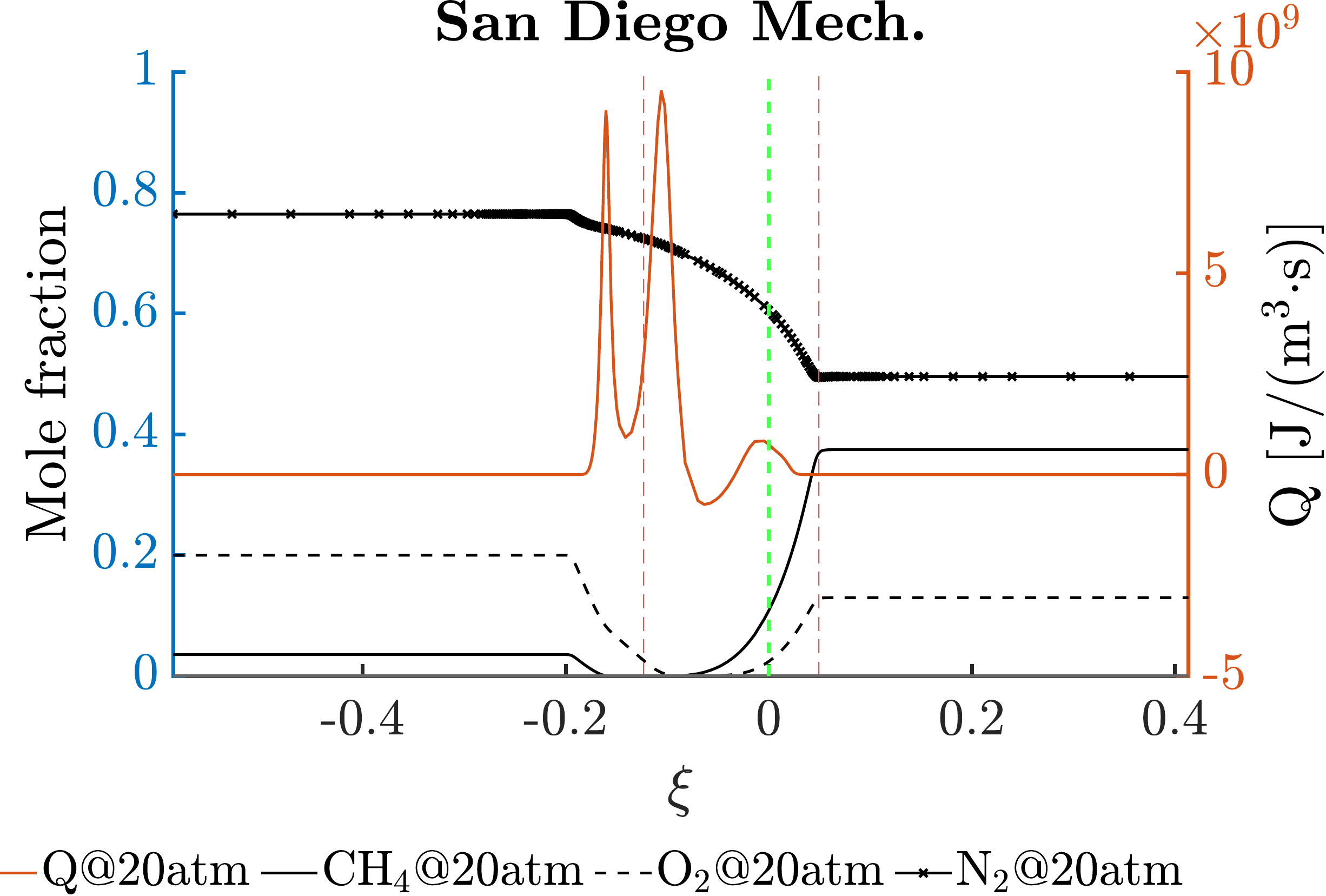}
	
		\caption{Comparison between two different strain rates for Case 3 at $1\:\text{atm}$, $10\:\text{atm}$ and $20\:\text{atm}$. $S = 10\text{\:s}^{-1}$ (left) and $50\text{\:s}^{-1}$ (right). San Diego Mechanism. Mole fractions of CH$_4$, O$_2$ and N$_2$ (black) and heat release rate (orange). Stagnation plane location (green) and the estimated mixing-layer edge (red). See the online version for color references.}
	\label{fig:SDMech_298K_case3_species}
\end{figure}

\begin{figure}[h!]
	\centering
	
	\includegraphics[width=0.8\textwidth]{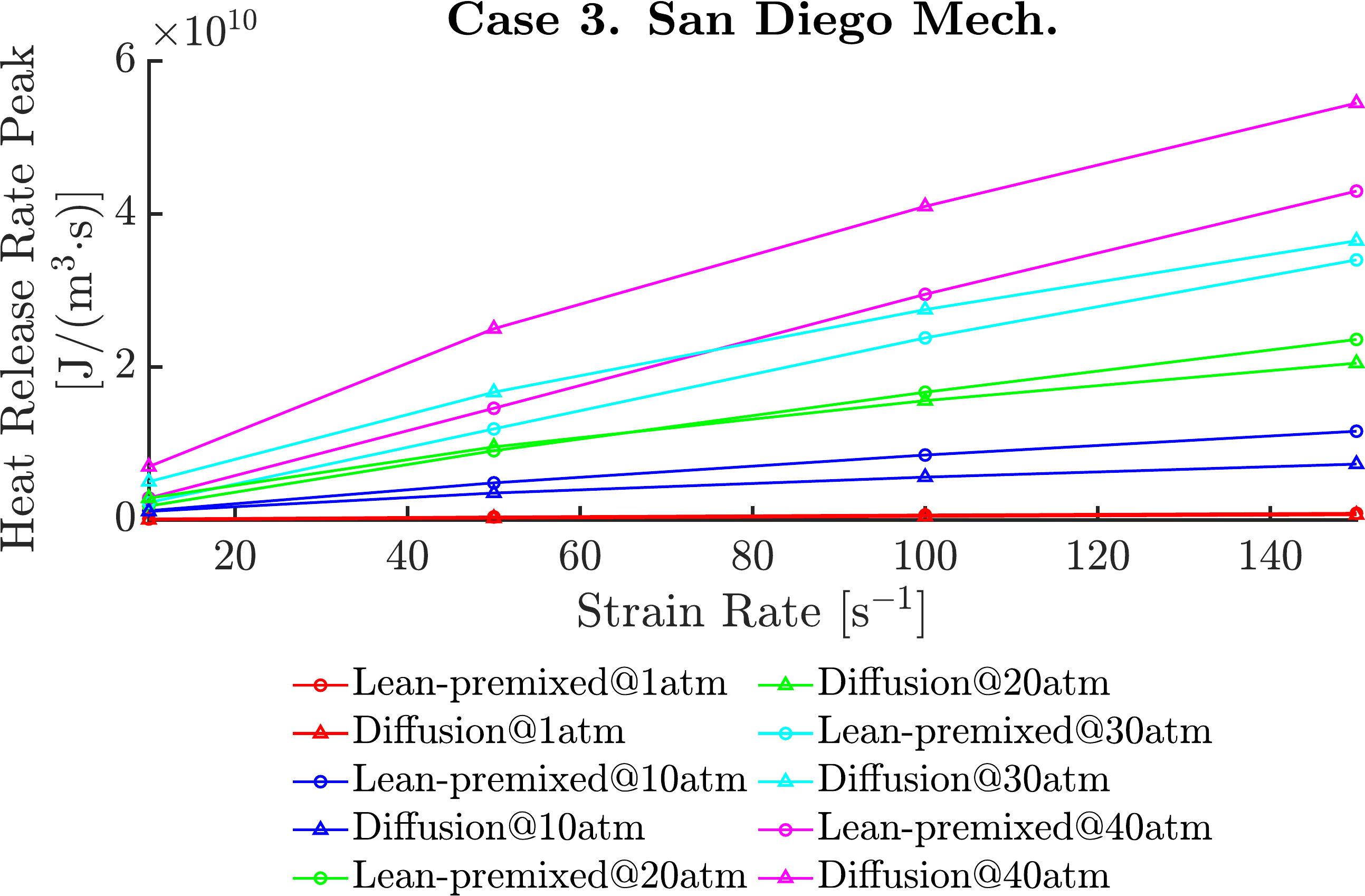}
	
	\caption{Case 3. Heat-release-rate peaks for different strain rates at pressures of $1\:\text{atm}$ (red), $10\:\text{atm}$ (dark blue), $20\:\text{atm}$ (green), $30\:\text{atm}$ (light blue), 40 atm (pink). Only lean-premixed flame (triangles) and diffusion flame (circles) peaks are represented. San Diego Mechanism. See the online version for color references.}
	\label{fig:SDMech_case3_maxQ_differentSP}
\end{figure}

\begin{figure}[ht!]
	\centering
	
	\includegraphics[width=.45\textwidth]{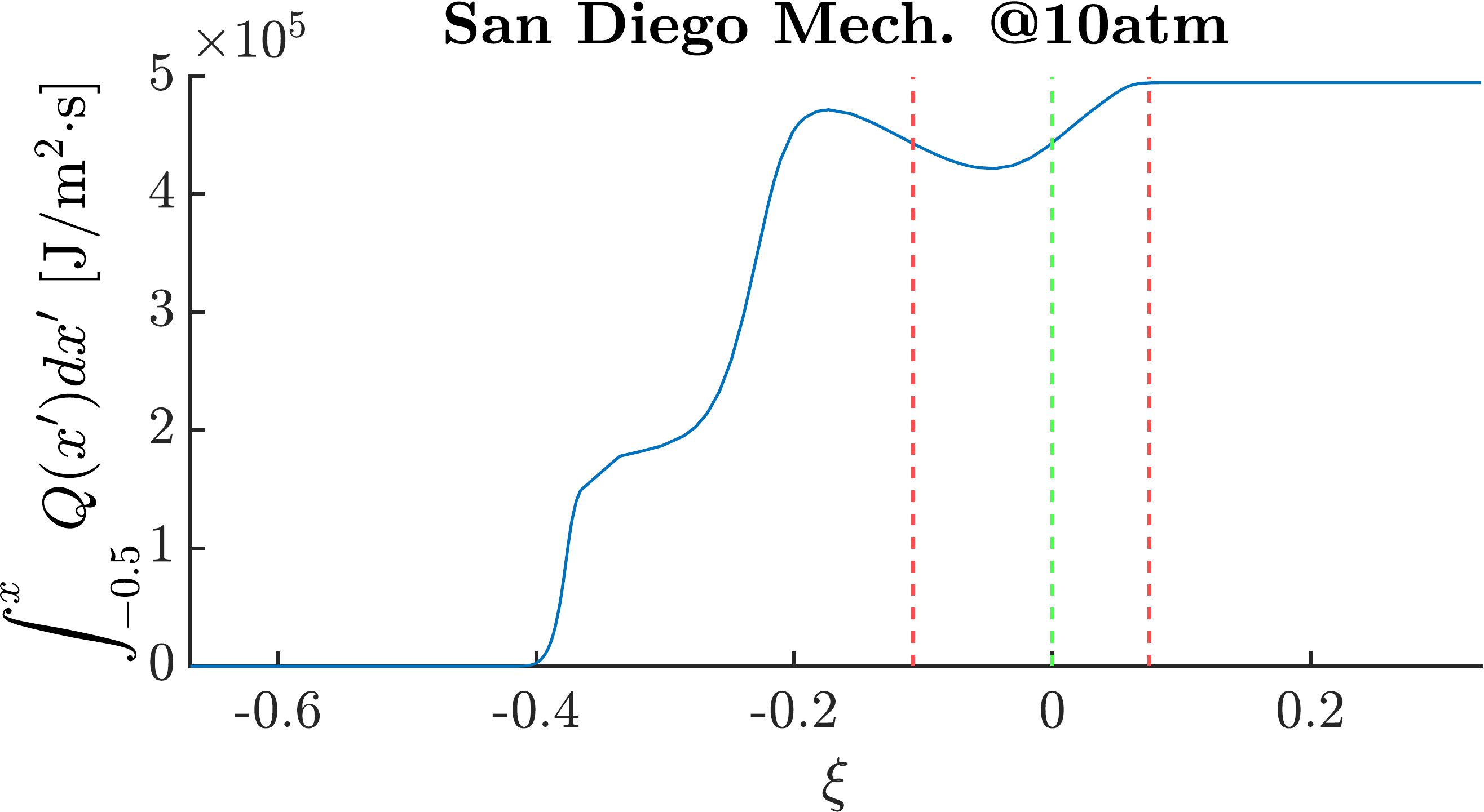}\quad
	\includegraphics[width=.45\textwidth]{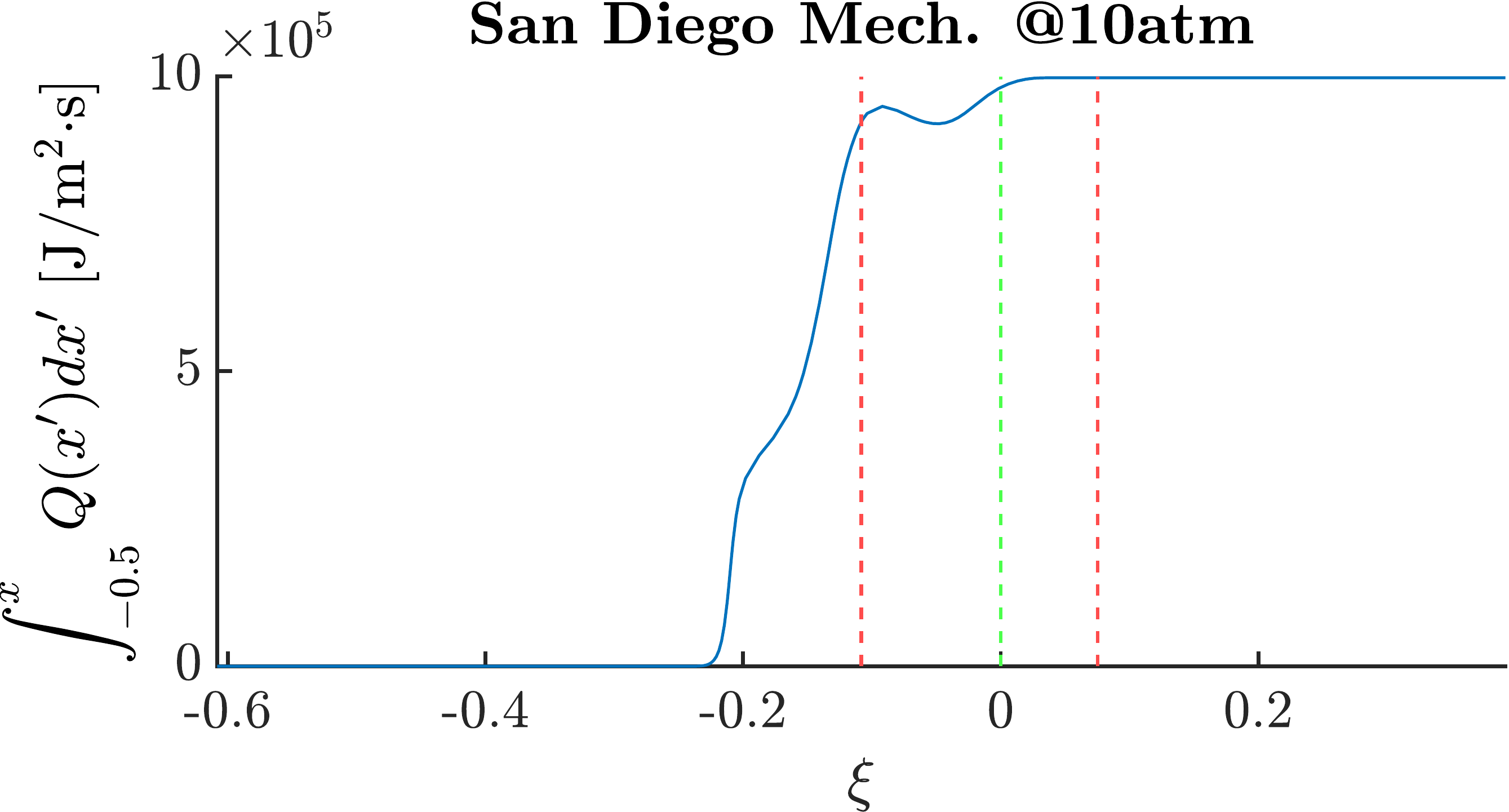}
	
	\medskip
	
	\includegraphics[width=.45\textwidth]{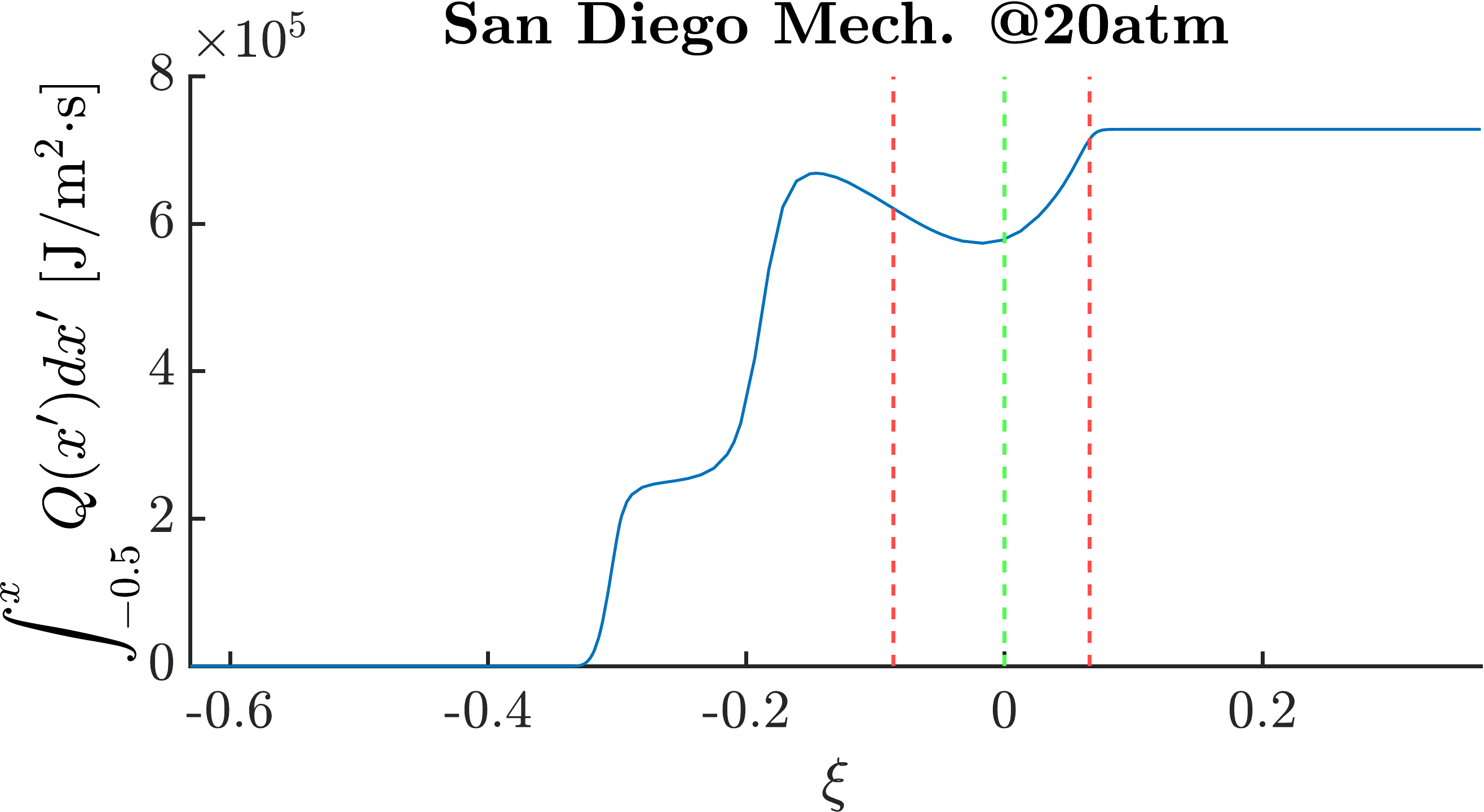}\quad
	\includegraphics[width=.45\textwidth]{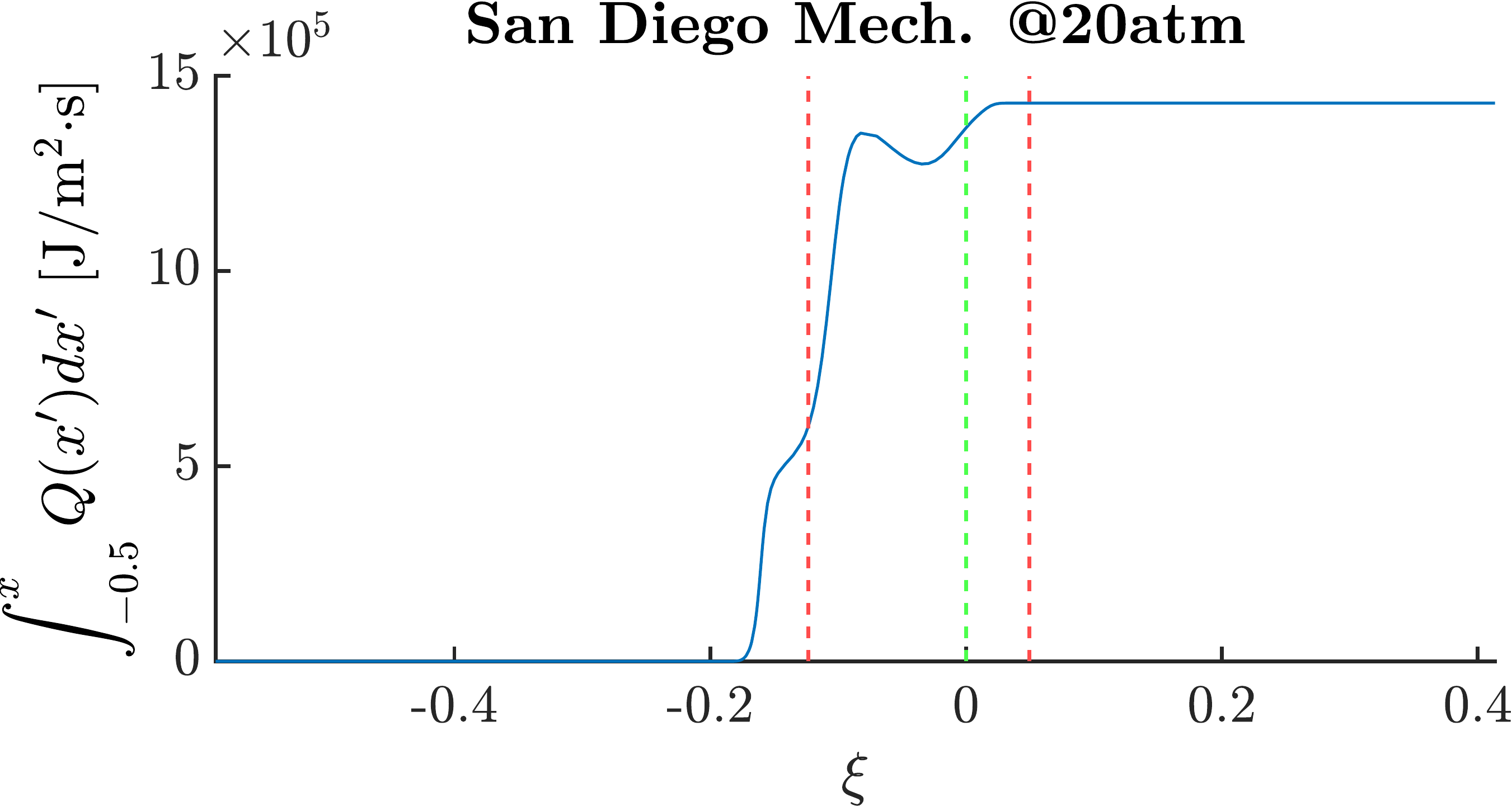}
	
		\caption{Comparison between two different strain rates for Case 3 at $10\:\text{atm}$ and $20\:\text{atm}$. $S = 10\text{\:s}^{-1}$ (left) and $50\text{\:s}^{-1}$ (right). San Diego Mechanism. Stagnation plane location (green) and the estimated mixing-layer edge (red). See the online version for color references.}
	\label{fig:SDMech_298K_case3_Q_int}
\end{figure}

The character of these flames can be classified, from left to right, as fuel-lean premixed and diffusion controlled, diffusion, and fuel-rich having both diffusion and premixed character (see Table \ref{tab:Characteristics_flames_Case3} for more details). For most configurations, the peak heat release rate is higher from left to right, with the fuel rich flame being the weakest of the three (see Figures \ref{fig:SDMech_298K_case3_species} and \ref{fig:SDMech_298K_case3_S50_S75}). However, at the highest pressure for the lower strain rates of $10\:\text{s}^{-1}$ and $50\:\text{s}^{-1}$ (Figure \ref{fig:SDMech_298K_case3_species}), the diffusion-flame reaction peak is the highest. See Figure \ref{fig:SDMech_case3_maxQ_differentSP} where this is highlighted; there is a crossover at a given strain rate where the lean-premixed flame switches from weaker to stronger compared to the diffusion flame. This crossover point occurs at higher strain rate with increasing pressure. Residence time for the diffusion flame is proportional to the reciprocal of strain rate. The diffusion flame weakens with decreasing residence time since a certain time is needed for complete reaction. That being said, the integral of the heat release rate versus the axial coordinate shown in Figure \ref{fig:SDMech_298K_case3_Q_int} clearly highlights how the jump corresponding to the diffusion flame is larger than those related to the other heat-release-rate peaks. Figure \ref{fig:SDMech_298K_case3_Q_int} also shows that the total heat release for the diffusion flame is dominant. Furthermore, profiles of mole fractions (Figure \ref{fig:SDMech_298K_case3_species}) indicate that the mass diffusion rate on major species at the premixed flame is determined by the demand created by the diffusion flame. Therefore, the diffusion flame produces more heat even for the cases where the fuel-lean premixed flame heat-release-rate peak is higher. That is explained as follows: for the case with a mass fraction of 0.02 for the fuel to be burned in the lean flame, only an O$_2$ fraction of a 0.04 will be consumed there, leaving a mass fraction of about 0.185 to be consumed in the diffusion flame. \hfill \break

Examining the case at the lowest pressure and strain rate, the lean and rich premixed flames are represented by the heat-release-rate peaks placed at $\xi$ = - 0.612 and $\xi$ = - 0.045, respectively. The premixed flame establishes its residence time based on pressure and mixture ratio without regard to strain rate. It simply relocates its position so that its propagation velocity matches the incoming stream velocity. The heat-release-rate peak observed at $\xi$ = - 0.340 refers to the zone where exothermic reactions \ref{reac:Production_CO2} and \ref{reac:Consumption_H2} are occurring. Therefore, the region from $\xi$ = - 0.340 to the end of the domain shows similar behavior to the one explained for Case 2 in Section \ref{Subsection:case2_increaseP}. In this case, this heat-release-rate peak is placed at the left side of the rich premixed flame peak and on the right of the diffusion flame peak, since the rich premixed mixture comes from the left side of the domain - i.e., opposite side than in Case 2.

\begin{center}
\LTcapwidth=\textwidth
\begin{longtable}{c|c|c|c|c}
	\caption{Characteristics of the heat-release-rate ($Q$) peaks for Case 3. $T$ stands for temperature [K], $P$ for pressure [atm], and $S$ to strain rate [$\text{\:s}^{-1}$].}
\label{tab:Characteristics_flames_Case3}\\

\hline  $P$ [atm] & $S$ [$\text{\:s}^{-1}$] & $\xi$ at $Q$ peak & Character of $Q$ peak & $T$ at $Q$ peak [K]\\ \hline 
\endfirsthead

\multicolumn{4}{c}%
{{\bfseries \tablename\ \thetable{} -- continued from previous page}} \\
\hline $P$ & $S$ & $\xi$ at $Q$ peak & Character of $Q$ peak & $T$ at $Q$ peak [K] \\ \hline 
\endhead

\hline \multicolumn{4}{|r|}{{Continued on next page}} \\ \hline
\endfoot

\hline \hline
\endlastfoot

 \multicolumn{1}{c|}{1} & \multicolumn{ 1}{|c|}{10} & -0.612     & Lean premixed flame with diffusive character& 1364.42 \\
\cline{3-5}
\multicolumn{ 1}{c|}{} & \multicolumn{ 1}{|c|}{} &  -0.340    &  Exothermic reactions (\textit{no flame})&2067.00\\
\cline{3-5}
\multicolumn{ 1}{c|}{} & \multicolumn{ 1}{|c|}{} &  -0.245    &  Diffusion flame & 2051.22\\
\cline{3-5}
\multicolumn{ 1}{c|}{} & \multicolumn{ 1}{|c|}{} &  -0.045    & Rich premixed flame with diffusive character&1606.51\\
\cline{2-5}
\multicolumn{ 1}{c|}{} & \multicolumn{ 1}{|c|}{50} &  -0.436    & Lean premixed flame with diffusive character&1380.28 \\
\cline{3-5}
\multicolumn{ 1}{c|}{} & \multicolumn{ 1}{|c|}{} &   -0.187   & Diffusion flame&2007.65\\
\cline{3-5}
\multicolumn{ 1}{c|}{} & \multicolumn{ 1}{|c|}{} &   -0.059   & Rich premixed flame with diffusive character&1670.89\\
\cline{2-5}
\multicolumn{ 1}{c|}{} & \multicolumn{ 1}{|c|}{100} &  -0.389    & Lean premixed flame with diffusive character& 1350.18\\
\cline{3-5}
\multicolumn{ 1}{c|}{} & \multicolumn{ 1}{|c|}{} &   -0.186  & Diffusion flame & 1979.44\\
\cline{2-5}
 \multicolumn{ 1}{c|}{} & \multicolumn{ 1}{|c|}{150} &  -0.361    & Lean premixed flame with diffusive character& 1369.71\\
\cline{3-5}
 \multicolumn{ 1}{c|}{} & \multicolumn{ 1}{|c|}{} &   -0.192   &Diffusion flame& 1960.90\\
\cline{1-5}
 \multicolumn{ 1}{c|}{10} & \multicolumn{ 1}{|c|}{10} &  -0.376    & Lean premixed flame with diffusive character& 1526.31\\
\cline{3-5}
 \multicolumn{ 1}{c|}{} & \multicolumn{ 1}{|c|}{} &  -0.230    & Diffusion flame &2255.97\\
\cline{3-5}
 \multicolumn{ 1}{c|}{} & \multicolumn{ 1}{|c|}{} &  0.000  &  Rich premixed flame with diffusive character&1520.22\\
\cline{2-5}
 \multicolumn{ 1}{c|}{} & \multicolumn{ 1}{|c|}{50} & -0.210   & Lean premixed flame with diffusive character &1593.89 \\
\cline{3-5}
 \multicolumn{ 1}{c|}{} & \multicolumn{ 1}{|c|}{} &  -0.135    &Diffusion flame & 2193.047\\
\cline{3-5}
\multicolumn{ 1}{c|}{} & \multicolumn{ 1}{|c|}{} &  -0.019   &  Rich premixed flame with diffusive character&1626.38\\
\cline{2-5}
 \multicolumn{ 1}{c|}{} & \multicolumn{ 1}{|c|}{100} & -0.177   &Lean premixed flame with diffusive character  & 1633.23\\
\cline{3-5}
 \multicolumn{ 1}{c|}{} & \multicolumn{ 1}{|c|}{} &  -0.126    & Diffusion flame& 2155.22\\
\cline{3-5}
\multicolumn{ 1}{c|}{} & \multicolumn{ 1}{|c|}{} &  -0.047  &  Rich premixed flame with diffusive character &1710.69 \\
\cline{2-5}
 \multicolumn{ 1}{c|}{} & \multicolumn{ 1}{|c|}{150} & -0.161   & Lean premixed flame with diffusive character&  1643.07\\
\cline{3-5}
 \multicolumn{ 1}{c|}{} & \multicolumn{ 1}{|c|}{} &  -0.120    &Diffusion flame & 2132.86\\
\cline{3-5}
\multicolumn{ 1}{c|}{} & \multicolumn{ 1}{|c|}{} &  -0.058   & Rich premixed flame with diffusive character & 1745.86\\
\cline{1-5}
\multicolumn{ 1}{c|}{20} & \multicolumn{ 1}{|c|}{10} &  -0.304  & Lean premixed flame with diffusive character&1519.78 \\
\cline{3-5}
\multicolumn{ 1}{c|}{} & \multicolumn{ 1}{|c|}{} &  -0.188    &Diffusion flame & 2283.75\\
\cline{3-5}
\multicolumn{ 1}{c|}{} & \multicolumn{ 1}{|c|}{} &  0.063    &  Rich premixed flame with diffusive character & 1134.40\\
\cline{2-5}
 \multicolumn{ 1}{c|}{} & \multicolumn{ 1}{|c|}{50} &  -0.160 &Lean premixed flame with diffusive character  & 1645.40\\
\cline{3-5}
\multicolumn{ 1}{c|}{} & \multicolumn{ 1}{|c|}{} &  -0.106    &Diffusion flame & 2232.40 \\
\cline{3-5}
\multicolumn{ 1}{c|}{} & \multicolumn{ 1}{|c|}{} &  -0.014   &  Rich premixed flame with diffusive character&1642.81 \\
\cline{2-5}
\multicolumn{ 1}{c|}{} & \multicolumn{ 1}{|c|}{100} &  -0.134 &Lean premixed flame with diffusive character & 1674.86\\
\cline{3-5}
\multicolumn{ 1}{c|}{} & \multicolumn{ 1}{|c|}{} &  -0.096    & Diffusion flame & 2206.48\\
\cline{3-5}
\multicolumn{ 1}{c|}{} & \multicolumn{ 1}{|c|}{} &  -0.032    & Rich premixed flame with diffusive character & 1664.28 \\
\cline{2-5}
 \multicolumn{ 1}{c|}{} & \multicolumn{ 1}{|c|}{150} &  -0.121    &Lean premixed flame with diffusive character &1698.14 \\
\cline{3-5}
\multicolumn{ 1}{c|}{} & \multicolumn{ 1}{|c|}{} &  -0.092    & Diffusion flame& 2182.94\\
\cline{3-5}
 \multicolumn{ 1}{c|}{} & \multicolumn{ 1}{|c|}{} &  -0.042  &  Rich premixed flame with diffusive character & 1713.78\\
\hline
\end{longtable}
\end{center}

%
%




\subsubsection{Strain Rate Effects}
At high strain rate ($100\:\text{s}^{-1}$ and $150\:\text{s}^{-1}$), the heat-release-rate peak of the flames increases from right to left, with the fuel rich premixed flame being the weakest of the three. This is also true at low strain rates. Note that the important factor is the integral of the reaction rate through the reaction zone rather than the peak value in terms of total heat release rate. Not many general flame behavioral differences were found between the high strain rates studied. However, the peaks of the flame are getting closer and further from the stagnation point, approaching the estimated mixing-layer edge. This behavior indicates a likely flame merging at still higher strain rates.\hfill \break



\section{Conclusions}
Flame merging has been observed numerically and experimentally for practical turbulent combustion situations. The novelty of the current numerical study is the in-depth analysis of the flame merging and separation processes entailing detailed chemistry and transport, and employing the counterflow canonical configuration commonly used in experimental studies. This is achieved by examining previously unreported mixture ratios in-flowing from each nozzle of the counterflow configuration.\hfill \break

This study describes the important roles that normal strain rate and pressure have with flames. An increase in the normal strain rate and/or a decrease in the pressure causes flames to merge while strain rate decreases and/or pressure increases lead to multiple flames.\hfill \break
Moreover, an endothermic region is observed for Case 1. This finding has to be further investigated. However, a preliminary explanation related to the production of ethylene (C$_2$H$_4$) is plausible.\hfill \break
At low strain rates and pressures, and only when a rich premixed mixture is injected, an extra heat-release-rate peak is obtained. After our investigation, this heat-release-rate peak is linked to high exothermic reactions producing CO$_2$ and H$_2$O, and consuming CO and H$_2$.\hfill \break
In the different multi-branched flames studied, unexpected character of the lean and rich premixed flames indicates that they are diffusion controlled rather than possessing a classical wave-like nature strongly dependent on chemical kinetic rates. \hfill \break

Further research should be conducted to characterize the extinction rate for the studied cases.

\section{Acknowledgements}
This work was supported by the Air Force Office of Scientific Research [FA9550-18-1-0392] with Dr. Mitat Birkan as the scientific officer.\hfill \break 
The authors would also like to thank Prof. Hai Wang (Stanford University) and Prof. Antonio L. Sanchez (UC San Diego) for the discussions and suggestions.

\bibliography{mybibfile}

\end{document}